# Hybrid nodal surface and nodal line phonons in solids

Wen-Han Dong,[1,2] Jinbo Pan,[1,2,3] Jia-Tao Sun,[4] and Shixuan Du[1,2,3,5*]

[1] *Beijing National Laboratory for Condensed Matter Physics and Institute of Physics, Chinese Academy of Sciences, Beijing 100190, China*

[2] *School of Physical Sciences, University of Chinese Academy of Sciences, Beijing 100049, China*

[3] *Songshan Lake Materials Laboratory, Dongguan 523808, China*

[4] *School of Integrated Circuits and Electronics, MIIT Key Laboratory for Low-Dimensional Quantum Structure and Devices, Beijing Institute of Technology, Beijing 100081, China*

[5] *CAS Center for Excellence in Topological Quantum Computation, Beijing 100190, China*

ORCiD: W.-H.D.: orcid.org/0000-0001-8356-276X; S.D.: orcid.org/0000-0001-9323-1307

*Correspondence to: sxdu@iphy.ac.cn

## ABSTRACT

Phonons have provided an ideal platform for a variety of intriguing physical states, such as non-Abelian braiding and the Haldane model. It is promising that phonons will realize the complicated nodal states accompanying unusual quantum phenomena. Here, we propose the hybrid nodal surface and nodal line (NS+NL) phonons beyond the single-genre nodal phonons. We categorize the NS+NL phonons into two- and four-band situations based on symmetry analysis and compatibility relationships. Combining database screening with first-principles calculations, we identify the ideal candidate materials for realizing all categorized NS+NL phonons. Our calculations and tight-binding models further demonstrate that the interplay between NS and NL induces unique phenomena. In space group (SG) 113, the quadratic NL acts as a hub of the Berry curvature between two NSs, generating ribbonlike surface states (SSs). In SG 128, the NS serves as the counterpart of the Weyl NL, in which NS-NL mixed topological SSs are observed. Our findings extend the scope of hybrid nodal states and enrich the phononic states in realistic materials.

## I. INTRODUCTION

Nodal semimetals have provided an attractive platform for exploring exotic quantum phenomena since their discovery in electronic systems [1-8]. These gapless states are generally classified as nodal points [9], nodal lines (NLs) [10] and nodal surfaces (NSs) [11,12] according to the degeneracy in their energy-momentum relationships. As compared with electrons, phonons have provided an ideal platform for realizing nodal semimetal states due to their spinless nature and no Fermi level limits, thus promoting the rising field of topological phononics [13,14]. Given the variety of space groups (SGs) and crystalline symmetries, multifarious symmetry-required nodal states have been discovered in phonons, for instance, multiband non-Abelian crossings [15,16], Weyl points [17-20], Dirac points [21], Weyl NL (WNL) [22], quadratic NLs (QNLs) [23], hybrid nodal rings [24], nodal nets, nodal chains and nodal boxes [25-27]. The class-II NS [12] phonons were also predicted, which are subdivided into 1NS, 2NS and 3NS according to the pairs of NS states [28-31]. Recently, Wang *et al.* [32] reviewed the rapidly expanding fields of NL and NS phonons with a systematic understanding of the classifications of NL phonons and their material realizations. Despite these advances of nodal phonons, however, few researchers have achieved hybrid phonons containing more than one genre of nodal points, NLs, or NSs so far. It remains elusive whether the hybrid systems host their unique physical phenomena.

Nodal states may exhibit distinctive surface states (SSs) regardless of whether they are topological. For example, the topologically trivial QNL is manifested by torus SSs that span over the surface Brillouin zone [33,34]. Unlike a nodal point or NL, the NS itself creates no SSs since its spatial codimension is zero [12]. One effective approach towards SSs of NS is to realize the chiral crystals with coexistent NS and Weyl points [35,36]. Guaranteed by the Nielsen-Ninomiya no-go theorem [37], the NS is topologically charged by Weyl points; hence, one can observe the surface-arc states connecting NSs and Weyl points [35,36]. It was also reported that the NS can possess chiral charges without an upper limit when wrapping multiple Weyl points [38]. Aside from the Weyl point, the NL is another choice to generate SSs of NSs. The NS and NL may coexist in many forms, such as 1NS linked by a straight WNL in the superprismane-carbon [39] and lanternlike phonons in $Li_6WN_4$ [40]. However, these studied systems revealed no SS of a NS but only the drumhead SSs from WNLs. It is of fundamental interest to answer whether NSs and NLs can interplay with each other and induce unique physical states, which is also crucial in understanding hybrid nodal states composed of nodal states with different classifications.

In this paper, we propose hybrid NS and NL (NS+NL) phonons, whose phonon dispersions simultaneously contain NS and NL within two bands. We address two main questions: one is the variety of NS+NL phonons, and the other is their distinctive SSs. Based on symmetry analysis and compatibility relationships, we first obtained a complete catalog of the NS+NL phonons in three-dimensional SGs. We subdivided them into two- and four-band situations according to the minimal bands required. Our database screening and first-principles calculations further evidenced that all categorized NS+NL phonons are feasible in realistic materials. Specifically, our calculations and spinless tight-binding (TB) models confirmed both QNLs and WNLs can induce SSs of NSs, reflecting the interplay between NSs and NLs. Our discoveries not only uncover the universality of NS+NL phonons in solids but also offer insight into hybrid nodal states.

## II. CATALOG AND MATERIAL SCREENING

As illustrated in Fig. 1(a), our approach toward NS+NL phonons was to search coexisting NLs in the 55 SGs with symmetry-enforced NSs [28-30]. We focused on NLs appearing along high-symmetric lines instead of nodal loops or accidentally degenerate NLs. We first inspected all possible irreducible representations of the 55 SGs at high-symmetric points and high-symmetric lines, then checked the compatibility relationships [41] of high-symmetric lines. These two steps ensured a complete consideration of the degeneracy and connectedness of phonon bands in the whole Brillouin zone, i.e., graph theory combinatorics [42]. We adopted the methods of Ref. [23] to determine whether each degenerate line is a WNL or QNL. Finally, we obtained a complete catalog of NS+NL phonons, where the two-band situations are listed in TABLE 1 and four-band situations are listed in Supplemental Material (SM) Tables S1-S3 [43]. Figure 1(b) shows the degenerate configurations of two-band NS+NL phonons. It notes that the NLs may constitute different nodal nets [48], such as the Weyl nodal net (WNN) in SG 128 and hybrid nodal net (HNN) composed of both a QNL and WNLs in SG 136. Hence the two-band NS+NL phonons take various forms of 1NS+WNN, 2NS+WNN, 2NS+HNN, etc. For four-band situations, we classified them by three different mechanisms, see SM Sec. IV [43] for details. Here, the NS+NL phonons are still identified by their two-band degeneracies, but the minimal bands required for hosting corresponding states, i.e., the dimensionality of elementary band representation [41], are four. Attractively, four-band situations can achieve NS+NL phonons beyond two-band ones, such as the 1NS+QNL+WNL phonons shown in Figs. S37-S41 in the SM [43].

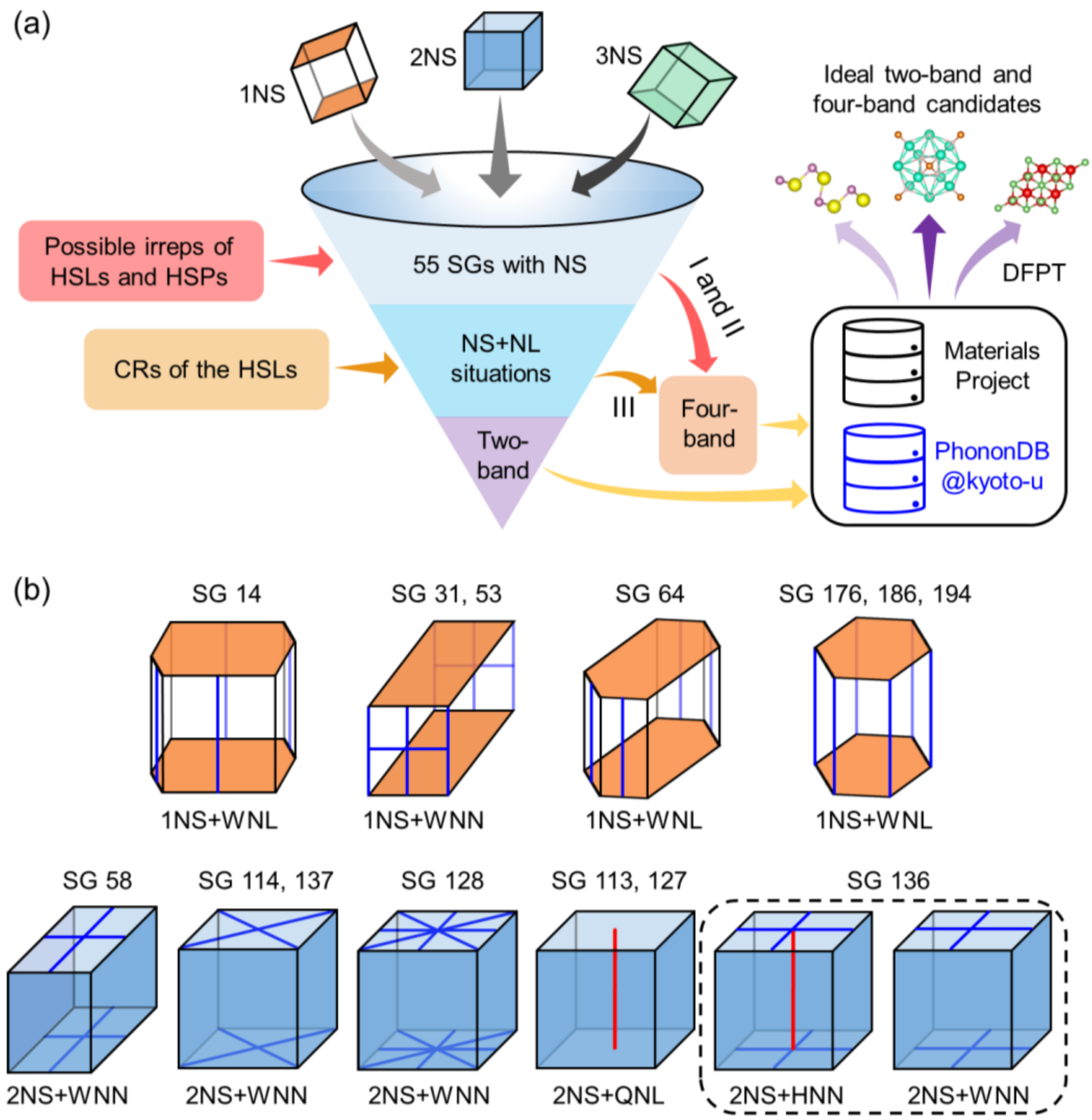

FIG 1. Catalog of nodal surface and nodal line hybrid (NS+NL) phonons. (a) Workflow for classifying NS+NL phonons and screening ideal candidates from material databases. Here, irrep is irreducible representation, HSL is high symmetric line, HSP is high symmetric point, SG is space group, CR is compatibility relationship. I, II and III stand for three mechanisms of four-band, see Supplemental Material [43] for details. (b) Classifications of two-band NS+NL phonons with their degenerate configurations in the Brillouin zones. The orange and blue surfaces stand for 1NS and 2NS, respectively. The blue and red lines are Weyl NL (WNL) and quadratic NL (QNL), respectively. WNN represents Weyl nodal net, HNN represents hybrid nodal net.

TABLE 1. Two-band situations of NS+NL phonons.#

| SG | Symbol | NS | WNL | QNL | Candidate |
|---|---|---|---|---|---|
| 14 | $P2_1$ | $NS_{ZCED}$ | BD $\{U_1^1 \oplus U_2^1\}$,<br>AE $\{V_1^1 \oplus V_2^1\}$ | | mp-4549 $SrSiN_2$ |
| 31* | $Pmn2_1$ | $NS_{ZURT}$ | XU $\{G_1^2\}$,<br>SR $\{Q_1^2\}$,<br>XS $\{D_1^1 \oplus D_2^1\}$ | | mp-755543 $HoHO_2$ |
| 53 | Pmna | $NS_{ZURT}$ | XU $\{G_1^2\}$,<br>SR $\{Q_1^2\}$,<br>XS $\{D_1^2\}$ | | mp-569052 $AgSe_3I$ |
| 58 | Pnnm | $NS_{UXS}$, $NS_{TYS}$ | ZU $\{A_1^2\}$,<br>ZT $\{B_1^2\}$ | | mp-19795 InS |
| 64 | Cmce | $NS_{ZRT}$ | SR $\{D_1^1 \oplus D_2^1\}$ | | mp-17667 $Sr_3AlSb_3$ |
| 113* | $P\text{-}42_1m$ | $2\times NS_{XRM}$ | | **ΓZ $\{LD_3^1 \oplus LD_4^1\}$** | $C_4N$ [52] |
| 114* | $P\text{-}42_1c$ | $2\times NS_{XRM}$ | ZA $\{S_1^1 \oplus S_2^1\}$ | | mp-27144 $TlPO_3$ |
| 127 | P4/mbm | $2\times NS_{XRM}$ | | **ΓZ $\{LD_5^2\}$** | mp-29138 $KAuSe_2$ |
| 128 | P4/mnc | $2\times NS_{XRM}$ | ZR $\{U_1^2\}$,<br>ZA $\{S_1^2\}$ | | mp-28247 $K_2PtI_6$ |
| 136 | $P4_2/mnm$ | $2\times NS_{XRM}$ | ZR $\{U_1^2\}$ | **ΓZ $\{LD_5^2\}$** | mp-3188 $ZnSb_2O_6$ |
| 136 | $P4_2/mnm$ | $2\times NS_{XRM}$ | ZR $\{U_1^2\}$ | | mp-662530 $KNiPS_4$ |
| 137 | $P4_2/nmc$ | $2\times NS_{XRM}$ | ZA $\{S_1^2\}$ | | mp-559639 $K_4UP_2O_{10}$ |
| 176 | $P6_3/m$ | $NS_{ALH}$ | **KH $\{P_2^1 \oplus P_3^1\}$** | | mp-27506 $K_3W_2Cl_9$ |
| 186* | $P6_3mc$ | $NS_{ALH}$ | **KH $\{P_3^2\}$** | | mp-561681 $CsSO_3$ |
| 194 | $P6_3/mmc$ | $NS_{ALH}$ | **KH $\{P_3^2\}$** | | mp-841 $Li_2O_2$ |

#The required SGs, positions of NS, irreducible representations of NLs and candidate materials with ideal two-band NS+NL phonons are given. Here, mp-xxx represents the Materials Project [49] ID. Notice the asterisks label non-centrosymmetric SGs. The NLs in bold lines depend on specific basis functions.

Next, we obtained candidate materials for realizing all categorized NS+NL phonons using high-throughput database screening and first-principles calculations. In detail, we first sought out nearly 200 materials hosting isolated phonon bands and desired SGs from over 10000 *ab initio* phonon records deposited in the Materials Project [49] and Phonon Database at Kyoto University (PhononDB@kyoto-u) [50]. We then checked the irreducible representations and compatibility relations of the database-downloaded phonon data of these materials and performed double-checks by redoing phonon calculations using density functional perturbation theory [51]. We eventually found most candidate materials after database screening and double-check. The remaining candidates were further identified in the literature and by performing phonon calculations of those materials from the Materials Project [49] that have no phonon data but host specific SGs. As listed in the last columns of TABLE 1 and Tables S1-S3 in the SM [43], almost all selected candidate materials have been experimentally synthesized and can be found in the Materials Project [49]. To highlight, these candidates all have ideal two-band or four-band NS+NL phonon dispersions within a specific frequency range. The two- NS+NL phonons of candidates listed in TABLE 1 are even robust against the longitudinal and transverse optical phonon splitting (LOTO). Collectively, these candidates evidence our proposed NS+NL phonons are feasible in realistic materials. The diversity of NS+NL phonons would induce rich physical states, which we will show in the examples below.

## III. TWO-BAND 2NS+QNL PHONON IN SG 113

When $\mathcal{P}$ is broken, nonzero Berry curvature can bridge two nodal states, such as the chirality-opposite Weyl points in TaAs [5]. In a similar way, if there only exists NS and QNL, the Berry curvature should connect them both. It motivated us to explore the 2NS+QNL phonon in noncentrosymmetric SG 113. As shown in Fig. 2(a), the studied material is a proposed carbonitride, $C_4N$ [52]. Here, $C_4N$ exhibits a diamond-like structure in SG 113, whose symmetry generators are twofold rotation $C_{2z}$, fourfold improper rotation $S_{4z} = \{C_{4z}\mathcal{P}|0,0,0\}$ and twofold screw rotation $S_{2y} = \{C_{2y}|\frac{1}{2},\frac{1}{2},0\}$. Our optimized lattice constants are $a = b = 3.50$ Å and $c = 4.88$ Å. The unit cell of $C_4N$ possesses two $sp^2$ C, six $sp^3$ C and two N atoms. We adopted the unit cell termination with $sp^2$ C and N atoms.

Figure 2(b) shows the phonon dispersion of $C_4N$, which is dynamically stable. Here, we focus on bands 27,28 with ideal 2NS+QNL dispersions [see Fig. 2(c)]. We distinguish the 2NS as NS1 at $k_x = \pi$ plane and NS2 at $k_y = \pi$ plane, as illustrated in Fig. 2(e). NS1 and NS2 are Kramer's degenerate under the joint operation of $S_{2x,y}$ and time-reversal $\mathcal{T}$, where $\mathcal{T}^2 = 1$ and $(S_{2x,y}\mathcal{T})^2 = e^{ik_{x,y}} = -1$ in the $k_{x,y} = \pi$ planes. In SM Sec. III [43], our $k \cdot p$ analysis and band expansions reveal that NS1 and NS2 are different from the free Weyl NS proposed in Ref. [53]. They have a linear dispersion near the central parts around the X-R path but a quadratic dispersion near the M-A path due to the intersection of two NSs. We then turn to the QNL along the Γ-Z path, whose twofold degeneracy is ensured by an irreducible representation of $LD_3 \oplus LD_4$. We have calculated the Berry phase along a closed loop surrounding the QNL: $\gamma_{\mathrm{L}} = \sum_{n\in \mathrm{occ}} \oint A_n(\mathbf{k}) \cdot d\mathbf{k} \bmod 2\pi$, where $A_n(\mathbf{k}) = i\langle u_n(\mathbf{k})|\nabla_{\mathbf{k}}|u_n(\mathbf{k})\rangle$ is the Berry connection of the $n$th band. The obtained Berry phase of zero indicates the QNL is topologically trivial. Additionally, our $k \cdot p$ analysis confirms the quadratic effective Hamiltonian near QNL, and the QNL is protected by $S_{4z}\mathcal{T}$ [43]. In Fig. S4 in the SM [43], we demonstrate the elementary band representation of bands 27,28 of $C_4N$ is B1@2$c$ or B2@2$c$. With this knowledge, we further constructed a two-band TB model using the $p_{xy}$ and $p_{-xy}$ orbitals locating at $2c$ Wyckoff positions, see Eqs. S2-S4 and Fig. S1 in the SM [43] for details. Figure 2(d) shows our TB model captures well the dispersion feature of 2NS+QNL phonon in $C_4N$.

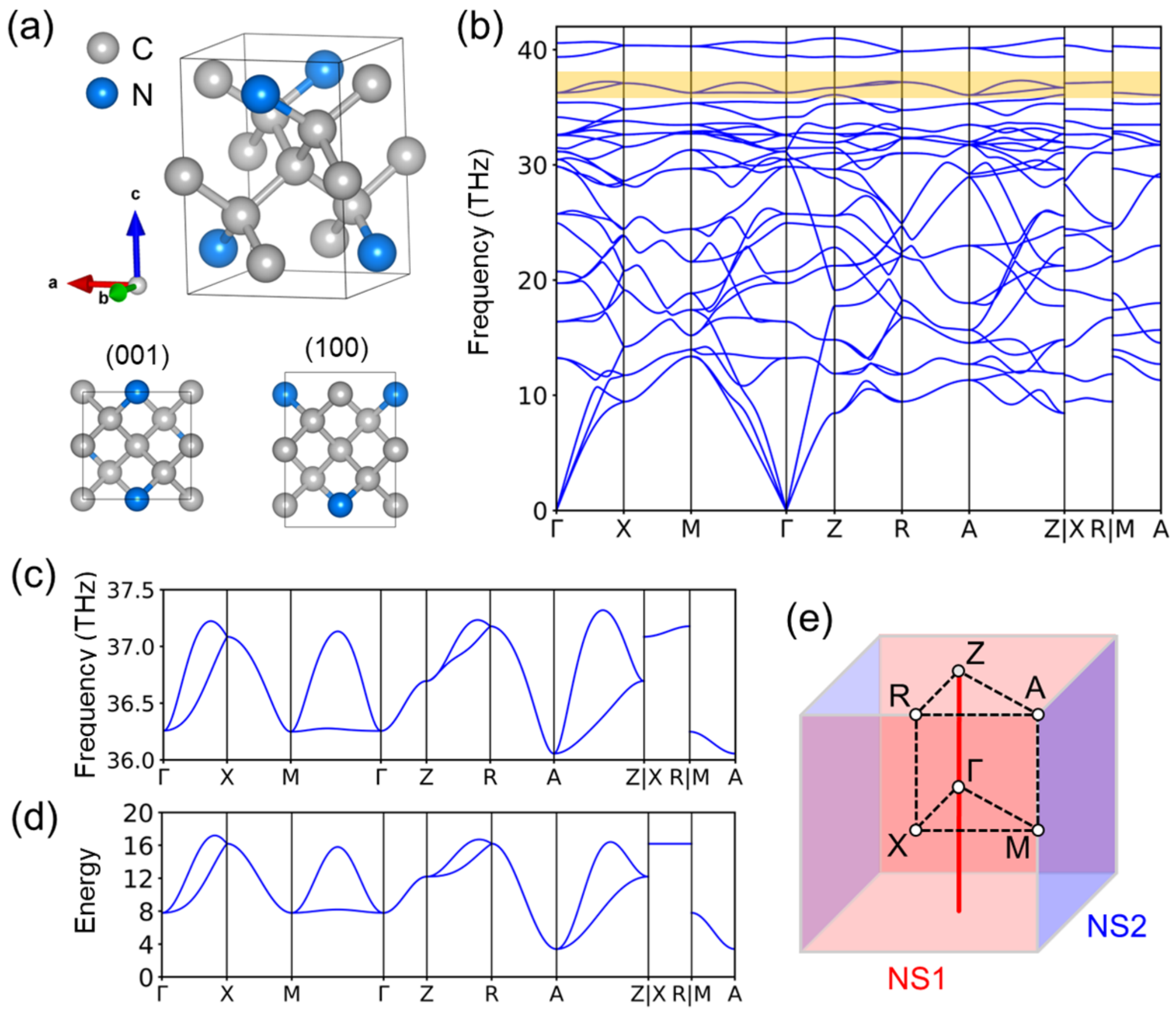

FIG 2. The 2NS+QNL phonon in SG 113 $C_4N$. (a) Crystal structure of diamond-like $C_4N$. (b) Phonon dispersions of $C_4N$ with LOTO correction. (c) Phonon bands 27,28, as marked yellow in (b). (d) Phonon dispersions of the two-band tight-binding (TB) model of SG113. The TB parameters are $\Delta = 12.0, t_1 = 0.55, t_{2\sigma} = -2.0, t_{2\pi} = -0.1, t_3 = 0.0, t_4 = -0.275, t_5 = -0.3,$ and $t_6 = -0.1$ in units of $t_0$, and $z = 0.1$. (e) Schematic of the Brillouin zone and the 2NS+QNL phonons. The 2NS are divided into NS1 at $k_x = \pi$ and NS2 at $k_y = \pi$.

Physically, the QNL is like a charge-neutral metal rod, and two NSs are analogous to metal plate capacitors. What will happen if one puts the QNL between two NSs? In Fig. 3, we reveal the Berry-curvature-driven SSs of the 2NS+QNL phonon. First, Fig. 3(a) shows the Berry curvature $\Omega(\mathbf{k})$ of band 27 of $C_4N$, which is almost confined in the $k_x$-$k_y$ plane. Since $\mathcal{P}$ is broken and $\mathcal{T}$ is reserved, $\Omega(\mathbf{k})$ is nonzero everywhere. Here, $S_{4z}$ guarantees NS1 and NS2 carry opposite Berry flux, so that NS1 acts like a source and NS2 acts like a sink. We estimated the Berry flux of $1.8\pi$/-$1.8\pi$ for NS1/NS2 by integrating the Berry curvature passing through left and right sides of NS1 (NS2). Interestingly, though the QNL carries a net flux of zero, it absorbs the Berry flux from the central part of NS1 and emits it to NS2. This hublike behavior of QNL blocks the direct exchange of the Berry flux between the central parts of NS1 and NS2, generating a cross-like barrier along [110] and $[\bar{1}10]$ directions near the Γ-Z path. In Fig. 3(b), our TB model reproduces the unique NS1-QNL-NS2 distribution of the Berry curvature. The minor difference between Figs. 3(a) and 3(b) may come from TB model simplification.

Next, we investigate the SSs induced by Berry curvature. Since NS1 and NS2 resemble a pair of source and sink, such SSs should appear in between the projections of two NSs. Figure 3(d) shows the (001) surface spectra of $C_4N$ along the NS1-to-NS2 paths, as arrowed in Fig. 3(c). One can find the torus SSs of QNL [33] emerge between 36.0 and 36.7 THz (see also Fig. S7 in the SM [43]). Our concerned SSs locate between the bulk gaps, consisting of two sections, i.e., sec1 and sec2. To distinguish, we term them as ribbonlike SSs and will explain below. Figure 3(e) shows the iso-frequency (001) surface contour of $C_4N$ at 37.12 THz. The ribbonlike SSs start from NS1, are then bridged by intermediate bulk states, and finally end at NS2. Figure 3(g) illustrates that the SSs are driven by nonzero the Berry flux of NS1 and NS2. The SSs follow the direction of the Berry curvature from NS1 to NS2. Notice each SS is separated by the intermediate bulk states into two sections, possibly because Berry curvature cannot directly connect the central parts of NS1 and NS2. We discuss in Fig. S6 in the SM [43] that the intermediate bulk states are crucial in stabilizing the SSs because NS1 and NS2 are connected without a global gap. In Fig. S6 in the SM [43], we also show the ribbonlike SSs appear in a wide range between 36.88 and 37.14 THz. With this knowledge, we illustrate that the SSs under different frequencies (momenta) collectively occupy the surface Brillouin zone like ribbons, i.e., these SSs are ribbonlike. We remark that the ribbonlike SSs do not come from topological charge of NS; instead, they are driven by the Berry curvature between 2NS and QNL. The QNL is essential in creating the indirectly connected SSs between 2NSs because the 2NS have a codimension of zero, and the zero-dimensional charge in bulk generates no SS itself [12,54]. We verify this point in Fig. S8 [43] that the SS will disappear if there only exist 2NS without a QNL. To further confirm the above results of $C_4N$, we have also obtained the ribbonlike SSs with two sections using the TB model, as shown in Fig. 3(f).

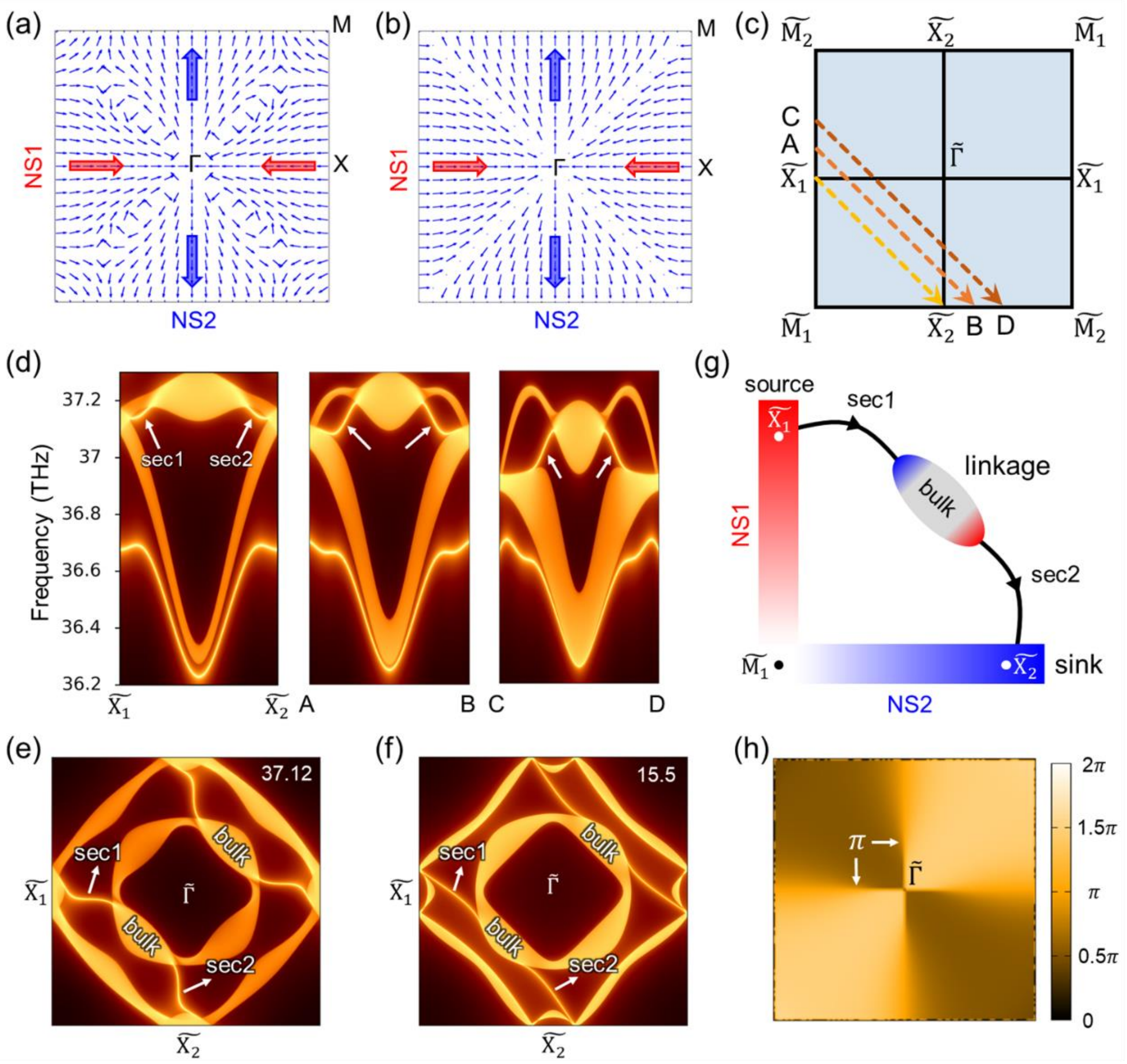

FIG 3. Ribbonlike surface states (SSs) of the 2NS+QNL phonons. (a) In-plane Berry curvature $\Omega_{x,y}(\mathbf{k})$ of band 27 of $C_4N$ at $k_z = 0$. (b) In-plane Berry curvature $\Omega_{x,y}(\mathbf{k})$ of the tight-binding (TB) model at $k_z = 0$. (c) Schematic of the (001) surface Brillouin zone. (d) Surface spectra of $C_4N$ along the paths labeled in (c). The arrows point to the two sections of SSs, namely sec1 and sec2, which are stabilized by intermediate bulk states. (e) The isofrequency (001) surface contour of $C_4N$ at 37.12 THz. (f) The iso-frequency (001) surface contour of TB model at E = $15.5t_0$. (g) Schematic of the ribbonlike SSs at a fixed frequency, where red (blue) represents positive (negative) Berry flux. If one considers the SSs at different frequencies collectively, they constitute the ribbon. (h) Calculated Zak phase $\mathcal{Z}(k_x, k_y)$ up to band 27 of $C_4N$.

Lastly, we revisit the ribbonlike SSs from the perspective of the Zak phase. The Zak phase of (001) surface is defined as the one-dimensional Berry phase along the $k_z$ axis [55] $\mathcal{Z}(k_x, k_y) = \sum_{n\in\text{occ}} \int_{-\pi}^{\pi} A_n^z(\mathbf{k}) dk_z$. Figure 3(h) shows the calculated Zak phase up to band 27 of $C_4N$. Notice $S_{4z}$ and glide mirrors $\widetilde{\mathcal{M}}_{\pm xy} = \{\mathcal{M}_{\pm xy}|\frac{1}{2},\frac{1}{2},0\}$ ensure: $\mathcal{Z}(k_x, k_y) = -\mathcal{Z}(-k_y, k_x) = \mathcal{Z}(k_y, k_x) = \mathcal{Z}(-k_y, -k_x) \mod 2\pi$, which gives the 0 or $\pi$ quantized Zak phase along $k_x = 0$ and $k_y = 0$. For $C_4N$ in Fig. 3(h), one can clearly see the Zak phase is nonzero in the whole surface Brillouin zone. In modern theory of electrical polarization, the nonzero Zak phase reveals accumulative surface polarization charge from the bulk [56]. Given the considerable Zak phase value between $0.45\pi$-$1.55\pi$, it indicates a wide distribution of SSs in (001) surface of $C_4N$ [57]. The $\pi$-quantized Zak phase along $k_x = 0$ and $k_y = 0$ helps explain why the stable SSs incline to appear around $\tilde{\Gamma}$-$\widetilde{X_1}$ and $\tilde{\Gamma}$-$\widetilde{X_2}$ paths. Moreover, we show the relation between Zak phase and Berry curvature: $\mathcal{Z}(k_x, k_y) - \mathcal{Z}(k_x', k_y') = \sum_{n\in\text{occ}} \oint A_n(\mathbf{k}) \cdot d\mathbf{C} = \sum_{n\in\text{occ}} \oiint \Omega(\mathbf{k}) \cdot d\mathbf{S} \neq 0 \mod 2\pi$, where $(k_x, k_y)$ and $(k_x', k_y')$ are arbitrary points, and $\mathbf{S}$ is the closed area with boundary $\mathbf{C}$: $(k_x, k_y, \pi) \rightarrow (k_x', k_y', \pi) \rightarrow (k_x', k_y', -\pi) \rightarrow (k_x, k_y, -\pi) \rightarrow (k_x, k_y, \pi)$. In this sense, the Berry flux from $\mathcal{P}$ breaking creates the considerable Zak phase in the whole surface Brillouin. The $\pi$-quantized Zak phase locally protects the ribbonlike SSs around $k_x = 0$ and $k_y = 0$. This Berry-curvature-based physics is quite different from that of Weyl points, revealing the uniqueness of ribbonlike SSs between NS and QNL.

## IV. TWO-BAND 2NS+WNN PHONON IN SG 128

In addition to generating SSs between NSs and QNLs, NSs may appear as the counterpart of WNLs and give rise to nontrivial SSs. Here we choose the example of a 2NS+WNN phonon in SG 128 to verify this point. The studied material is $K_2PtI_6$, which was synthesized using solution methods [58]. Here, $K_2PtI_6$ is crystallized in SG 128, whose structure is shown in Fig. 4(a). The symmetry generators are fourfold rotation $C_{4z}$, twofold screw rotation $S_{2y} = \{C_{2y}|\frac{1}{2},\frac{1}{2},\frac{1}{2}\}$, and $\mathcal{P}$. Our optimized lattice constants are $a = b = 7.73$ Å and $c = 12.13$ Å. Figure 4(b) shows the phonon dispersion of $K_2PtI_6$ with LOTO correction, within which bands 43,44, bands 45,46 and bands 47, 48 exhibit ideal 2NS+WNN phonons. It is noted that any isolated two bands can host such 2NS+WNN phonon since it is symmetry-enforced.

We take bands 43,44 of $K_2PtI_6$ for investigating the 2NS+WNN phonon [see Fig. 4(c)]. Figure 4(d) displays that bands 43,44 have 2NS in the $k_{x,y} = \pi$ planes and WNN in the $k_z = \pi$ plane. From previous discussions, the two NSs are enforced by $S_{2x,y}\mathcal{T}$. One cannot directly define the topological charge of NS due to its connection with the WNN. The WNN is formed by two WNLs along Z-A paths and two WNLs along Z-R paths. The WNLs along Z-R paths are protected by the anticommutation relations: $\{\mathcal{M}_z, \widetilde{\mathcal{M}}_{x,y}\} = 0$ and $\{\mathcal{M}_z, S_{2x,y}\} = 0$, where $\mathcal{M}_z$ is the perpendicular mirror and $\widetilde{\mathcal{M}}_{x,y} = \{\mathcal{M}_{x,y}|\frac{1}{2},\frac{1}{2},\frac{1}{2}\}$ are glide mirrors. The WNLs along Z-A paths are protected by: $\{\mathcal{M}_z, \widetilde{\mathcal{M}}_{\pm xy}\} = 0$ and $\{\mathcal{M}_z, S_{2(\mp xy)}\} = 0$, where $\widetilde{\mathcal{M}}_{\pm xy} = \{\mathcal{M}_{\pm xy}|\frac{1}{2},\frac{1}{2},\frac{1}{2}\}$ and $S_{2(\mp xy)} = \{C_{2(\mp xy)}|\frac{1}{2},\frac{1}{2},\frac{1}{2}\}$. Additionally, we have determined the topological charge of these WNLs by calculating the Berry phases $\gamma_{\mathrm{L}}$ using closed loops away from the Z point and 2NS. We obtained the nontrivial $\gamma_{\mathrm{L}} = \pi$, which is quantized by $\mathcal{M}_z$.

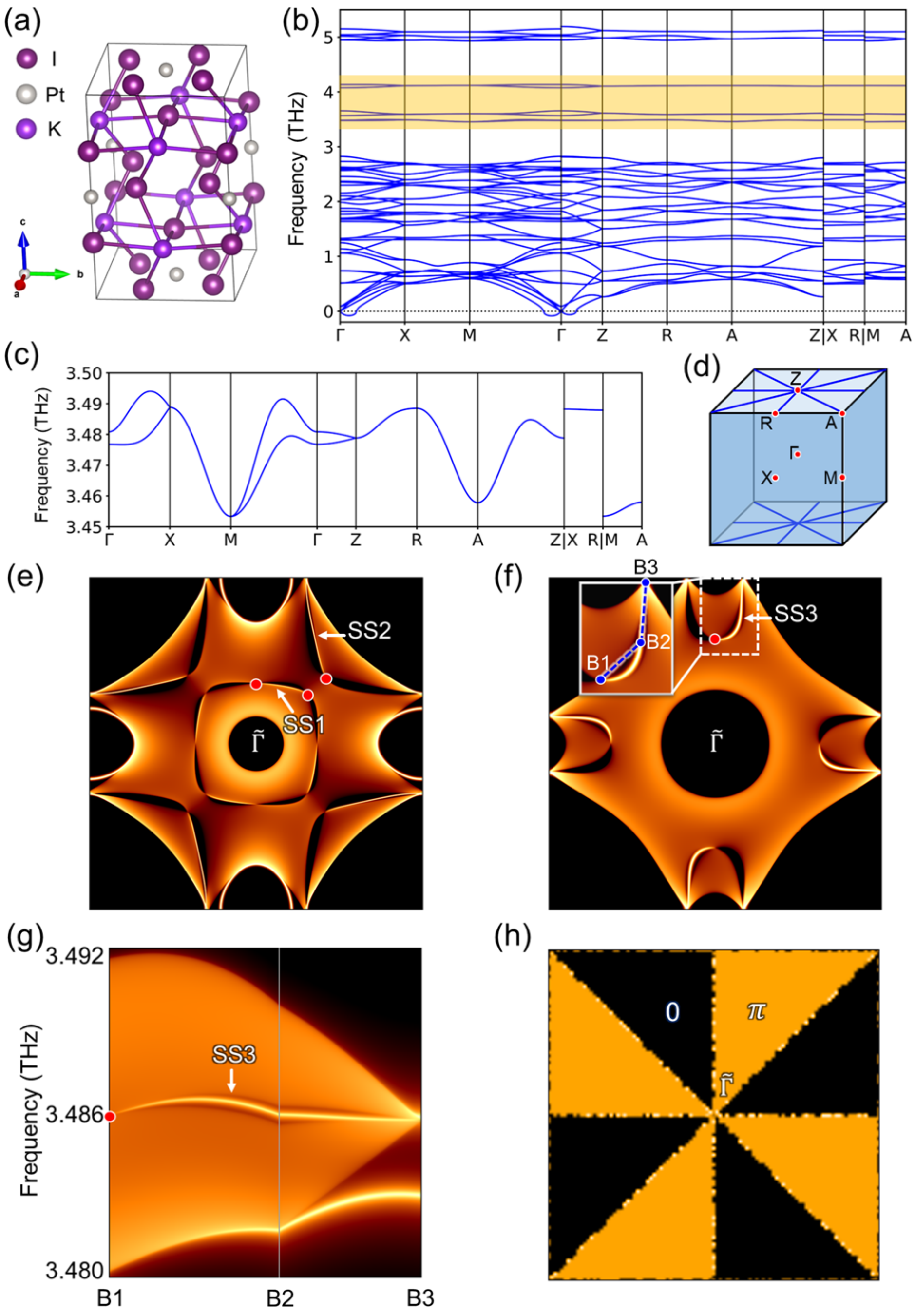

FIG 4. The 2NS+WNN phonons in SG 128 $K_2PtI_6$. (a) Crystal structure of $K_2PtI_6$ (mp-28247). (b) Phonon dispersions of $K_2PtI_6$ with LOTO correction. Bands 43,44, bands 45,46 and bands 47,48 are marked by yellow, which respectively exhibits ideal two-band 2NS+WNN phonons. (c) Zoom-in view of bands 43,44. (d) Schematic of the Brillouin zone and the 2NS+WNN phonon of bands 43,44. The iso-frequency contours of (001) surface at (e) 3.4822 THz and (f) 3.486 THz. SS1, SS2, and SS3 are topological SSs from either WNL to WNL or NS to WNL. (g) (001) surface spectrum of $K_2PtI_6$ along projections of WNL to NS. The path is shown in inset of (f). (h) Calculated Zak phase $\mathcal{Z}(k_x, k_y)$ of (001) surface up to band 43.

Figures 4(e) and 4(f) exhibit the SSs of $K_2PtI_6$ at different frequencies. We find three kinds of topological SSs, i.e., SS1 from WNL of the Z-A path to WNL of Z-R path, SS2 from WNL of the Z-A path to one NS and SS3 from WNL of Z-R path to one NS. Figure 4(g) shows the (001) surface spectrum that SS3 starts from a WNL and ends into NS projections. To our knowledge, SS2 and SS3 are the first discovered topological SSs between NS and WNL. SS2 and SS3 are quite different from SS1 in that their terminations near the NS projections are not simple points. Considering the NS cannot generate topological charge by itself, it means such NS-WNL SSs have their unique formation mechanism, which we will discuss below. Then, we demonstrate the topological nature of SS1, SS2 and SS3. Figure 4(h) displays the (001) Zak phase distribution up to band 43 of $K_2PtI_6$. The Zak phase undergoes a jump of $\pi$ once it crosses the projection of WNL. Here, $\mathcal{P}$ ensures that $\mathcal{Z}(k_x, k_y)$ is quantized to either 0 or $\pi$. Eventually, the $\mathcal{Z}(k_x, k_y) = \pi$ and $\mathcal{Z}(k_x, k_y) = 0$ regions each occupy half of the (001) surface. Here, SS1, SS2, and SS3 are topologically nontrivial because they appear in the $\mathcal{Z}(k_x, k_y) = \pi$ regions. Furthermore, we have constructed a two-band TB model to describe the 2NS+WNN phonon in SG 128, see Eqs. S7-S10 in the SM [43] for details. In Fig. S2 of the SM [43], we reproduce the WNL-WNL SS1 and NS-WNL mixed SS2, SS3 using our TB model, which is consistent with the results of $K_2PtI_6$.

## V. ORIGIN OF NS-WNL MIXED TOPOLOGICAL SS

Most importantly, we discuss about the formation mechanism of NS-WNL mixed topological SSs. Figure 5 shows the schematics of Zak phases of all possible two-band NS+WNL phonons in centrosymmetric SGs. The Zak phase is defined in (001) surface for 2NS+WNN phonons and (100) surface for 1NS+WNL or 1NS+WNN phonons. Notice the (100) Zak phase $\mathcal{Z}(k_y, k_z)$ is integrated along the $k_x$ axis. Due to $\mathcal{P}$, the Zak phase is quantized to 0 or $\pi$. We define $\mathcal{N}$ as the number of times to pass through the projection of WNL when travelling from one side of NS to the other. We begin with aforementioned 2NS+WNN phonon in SG 128. Figure 5(a) shows the Zak phase must have a jump of $\pi$ when it crosses the projection of the NS at the surface Brillouin zone boundaries. It is because the Zak phase has already been altered by $\pi$ three times due to $\mathcal{N}$ = 3. In this sense, the NS plays the same role as the WNL in the (001) surface and exists as the counterpart of WNL. The NS at $k_x = \pi$ ($k_y = \pi$) equivalently carries a Berry phase of $\pi$. Hence, the NS-WNL mixed SSs are topologically protected in SG 128, as observed in $K_2PtI_6$. Figure 5(b) shows such a physical image is also applicable for $\mathcal{N}$ = 1 cases, e.g., the 1NS+WNN phonon in SG 53 and 2NS+WNN phonons in SG 58 and SG 136. Taking SG 53 as an example, we have investigated the 1NS+WNN phonon using both a TB model and realistic material $AgSe_3I$ [59]. The results evidence the existence of NS-WNL mixed SSs, as well as WNL-WNL SSs (see Figs. S3, S11, and S12 in the SM [43]). In addition, we demonstrate the $\mathcal{N}$ = 2 and 0 cases cannot host NS-WNL mixed SSs because the NS equivalently carries no topological charge [see Figs. 5(c) and 5(d)]. This explains why previous researchers exploring, e.g., superprismane-carbon in SG 194 ($\mathcal{N}$ = 0) [39] and $Li_6WN_4$ in SG 137 ($\mathcal{N}$ = 2) [40], only found topological SSs between WNLs. In our example of $Li_2O_2$ in SG 194 ($\mathcal{N}$ = 0), we illustrate the NS cannot form SSs, and there only exist drumhead SSs between WNLs (see Fig. S25 in the SM [43]). In short, all these demonstrate the topological nature of the NS is determined by the arrangements of WNLs. The NS-WNL mixed SSs appear when $\mathcal{N}$ is odd.

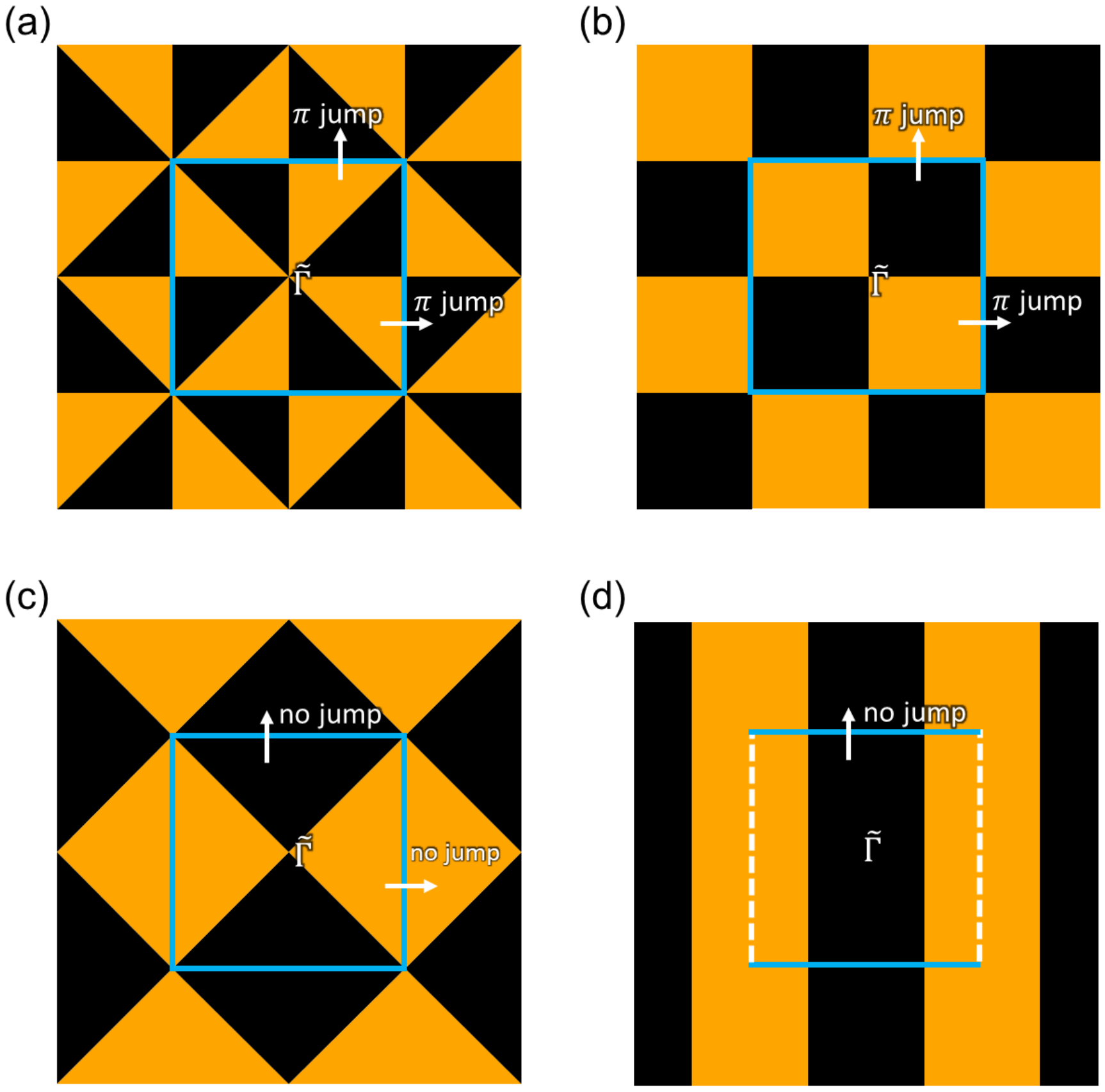

FIG 5. Determining the NS-WNL mixed topological SSs in centrosymmetric SGs by Zak phase. (a) (001) Zak phase of 2NS+WNN phonon in SG 128. (b) (001) Zak phase of 2NS+WNN phonons in SG 58 and SG 136. The (100) Zak phase of 1NS+WNN phonon in SG 53 is like (b). (c) (001) Zak phase of 2NS+WNN phonon in SG 137. (d) (100) Zak phase of 1NS+WNL phonons in SG 14, SG 64, SG 176, and SG 194. Here, the cyan boxes (lines) represent projections of NS. The gold and black regions are separated from either other by the projection of WNL or NS, exhibiting quantized Zak phase of 0 and $\pi$, respectively.

For better understanding the interplay between NSs and WNLs, we provide a more intrinsic viewpoint that the NS-WNL interaction depends on whether the NS is physically equivalent to the WNL in subspace, in other words, whether the NS can degenerate into a WNL under certain perturbations. For the cases with NS-WNL interaction, e.g., the 2NS+WNN phonon of SG 128 ($\mathcal{N}$ = 3), supposing the perturbations with breaking twofold screw rotations, one can find the NS degenerates into the WNL locating at the A-R path since the anticommutation relation $\{\mathcal{M}_z, \widetilde{\mathcal{M}}_{x,y}\} = 0$ is still reserved. It means the NS is equivalent to the WNL in subspace. Therefore, the interaction between NSs and WNLs are allowed in these cases. For the cases without NS-WNL interaction, e.g., the 2NS+WNN phonon of SG 137 ($\mathcal{N}$ = 2), one can check the NS is completely gapped under the perturbation that breaks twofold screw rotations. It means the NS is not equivalent to any WNL in subspace, so that the NS-WNL interaction is not allowed in these cases.

## VI. FOUR-BAND NS+NL PHONONS

In SM Sec. IV [43], we discuss the four-band NS+NL phonons as categorized in Fig. 1(a). The four-bands are caused by three mechanisms: (I) symmetry-enforced fourfold Dirac points or Dirac NLs, (II) Dirac points induced by paired NLs, and (III) hourglass phonons caused by the violation of compatibility relationship. In four-band situations, the NS+NL phonons appear in either upper or lower two bands of the four bands, i.e., the NS+NL phonons are valid at 1/4 filling or 3/4 filling of the four bands. Additionally, the Dirac points, Dirac NLs, or hourglass phonons are at 1/2 filling. We show the material candidates in Figs. S26-S44 in the SM [43]. For example, in Fig. S39 of the SM [43], $Ta_3SeI_7$ of SG 186 [60] hosts a pair of 1NS+QNL+WNL phonons in bands 61,62 and bands 63,64. The paired QNLs further induce a fourfold Dirac point at the A point (mechanism II), exhibiting double surface arcs at the (100) surface. In brief, these four-band cases not only enrich the genres of NS+NL phonons, but also intertwine NS+NL phonons with Dirac points [21,61], Dirac NLs [62], and hourglass phonons [63-65], which are promising for realizing much more intriguing nodal states.

## VII. DISCUSSION AND CONCLUSIONS

Before closing, we would like to make some important remarks on our work. First, our symmetry analysis and candidate materials reveal the discovered NS+NL phonons are ubiquitous in solids. If satisfying the symmetries, these NS+NL states are obtainable in other spinless systems, such as acoustic metamaterials and photonic lattices. It means, in this paper, we offer broad opportunities to investigate the hybrid nodal states and their SSs. Second, we reveal the distinctive interplay/interaction between NSs and NLs. For NS and QNL, their interplay is reflected by the Berry-curvature-driven ribbonlike SSs. For NS and WNL, their interplay gives rise to the NS-WNL mixed topological SSs. These findings go beyond the understandings of single-genre nodal points, NLs, or NSs, thus distinguish our work from recent high-throughput studies on topological/nodal phonons [66,67]. Third, NS+NL phonons may have interesting physical effects and practical applications. For bulk NS+NL phonons, they can exhibit excellent phonon transport properties due to the high degeneracy. In a wide frequency range near the NS or NL, these gapless phonons can greatly increase the scattering centers and strengthen the phonon-phonon scattering processes [68], which reduces the lattice thermal conductivity ($\kappa_{\mathrm{L}}$) by lowering phonon relaxation time [69]. The low $\kappa_{\mathrm{L}}$ means materials hosting NS+NL phonons are promising for high-performance thermoelectric applications. For surface NS+NL phonons, they have large density of states at boundaries, which may enhance the electron-phonon-interaction-related phenomena [70,71], such as interfacial high-$T_{\mathrm{C}}$ superconductivity [72]. Particularly, the NS-related SSs have open constant-frequency contours, which are desirable for realizing negative refraction at interfaces [73]. For $\mathcal{P}$-breaking systems, their SSs are also applicable for valley-based interfacial phononic devices, such as waveguides [74] and antennas [75]. Overall, the various NS+NL phonons and their SSs offer an attractive platform for future research and applications of nodal phonons.

Despite the broad prospects, one should notice it remains challenging to experimentally observe these NS+NL phonons. First, it requires high-resolution techniques to detect bulk NS+NL phonons. Authors of previous studies have applied inelastic x-ray scattering to detect nodal phonons in FeSi and $MoB_2$ [76,77]. This meV-resolved technique is not suitable for detecting NS+NL phonons if their band degeneracies appear within a narrow frequency range. Instead, methods like ultrasensitive infrared spectroscopy [78] and high-resolution resonant impulsive stimulated Raman spectroscopy [79] should be much more effective. Second, the intrinsic SSs of NS+NL phonons may be buried in bulk states or exhibit weak transport signals due to surface defects and other material-specific trivial states. This issue can be overcome by fabricating material samples with high-quality surfaces or constructing artificial acoustic lattices as an alternative. We suggest the atomic-resolved techniques like phonon laser [80] and electron energy loss spectroscopy [81] to be applied for detecting the SSs. We expect these discussions can promote the experimental realizations of NS+NL phonons in the near future.

In summary, we propose the NS+NL phonons in three-dimensional SGs and provide a complete classification based on symmetry analysis and compatibility relationships. We identify the candidate materials with ideal two- and four-band NS+NL phonons through database screening and first-principles calculations. Particularly, the two-band candidates are robust against LOTO. Our calculations and spinless TB models together confirm the unique interactions between NSs and NLs, including ribbonlike SSs of the 2NS+QNL phonon in SG 113, and NS-WNL mixed topological SSs in SGs represented by SG 128. Our results not only theoretically extend the scope of hybrid nodal states with a deeper understanding, but also offer plentiful realistic materials for experimental applications.

## ACKNOWLEDGMENTS

We thank Dr. Zhongjia Chen, Prof. Hongming Weng and Prof. Feng Liu for their helpful discussions. This work was supported by the National Key Research and Development Program of China (Grant Nos. 2020YFA0308800), the National Natural Science Foundation of China (Grant No. 11974045, 61888102), Chinese Academy of Sciences (Grant Nos. XDB30000000). We acknowledge computational support from the National Superconducting Center in Guangzhou.

Supplemental Material for

# "Hybrid nodal surface and nodal line phonons in solids"

Wen-Han Dong,[1,2] Jinbo Pan,[1,2,3] Jia-Tao Sun,[4] and Shixuan Du[1,2,3,5*]

[1] *Beijing National Laboratory for Condensed Matter Physics and Institute of Physics, Chinese Academy of Sciences, Beijing 100190, China*

[2] *School of Physical Sciences, University of Chinese Academy of Sciences, Beijing 100049, China*

[3] *Songshan Lake Materials Laboratory, Dongguan 523808, China*

[4] *School of Integrated Circuits and Electronics, MIIT Key Laboratory for Low-Dimensional Quantum Structure and Devices, Beijing Institute of Technology, Beijing 100081, China*

[5] *CAS Center for Excellence in Topological Quantum Computation, Beijing 100190, China*

*Correspondence to: sxdu@iphy.ac.cn

## Abbreviations

TB: tight binding; NS: nodal surface; NL: nodal line; NS+NL: hybrid nodal surface and nodal line; QNL: quadratic nodal line; WNL: Weyl nodal line; WNN: Weyl nodal net; HNN: hybrid nodal net; DP: Dirac point; DNL: Dirac nodal line; SS: surface state; CR: compatibility relationship; SG: space group; LOTO: longitudinal and transverse optical phonon splitting

## This file contains:

# I. Computational Methods

Based on our two-band and four-band classifications (see Table 1 and Tables S1-3), we have identified most of the candidate materials for NS+NL phonons from two databases, PhononDB@kyoto-u and Materials Project. We performed first-principles calculations via Vienna ab initio simulation package [1] with Perdew-Bruke-Ernzerhof exchange-correlation functional under generalized gradient approximation [2]. We employed a uniform plane-wave energy cutoff of 500 eV, and the *k*-mesh size were chosen by the criterion that $k_i a_i > 40$ Å ($i$ = 1, 2, 3). The convergence conditions were $10^{-8}$ eV for total energy and 0.001 eV/Å for structural relaxations. The Born effective charge for LOTO corrections were selectively calculated for systems with less than 40 atoms per unit cell. The phonon dispersions were obtained by PHONOPY package [3] within density functional perturbation theory, with supercell size larger than 10 Å along each direction. We generated the phononic TB Hamiltonian by phonopyTB code, as implemented in WannierTools package [4]. The SSs were calculated by the iterative Green's function technique [5].

## II. Spinless Two-Band TB Models

### 1. SG 113 with 2NS+QNL phonon

Generally, the two-band phonons are described by a spinless 2 × 2 Hamiltonian, with $p_x/p_y/p_z$ orbitals as orbital basis:

$$H(k_x, k_y, k_z) = \begin{bmatrix} H11 & H12 \\ H21 & H22 \end{bmatrix}, \ H21 = H12^*. \qquad \text{(S1)}$$

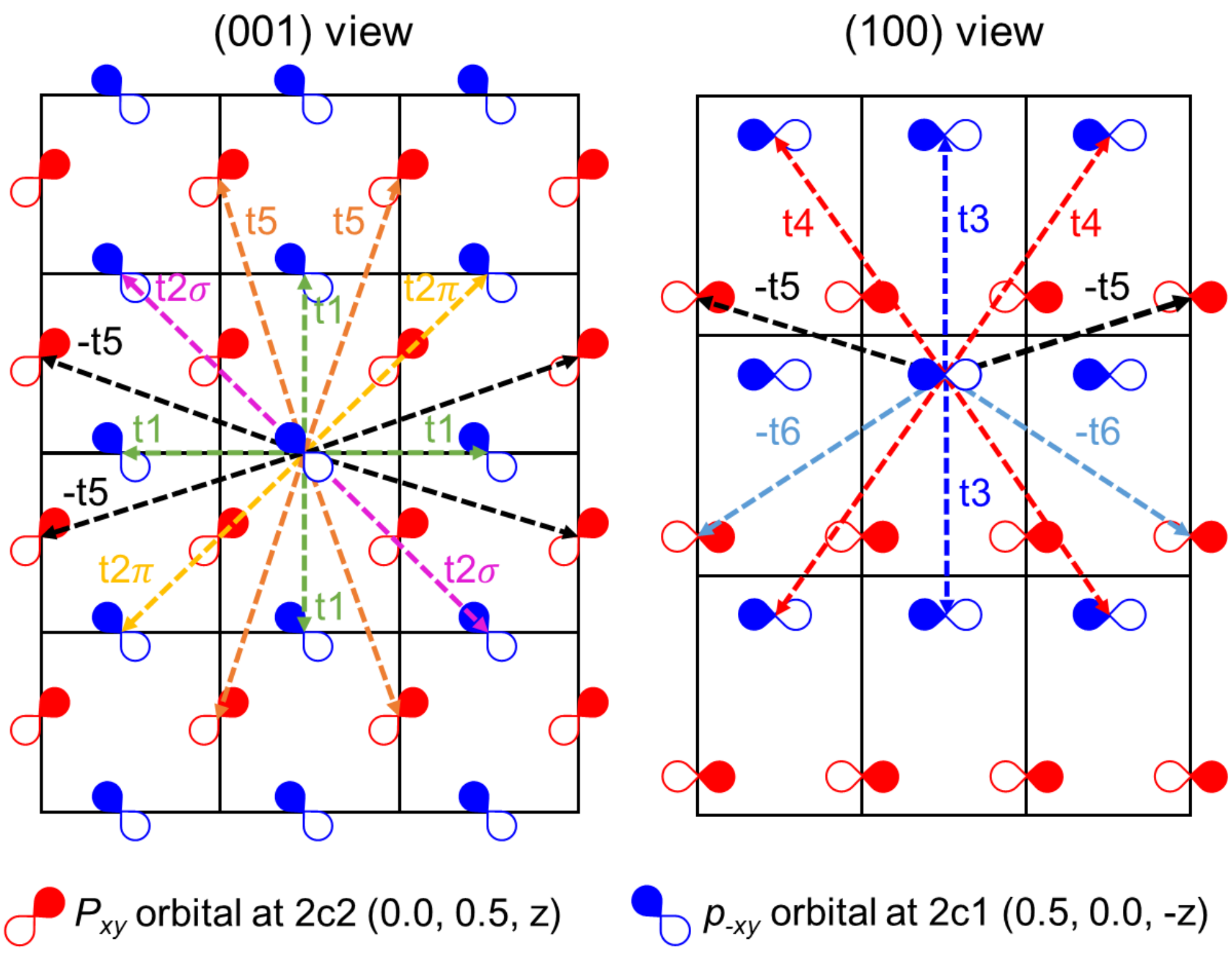

**Figure S1.** Schematic of the TB model of SG 113.

For SG 113, the unit cell is $\vec{a} = (a, 0,0)$ and $\vec{c} = (0,0, c)$. The 2NS+QNL phonon satisfies an elementary band representation [6] of either $B_1$@2c or $B_2$@2c. Figure S1 shows the parameter setup of corresponding TB model. Here, the 2c Wyckoff positions locate at 2c1 (0.5, 0.0, -*z*) and 2c2 (0.0, 0.5, *z*), where *z* values between 0 and 1. The orbital basis can be chosen as $p_{xy}$@2c1 and $p_{-xy}$@2c2. Hence, the 2 × 2 Hamiltonian writes as:

$$H11 = \Delta + 2t_1(\cos k_x + \cos k_y) + 2t_{2\pi}\cos(k_x + k_y) + 2t_{2\sigma}\cos(k_x - k_y) +$$

$$2t_3 \cos k_z + 4t_4(\cos k_x \cos k_z + \cos k_y \cos k_z), \qquad \text{(S2)}$$

$$H22 = \Delta + 2t_1(\cos k_x + \cos k_y) + 2t_{2\sigma}\cos(k_x + k_y) + 2t_{2\pi}\cos(k_x - k_y) + 2t_3\cos k_z + 4t_4(\cos k_x \cos k_z + \cos k_y \cos k_z), \quad \text{(S3)}$$

$$H12 = 4(e^{i2zk_z}t_5 - 4e^{i(1-2z)k_z}t_6) \times (\cos\left(\frac{k_x}{2}\right)\cos\left(\frac{3k_y}{2}\right) - \cos\left(\frac{3k_x}{2}\right)\cos\left(\frac{k_y}{2}\right)). \quad \text{(S4)}$$

Since $t_{2\pi} \neq t_{2\sigma}$, the double degeneracy of 2NS+QNL happens when:

$$\cos(k_x + k_y) = \cos(k_x - k_y), \text{ and } \cos\left(\frac{k_x}{2}\right)\cos\left(\frac{3k_y}{2}\right) - \cos\left(\frac{3k_x}{2}\right)\cos\left(\frac{k_y}{2}\right) = 0. \quad \text{(S5)}$$

One can also check such Hamiltonian satisfies arbitrary symmetry operations $Q = \{R|t\}$ of SG 113. For instance, as for $Q = S_{4z}$, its matrix representation near the Γ-Z path is:

$$D(S_{4z}) = \begin{bmatrix} 0 & -1 \\ 1 & 0 \end{bmatrix}, \text{ and } D(Q)H(\mathbf{k})D^{-1}(Q) = H(R\mathbf{k}). \quad \text{(S6)}$$

As shown in Figure 2(d) and Figure 3(b,f), our TB model (with $\Delta = 12.0, t_1 = 0.55, t_{2\sigma} = -2.0, t_{2\pi} = -0.1, t_3 = 0.0, t_4 = -0.275, t_5 = -0.3, t_6 = -0.1$ in unit of $t_0$, and $z = 0.1$) evidences the discovery of 2NS+QNL phonon and corresponding ribbon-like SSs in $C_4N$. Notice the SSs appear when $|t_5| > |t_6|$.

## 2. SG 128 with 2NS+WNN phonon

For SG 128, the unit cell is $\vec{a} = (a, 0,0)$ and $\vec{c} = (0,0, c)$. The two-band 2NS+WNN phonon is symmetry-enforced, and the Wyckoff positions are either 2a or 2b. Here, the $p_z$ orbitals locating at 2a positions ((0.0, 0.0, 0.0) and (0.5, 0.5, 0.5)) are chosen as orbital basis. Similarly, the 2 × 2 Hamiltonian is written as:

$$H11 = \Delta + 2t_1(\cos k_x + \cos k_y) + 4t_2 \cos k_x \cos k_y + 2t_5(\cos(2k_x - k_y) + \cos(k_x + 2k_y)) + 2t_6(\cos(k_x - 2k_y) + \cos(2k_x + k_y)) + 2t_4\cos k_z, \quad \text{(S7)}$$

$$H22 = \Delta + 2t_1(\cos k_x + \cos k_y) + 4t_2 \cos k_x \cos k_y + 2t_6(\cos(2k_x - k_y) + \cos(k_x + 2k_y)) + 2t_5(\cos(k_x - 2k_y) + \cos(2k_x + k_y)) + 2t_4\cos k_z, \quad \text{(S8)}$$

$$H12 = 8t_3\cos\left(\frac{k_x}{2}\right)\cos\left(\frac{k_y}{2}\right)\cos\left(\frac{k_z}{2}\right). \quad \text{(S9)}$$

Notice $t_5 \neq t_6$, so the double degeneracy of 2NS+WNN happens when:

$$\cos\left(\frac{k_x}{2}\right)\cos\left(\frac{k_y}{2}\right)\cos\left(\frac{k_z}{2}\right) = 0, \text{ and}$$

$$\cos(2k_x - k_y) + \cos(k_x + 2k_y) = \cos(k_x - 2k_y) + \cos(2k_x + k_y). \quad \text{(S10)}$$

One can check the above Hamiltonian satisfies arbitrary symmetry operations $Q$ of SG 128, therefore gives the 2NS+WNN phonon dispersion.

Figure S2 shows the phonon dispersions, Zak phase and SSs using TB model of SG 128. Notice we chose the on-site energy $\Delta = 8t_0 > 0$ just to ensure all phonons with a reasonable positive value. In particular, these TB results reproduces the discoveries of $K_2PtI_6$ in Figure 4, i.e., Zak phase alternating between 0 (trivial) and π (nontrivial), the SS1 connecting two WNLs, and SS2/SS3 which connect the WNLs and NSs. Such consistency evidences the realization of a variety of topological SSs, especially the WNL-NS mixed topological SSs which have not been reported before.

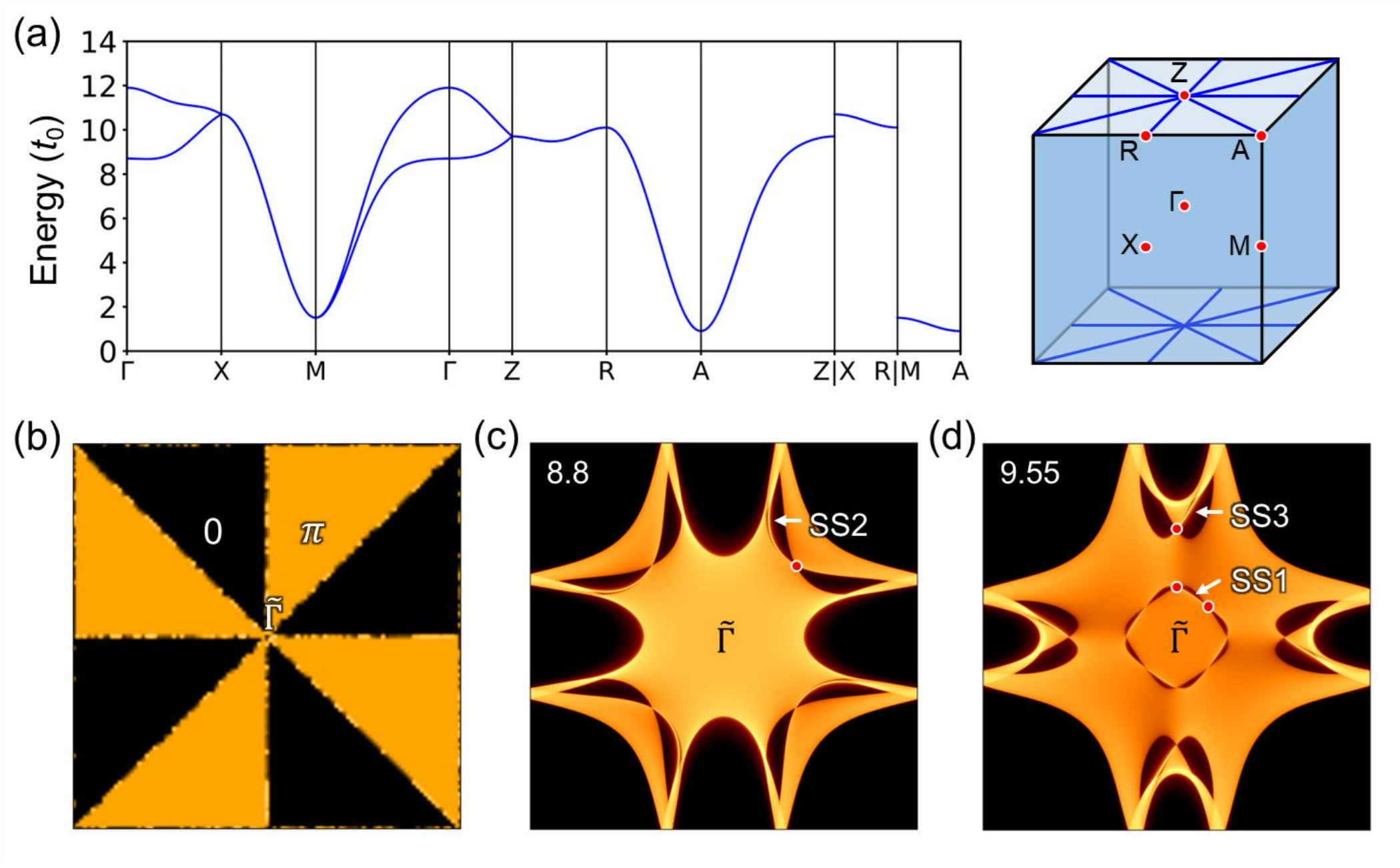

**Figure S2.** TB model of SG 128. (a) Phonon dispersions with $\Delta = 8.0$, $t_1 = 1.0$, $t_2 = -0.6$, $t_3 = -0.2$, $t_4 = 0.15$, $t_5 = 0.1$ and $t_6 = 0.0$, in unit of $t_0$. (b) Zak phase $\mathcal{Z}(k_x, k_y)$ of band 1. The right inset shows the 2NS+WNN phonon. The iso-frequency contours of (001) surface at (c) E = $8.8t_0$ and (d) E = $9.55t_0$.

### 3. SG 53 with 1NS+WNN phonon

For SG 53, the unit cell is $\vec{a} = (a, 0, 0)$, $\vec{b} = (0, b, 0)$ and $\vec{c} = (0, 0, c)$. The 1NS+WNN phonon is symmetry-enforced. We chose the $p_z$ orbitals locating at 2a positions ((0.0, 0.0, 0.0) and (0.5, 0.0, 0.5)) as the orbital basis. The 2 × 2 TB Hamiltonian is:

$$H11 = \Delta + 2t_2\cos k_y + 2t_3\cos k_x + 2t_4\cos(k_y + k_z) + 2t_5\cos(k_y - k_z), \quad \text{(S11)}$$

$$H22 = \Delta + 2t_2\cos k_y + 2t_3\cos k_x + 2t_5\cos(k_y + k_z) + 2t_4\cos(k_y - k_z), \quad \text{(S12)}$$

$$H12 = 4t_1\cos\left(\frac{k_x}{2}\right)\cos\left(\frac{k_z}{2}\right). \quad \text{(S13)}$$

Notice $t_4 \neq t_5$, so the double degeneracy of 2NS+WNN happens when:

$$\cos\left(\frac{k_x}{2}\right)\cos\left(\frac{k_z}{2}\right) = 0, \text{ and } \cos(k_y + k_z) = \cos(k_y - k_z). \quad \text{(S14)}$$

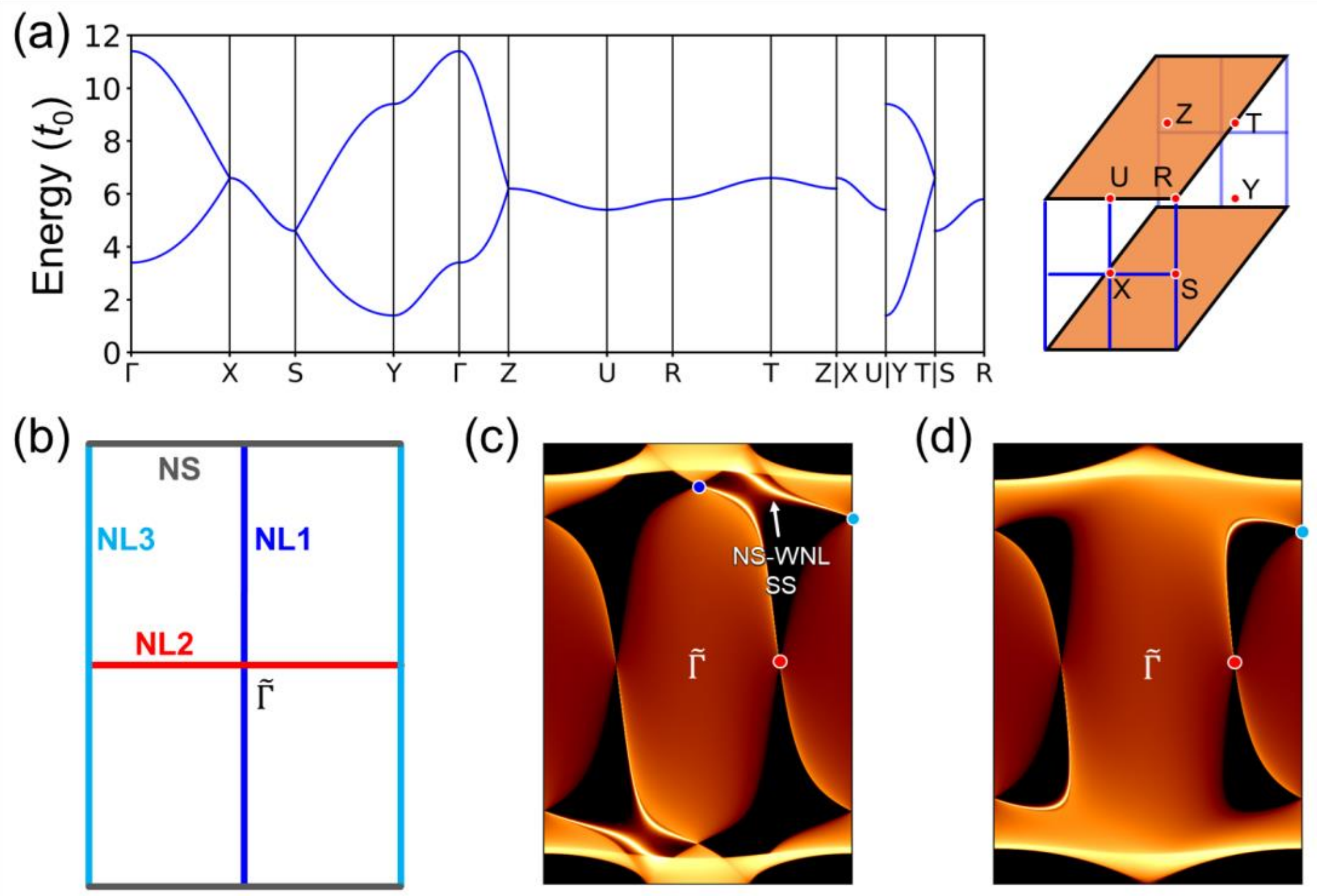

**Figure S3.** TB model of SG 53. (a) Phonon dispersions with Δ = 6.0, $t_1$ = 1.0, $t_2$ = 0.2, $t_3$ = 0.1, $t_4$ = 0.2 and $t_5$ = 0.1, in unit of $t_0$. The right inset shows the 1NS+WNN phonon. (b) schematic of the (100) surface and projected NLs or NSs. The iso-frequency contours of (100) surface at (c) E = 5.5$t_0$ and (d) E = 5.4$t_0$.

Here, the WNL along X-S path (0.5, u, 0.0) is guaranteed by $\{\mathcal{M}_x, \widetilde{\mathcal{M}}_z\} = 0$, where $\widetilde{\mathcal{M}}_z = \{\mathcal{M}_z | \frac{1}{2}, 0, \frac{1}{2}\}$. The WNL along X-U path (0.5, 0.0, u) is guaranteed by $\{\mathcal{M}_x, \widetilde{\mathcal{M}}_y\} = 0$, where $\widetilde{\mathcal{M}}_y = \{\mathcal{M}_y | \frac{1}{2}, 0, \frac{1}{2}\}$.

Figure S3 shows the phonon dispersions and SSs using TB model of SG 53. Similar to SG 128, the topological SSs are also observed, including WNL-WNL SSs and WNL-NS mixed SSs (see Figure S3(c,d)). These agree well with the example of 1NS+WNN phonon in SG53 $AgSe_3I$ (see Figure S11 and Figure S12).

In short, the cases of SG 128 and SG 53 together confirm a variety of topological SSs in the NS and WNN hybrid systems, offering insights into the interplay between NSs and WNLs.

## III. Two-Band NS+NL Phonons

### 1. Analysis of the 2NS+QNL phonon in SG113 $C_4N$

As a supplement to Figure 2 and Figure 3, this subsection briefly discusses the 2NS+QNL phonon in $C_4N$. First, Figure S4 confirms the Γ-Z path is a QNL with the irreducible representation of $LD_3 \oplus LD_4$, and the elementary band representation [6] of bands 27,28 should be $B_1$@2c or $B_2$@2c.

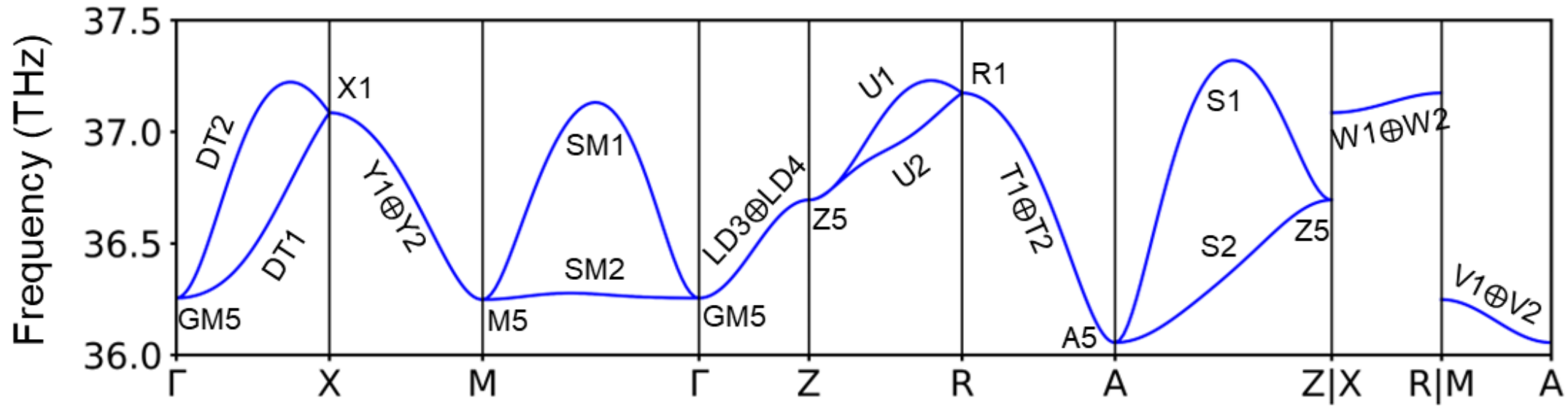

**Figure S4.** Irreducible representations of phonon bands 27,28 of $C_4N$, where the convention follows Bilbao crystallographic server https://www.cryst.ehu.es/.

Next, the $k \cdot p$ effective Hamiltonian is demonstrated. For an arbitrary point around the NS or QNL of $C_4N$, the 2 × 2 Hamiltonian is expressed as:

$$H(\mathbf{q}) = d_x(\mathbf{q})\sigma_x + d_y(\mathbf{q})\sigma_y + d_z(\mathbf{q})\sigma_z, \qquad \text{(S15)}$$

where $\sigma_{x,y,z}$ are Pauli matrices and the $\sigma_0$ terms are neglected, $d_{x/y/z}(\mathbf{q})$ present real functions.

(1) Near the central part of NS1 ($k_x = \pi$), an arbitrary point around X-R path has $C_{2z}$ and $S_{2x}\mathcal{T}$ as generators, where $\mathcal{T}$ is the time-reversal operator. Here, $\mathcal{T}^2 = 1$, $C_{2z}^2 = 1$ and $(S_{2x}\mathcal{T})^2 = -1$. The matrix representations are:

$$D(C_{2z}) = \begin{bmatrix} 1 & 0 \\ 0 & -1 \end{bmatrix},\ D(S_{2x}\mathcal{T}) = \begin{bmatrix} 0 & -1 \\ 1 & 0 \end{bmatrix} K, \tag{S16}$$

where $K$ is the complex conjugation operator. Then the symmetry restrictions on $d_{x/y/z}(\mathbf{q})$ are:

$$d_x(q_x, q_y, q_z) = -d_x(-q_x, -q_y, q_z) = -d_x(-q_x, q_y, q_z), \tag{S17}$$

$$d_y(q_x, q_y, q_z) = -d_y(-q_x, -q_y, q_z) = -d_y(-q_x, q_y, q_z), \tag{S18}$$

$$d_z(q_x, q_y, q_z) = d_z(-q_x, -q_y, q_z) = -d_z(-q_x, q_y, q_z). \tag{S19}$$

To the leading order, it is written as:

$$H_{\mathrm{XR}}^{\mathrm{eff}}(\mathbf{q}) = (a_1\sigma_x + a_2\sigma_y)q_x. \tag{S20}$$

(2) Near the boundary part of NS1, an arbitrary point along M-A path has $C_{2z}$, $S_{4z}\mathcal{T}$ and $\widetilde{\mathcal{M}}_{x,y} = \left\{\mathcal{M}_{x,y} | \frac{1}{2}, \frac{1}{2}, 0\right\}$ as generators. Under an irreducible representation of $\mathrm{V}_1 \oplus \mathrm{V}_2$ (see Figure S4), their matrix representations are:

$$D(C_{2z}) = \begin{bmatrix} 1 & 0 \\ 0 & 1 \end{bmatrix},\ D(S_{4z}\mathcal{T}) = \begin{bmatrix} 0 & 1 \\ 1 & 0 \end{bmatrix} K,\ D(\widetilde{\mathcal{M}}_{x,y}) = \begin{bmatrix} -1 & 0 \\ 0 & 1 \end{bmatrix}, \tag{S21}$$

These symmetries constrain:

$$d_x(q_x, q_y, q_z) = d_x(-q_x, -q_y, q_z) = d_x(-q_y, q_x, q_z) = -d_x(q_y, q_x, q_z), \tag{S22}$$

$$d_y(q_x, q_y, q_z) = d_y(-q_x, -q_y, q_z) = d_y(-q_y, q_x, q_z) = -d_y(q_y, q_x, q_z), \tag{S23}$$

$$d_z(q_x, q_y, q_z) = d_z(-q_x, -q_y, q_z) = -d_z(-q_y, q_x, q_z) = d_z(q_y, q_x, q_z). \tag{S24}$$

To the leading order, it gives:

$$H_{\mathrm{MA}}^{\mathrm{eff}}(\mathbf{q}) = b_1\sigma_z q_x q_y. \tag{S25}$$

(3) For an arbitrary point around the QNL of Γ-Z path, it has $C_{2z}$, $S_{4z}\mathcal{T}$ and $\widetilde{\mathcal{M}}_{x,y}$

as generators. Noted that $C_{2z}^2 = 1$, $(S_{4z}\mathcal{T})^2 = -1$ and $\widetilde{\mathcal{M}}_{x,y}^2 = 1$. Their matrix representations are:

$$D(C_{2z}) = \begin{bmatrix} -1 & 0 \\ 0 & -1 \end{bmatrix},\ D(S_{4z}\mathcal{T}) = \begin{bmatrix} 0 & -1 \\ 1 & 0 \end{bmatrix}K,\ D(\widetilde{\mathcal{M}}_{x,y}) = \begin{bmatrix} 1 & 0 \\ 0 & -1 \end{bmatrix}, \quad \text{(S26)}$$

These symmetries constrain:

$$d_x(q_x, q_y, q_z) = d_x(-q_x, -q_y, q_z) = -d_x(-q_y, q_x, q_z) = -d_x(q_y, q_x, q_z), \quad \text{(S27)}$$

$$d_y(q_x, q_y, q_z) = d_y(-q_x, -q_y, q_z) = -d_y(-q_y, q_x, q_z) = -d_y(q_y, q_x, q_z), \quad \text{(S28)}$$

$$d_z(q_x, q_y, q_z) = d_y(-q_x, -q_y, q_z) = -d_z(-q_y, q_x, q_z) = d_z(q_y, q_x, q_z). \quad \text{(S29)}$$

Thus, the effective Hamiltonian is:

$$H_{\Gamma Z}^{\text{eff}}(\mathbf{q}) = (c_1\sigma_x + c_2\sigma_y)(q_x^2 - q_y^2) + c_3\sigma_z q_x q_y. \quad \text{(S30)}$$

Since the eigenstates $|\pm 1\rangle$ of $\widetilde{\mathcal{M}}_{x,y}$ (in the form of $|p_{xy}\rangle$ and $|p_{-xy}\rangle$) can be chosen as the basis states, one observes that:

$$\widetilde{\mathcal{M}}_{x,y} S_{4z}\mathcal{T}|\pm 1\rangle = -S_{4z}\mathcal{T}\widetilde{\mathcal{M}}_{x,y}|\pm 1\rangle = \mp(S_{4z}\mathcal{T}|\pm 1\rangle). \quad \text{(S31)}$$

which gives $S_{4z}\mathcal{T}|\pm 1\rangle = |\mp 1\rangle$ and hence the two states $|+1\rangle$ and $|-1\rangle$ are degenerate.

Such linear dispersions near X-R path and quadratic dispersions near M-A and Γ-Z paths are also captured by the first-principles results, as shown in Figure S5.

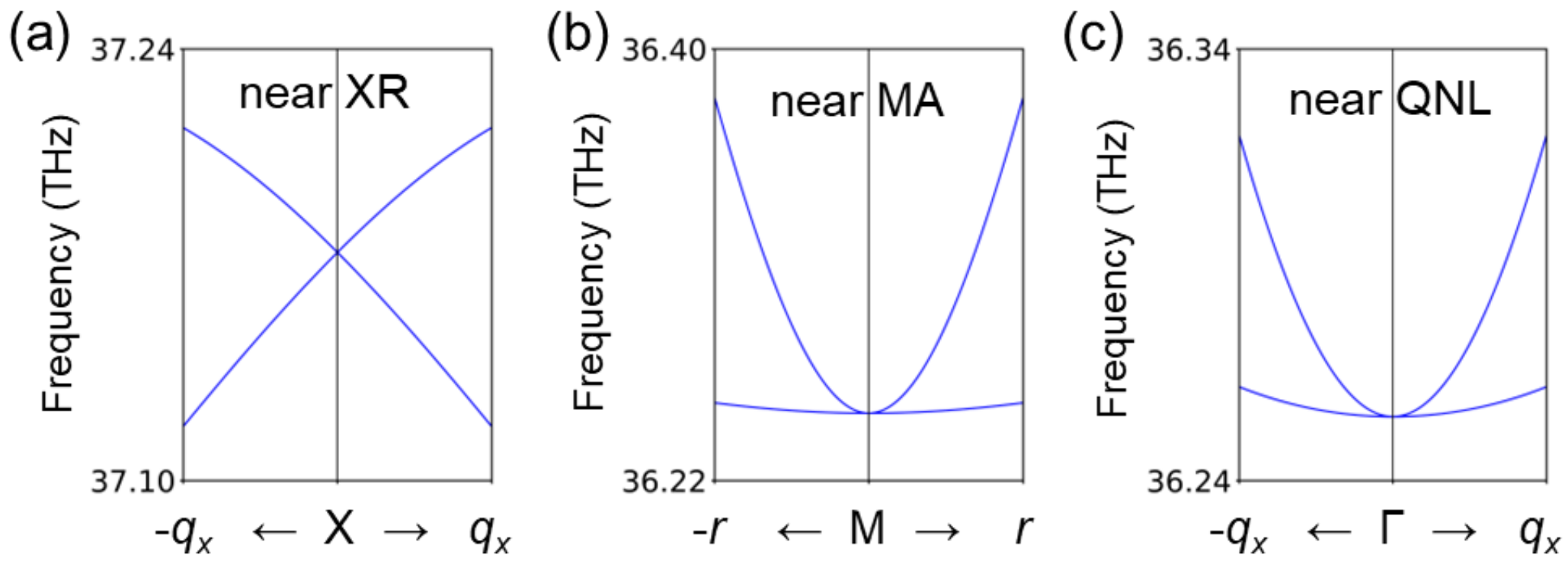

**Figure S5.** Expansions of 2NS+QNL phonon of $C_4N$ near (a) central part of NS, i.e., the sections around X-R path; (b) boundary part of NS, i.e., the sections around M-A path and (c) the QNL. In (a-c), $q_x = q_y = 0.1\pi$, and $\boldsymbol{r} = (q_x, q_y)$.

Figure S6 gives the wide frequency range (36.88-37.14 THz) within which the ribbon-like SSs emerge. The ribbon-like feature is reflected by the shift of SSs when changing frequencies. Notice the central intermediate bulk states, which are detached from NS projections, are essential for supporting the SSs. As a general discussion, if there are no such intermediate bulk states to indirectly link the source and sink at two NSs, such SSs should annihilate because the two NSs are connected without a global gap. On the other hand, Figure 3a shows the central parts of two NSs are indirectly connected by the Berry curvature, which require another bulk state to stabilize the SSs.

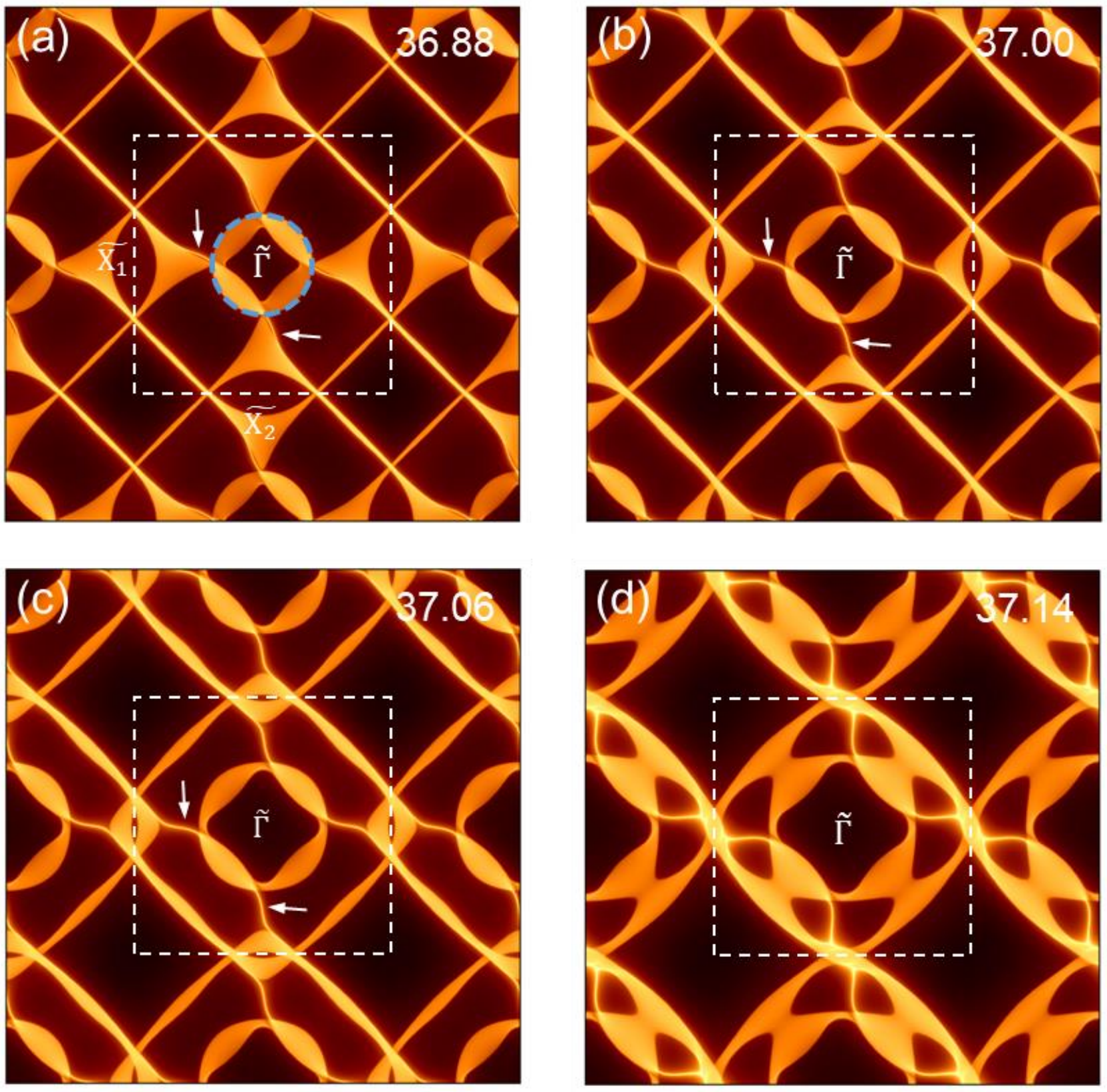

**Figure S6.** Ribbon-like SSs from one NS to another. (a) The SSs begin to emerge at 36.88 THz upon the central intermediate bulk states (light blue circle) are gapped away from the projections of two NSs. The iso-frequency (001) surface contours at (b) 37.00 THz; (c) 37.06 THz and (d) 37.14 THz. In (d), The SSs begin to disappear when the projections of two NSs and central intermediate bulk states are connected.

We have also observed torus SSs [7,8] spanning over the surface Brillouin zone in Figure S7. Such torus SSs originate from the QNL, which are topologically trivial.

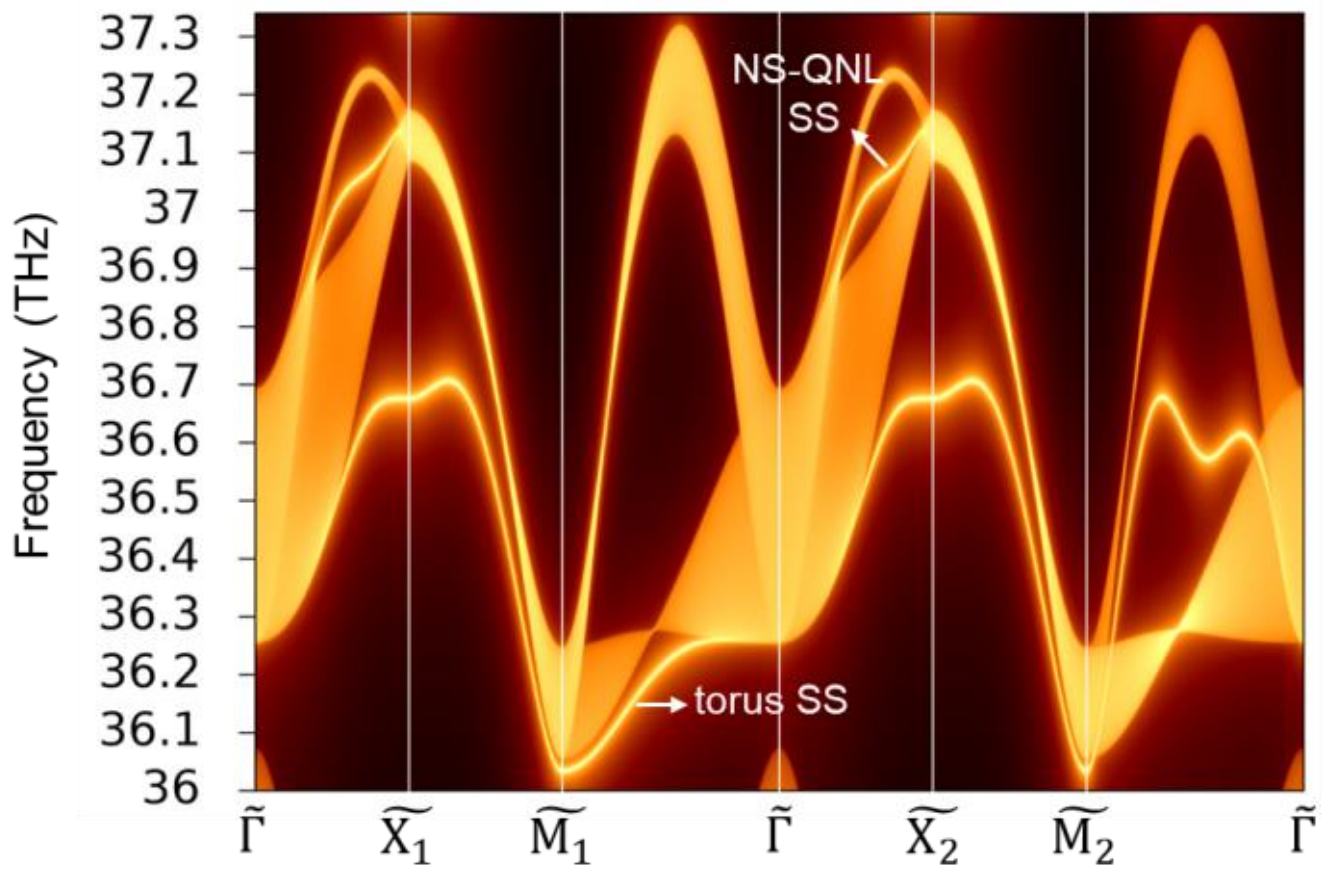

**Figure S7.** Surface spectrum of $C_4N$ along high-symmetry lines of the (001) projected Brillouin zone. The torus SSs from QNL and in-gap ribbon-like SSs are marked.

Figure S8 shows the 2NS phonon of $C_4N$. As compared with the 2NS+QNL phonon, the Zak phase is almost zero due to the cancellation of Berry curvature at $k_z = 0$ and $k_z = \pi$. It reveals the crucial role of QNL in generating the ribbon-like SSs.

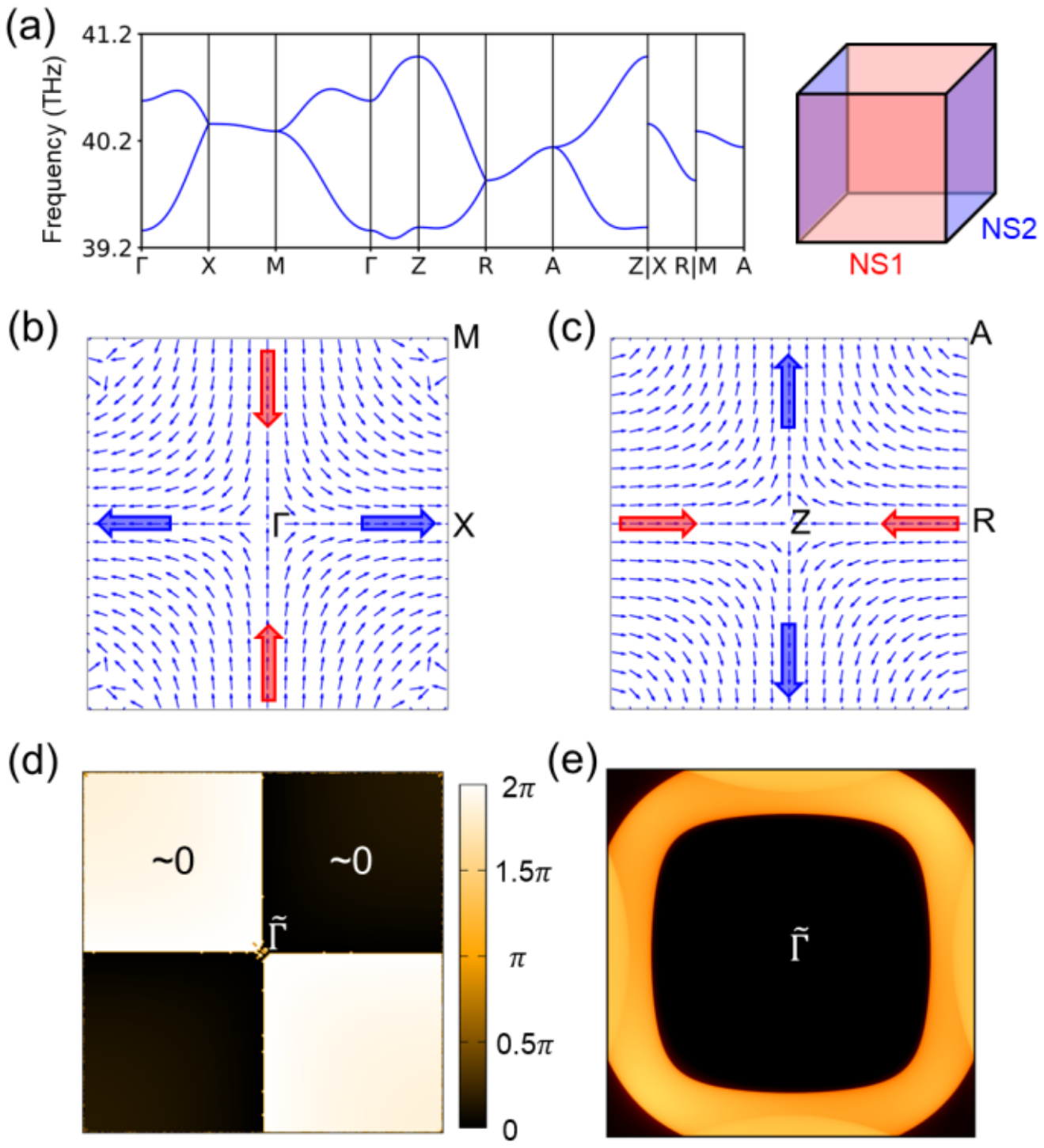

**Figure S8.** (a) The 2NS phonon in bands 29,30 of $C_4N$, without LOTO. In-plane Berry curvature $\Omega_{x,y}(\mathbf{k})$ of band 29 at (b) $k_z = 0$ and (c) $k_z = \pi$. (d) Calculated $\mathcal{Z}(k_x, k_y)$. (e) The iso-frequency (001) surface contour of $C_4N$ at 40.0 THz. There exists no SS.

## 2. Material candidates for ideal two-band NS+NL phonons

Following the classifications in Table 1, this subsection gives material candidates for all possible two-band NS+NL phonons. In addition to $C_4N$ and $K_2PtI_6$ shown in the main text, we have provided other candidates of SG 113 (see Figure S15) and SG 128 (see Figure S18), although their NS+NL phonons are not robust against LOTO.

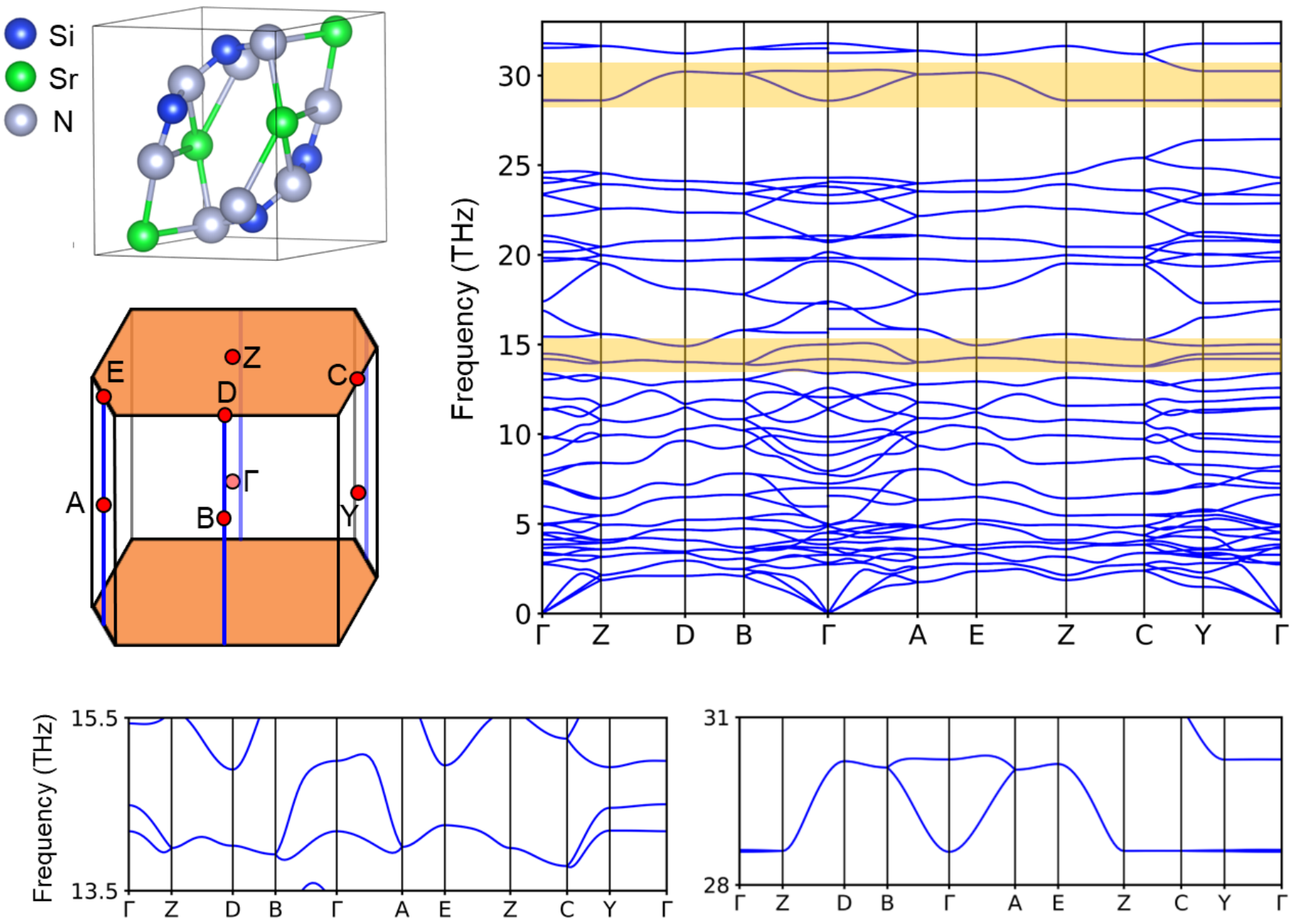

**Figure S9.** The 1NS+WNL phonons of mp-4549 $SrSiN_2$ (SG 14), with LOTO.

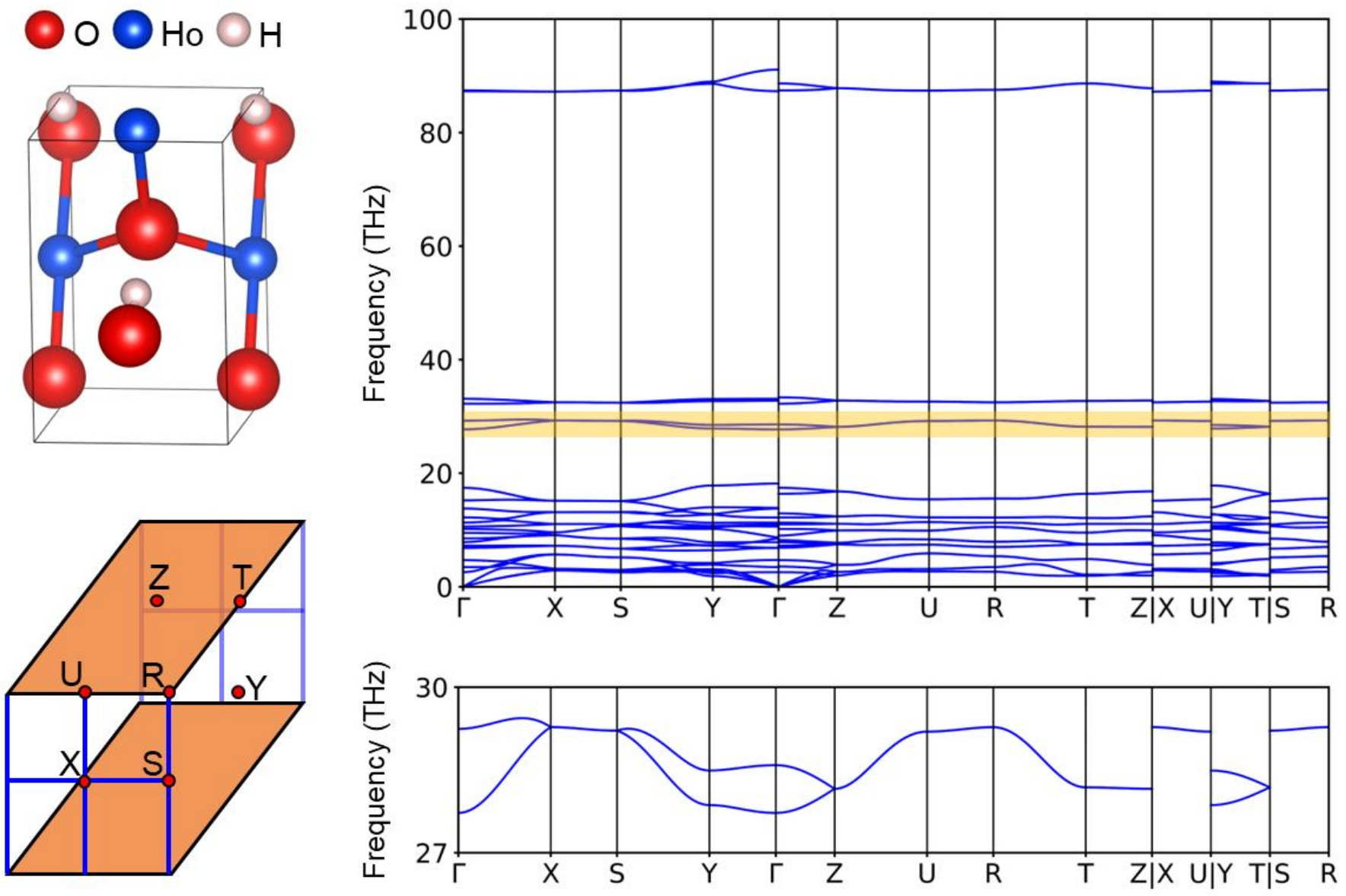

**Figure S10.** The 1NS+WNN phonon of mp-755543 $HoHO_2$ (SG 31), with LOTO.

Figure S11 and Figure S12 show the 1NS+WNN phonon, Zak phase and SSs of SG 53 $AgSe_3I$. In Figure S12c-e, a variety of topological SSs are found, including the WNL-WNL SS2 and WNL-NS mixed SS1, SS3.

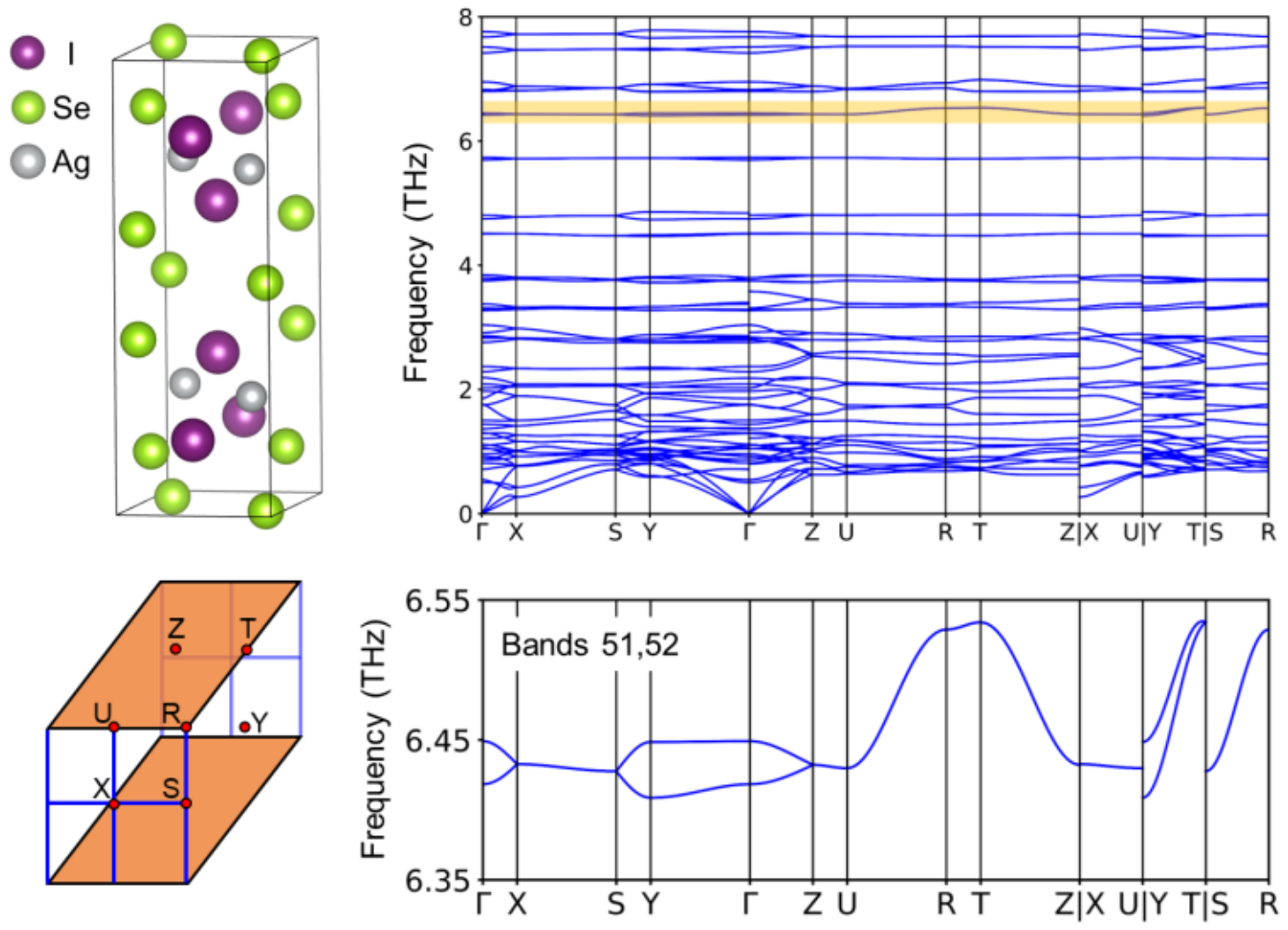

**Figure S11.** The 1NS+WNN phonon of mp-569052 $AgSe_3I$ (SG 53), with LOTO.

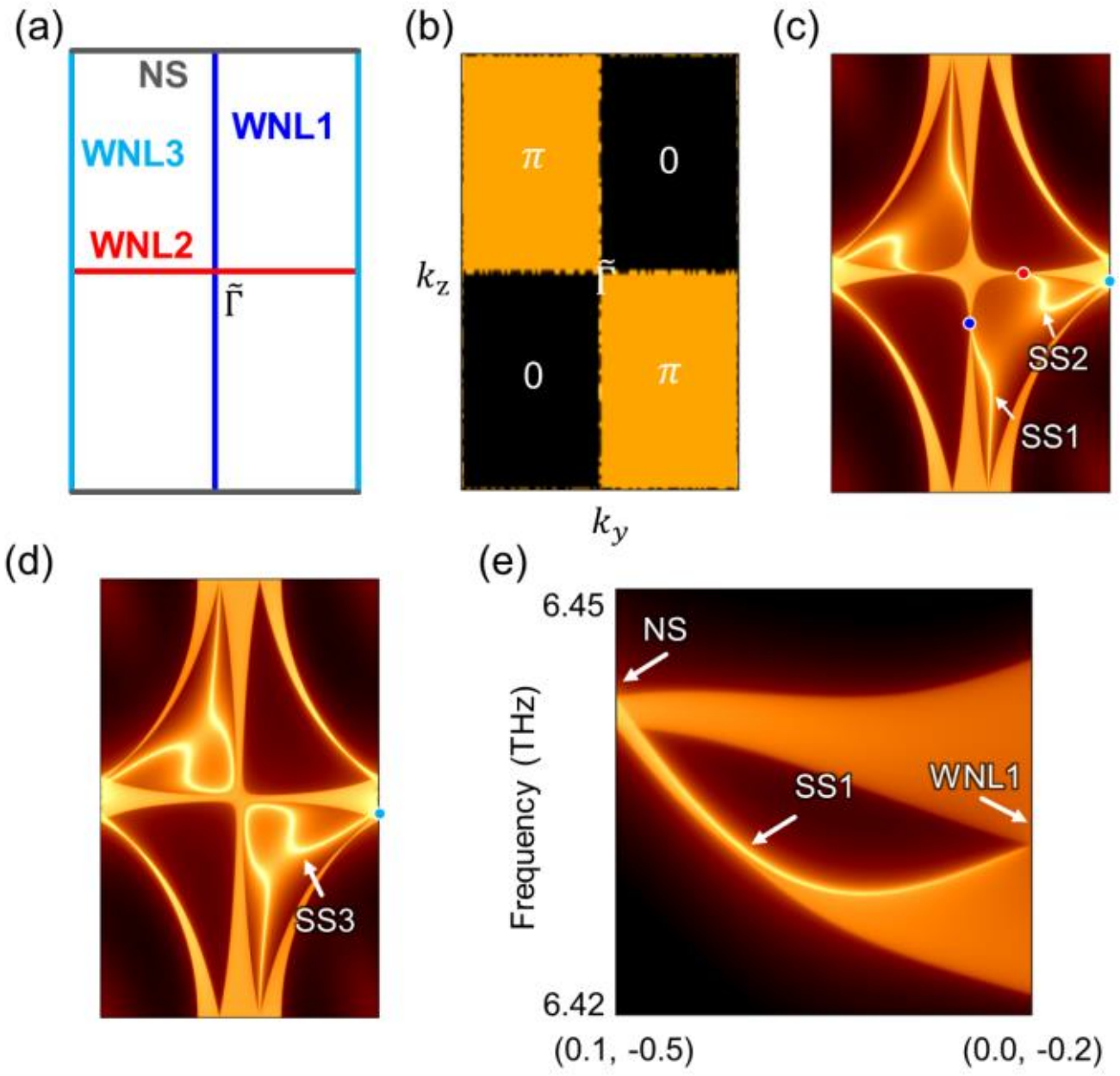

**Figure S12.** (a) Schematic of the (100) surface Brillouin zone of $AgSe_3I$, with the projections of NS and three types of NLs (i.e., NL1, NL2 and NL3). (b) Calculated Zak phase $\mathcal{Z}(k_y, k_z)$ of (100) plane up to band 51. The iso-frequency surface contours at (c) 6.432 THz and (d) 6.433 THz. (e) (100) surface spectrum of WNL-NS mixed SS1.

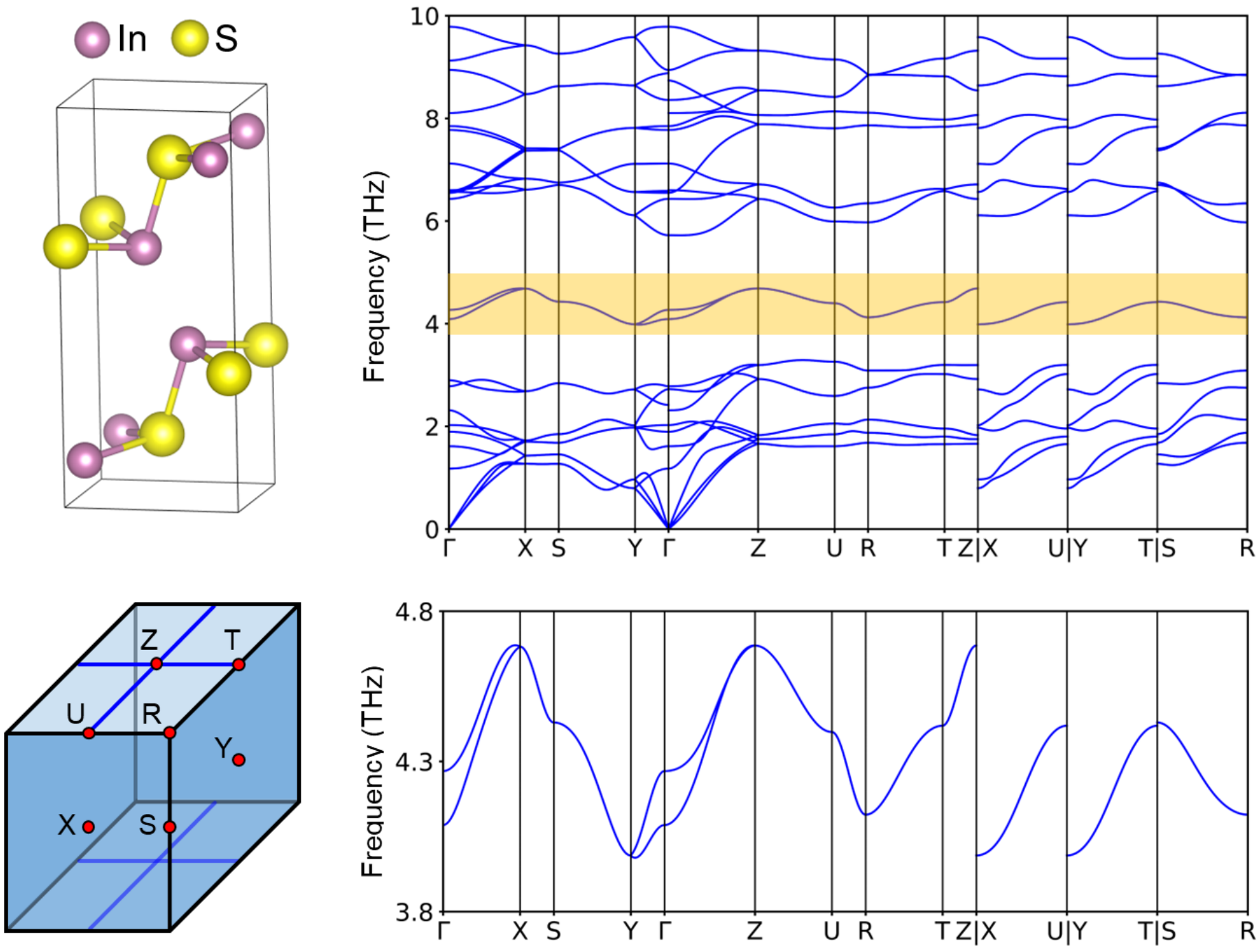

**Figure S13.** The 2NS+WNN phonon of mp-19795 InS (SG 58), with LOTO.

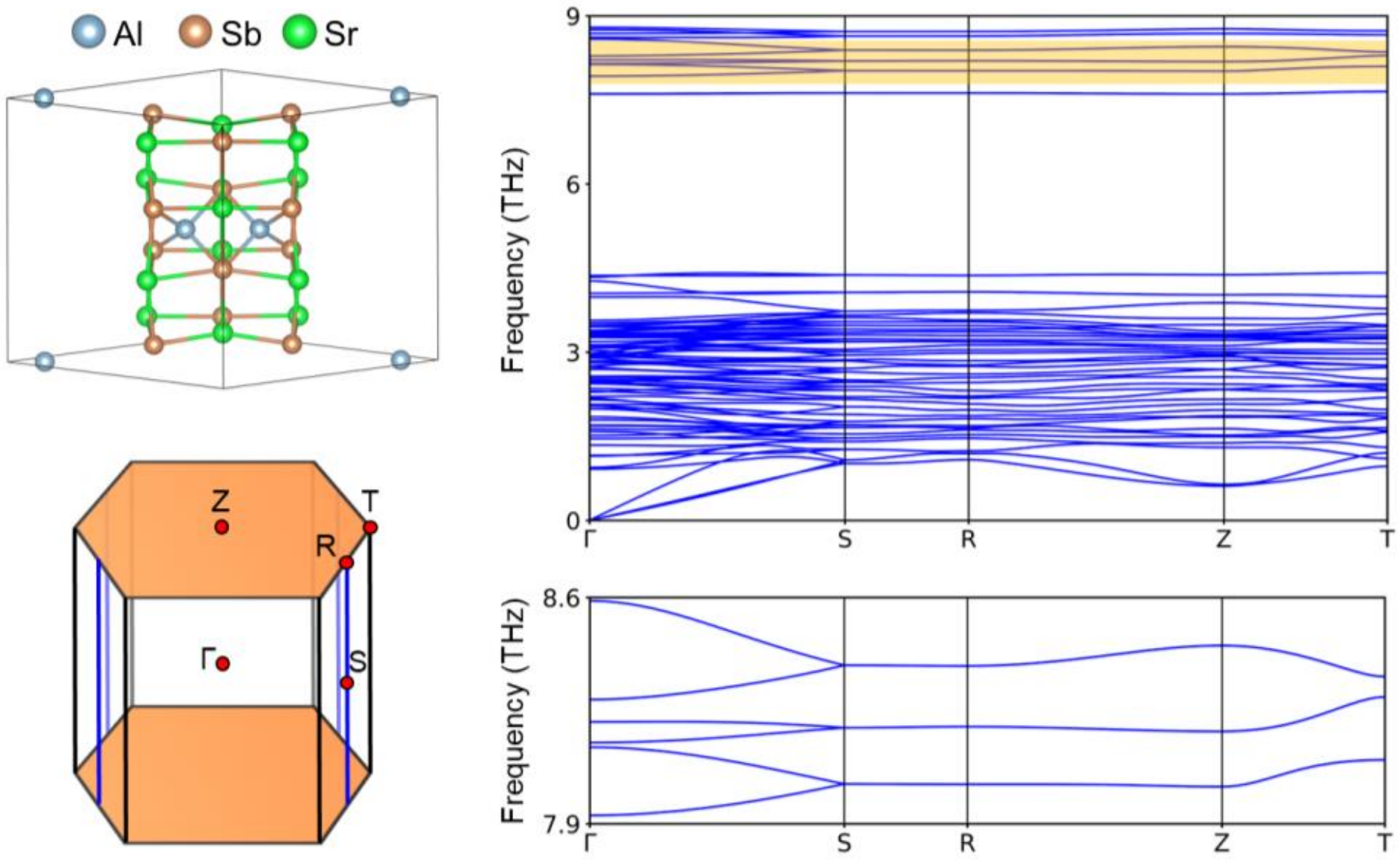

**Figure S14.** The 1NS+WNL phonons of mp-17667 $Sr_3AlSb_3$ (SG 64), with LOTO.

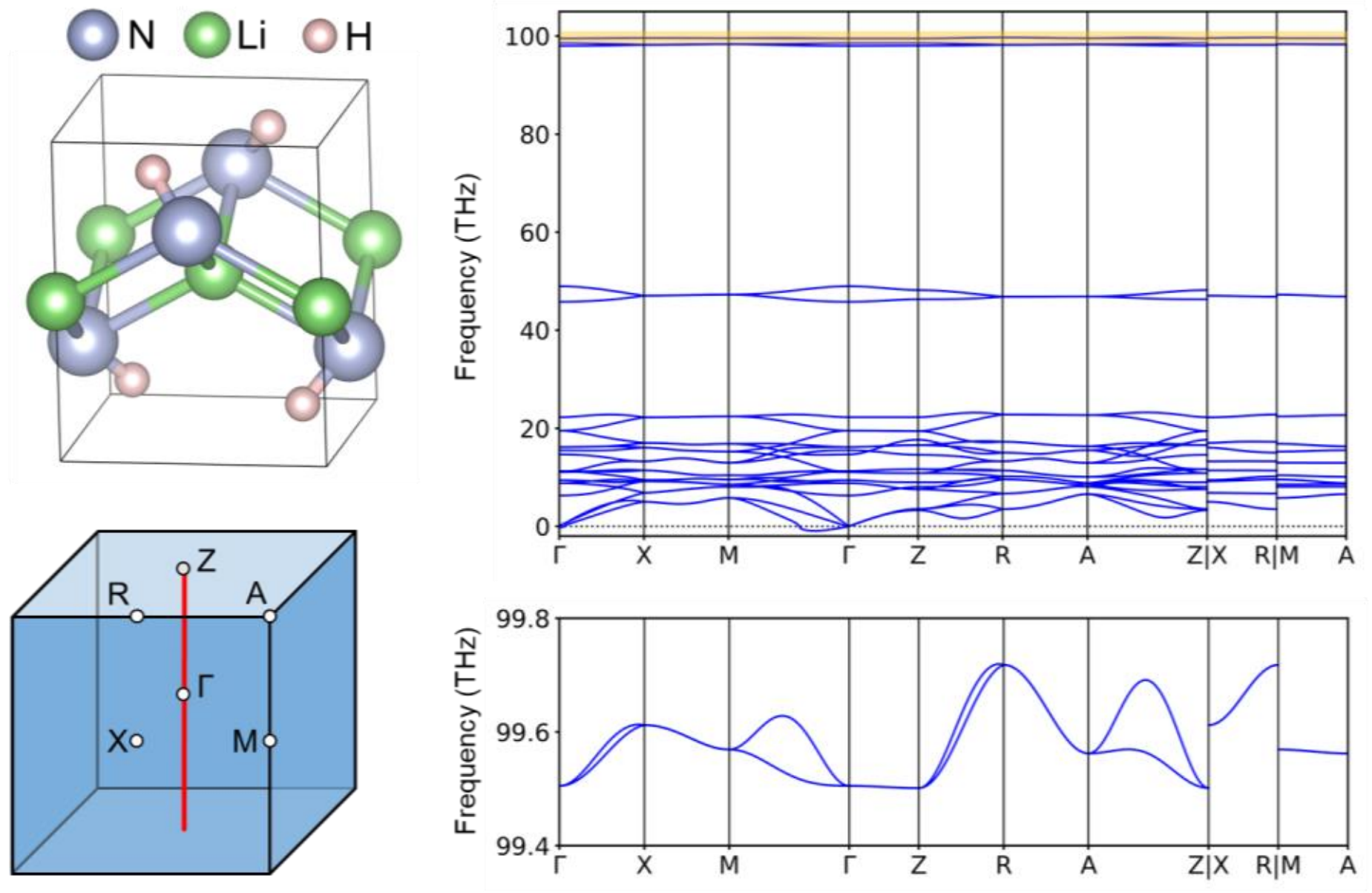

**Figure S15.** The 2NS+QNL phonon of mp-1079418 $LiH_2N$ (SG 113), without LOTO.

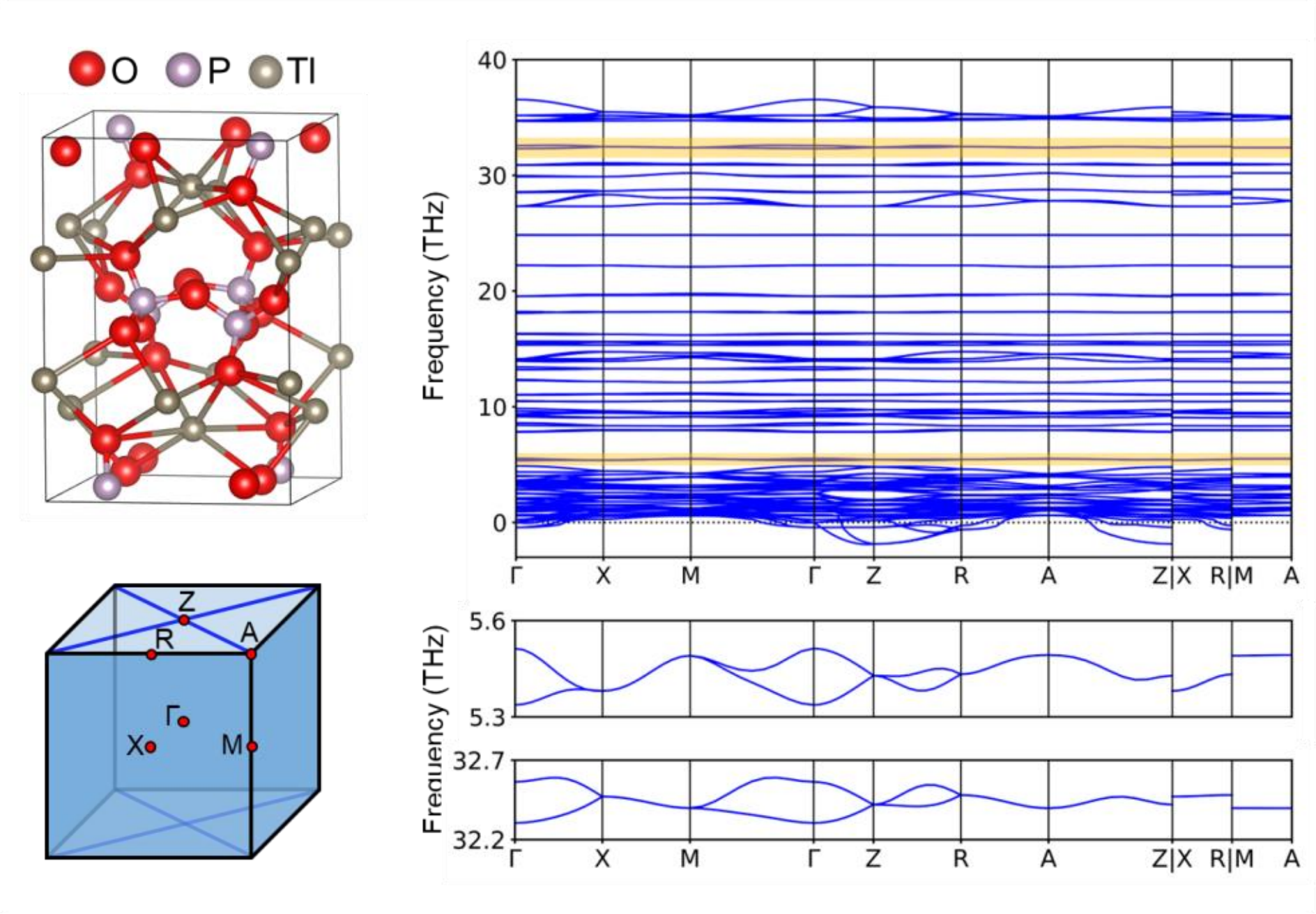

**Figure S16.** The 2NS+WNN phonons of mp-27144 $TlPO_3$ (SG 114), without LOTO.

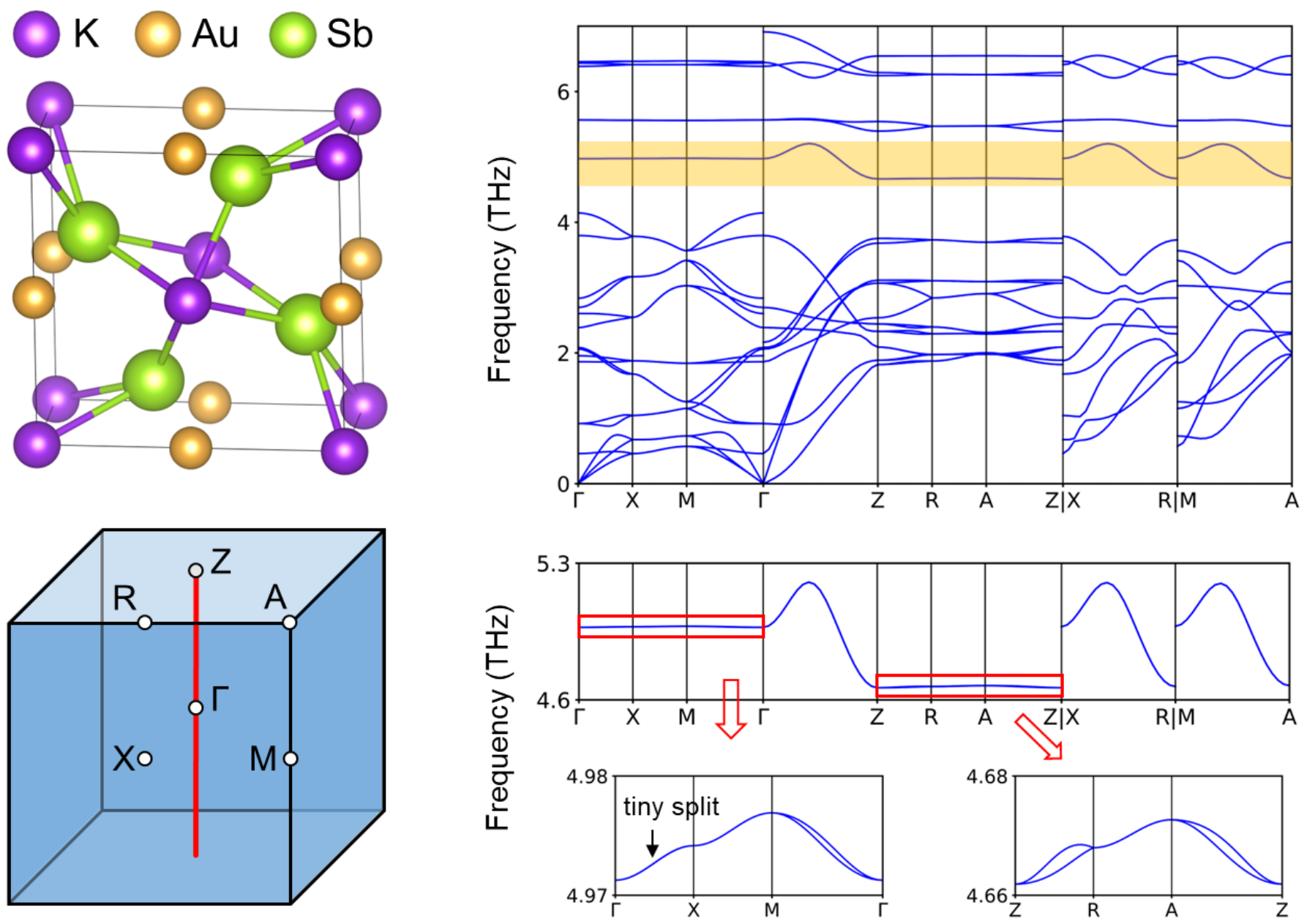

**Figure S17.** The 2NS+QNL phonon of mp-29138 $KAuSe_2$ (SG 127), with LOTO.

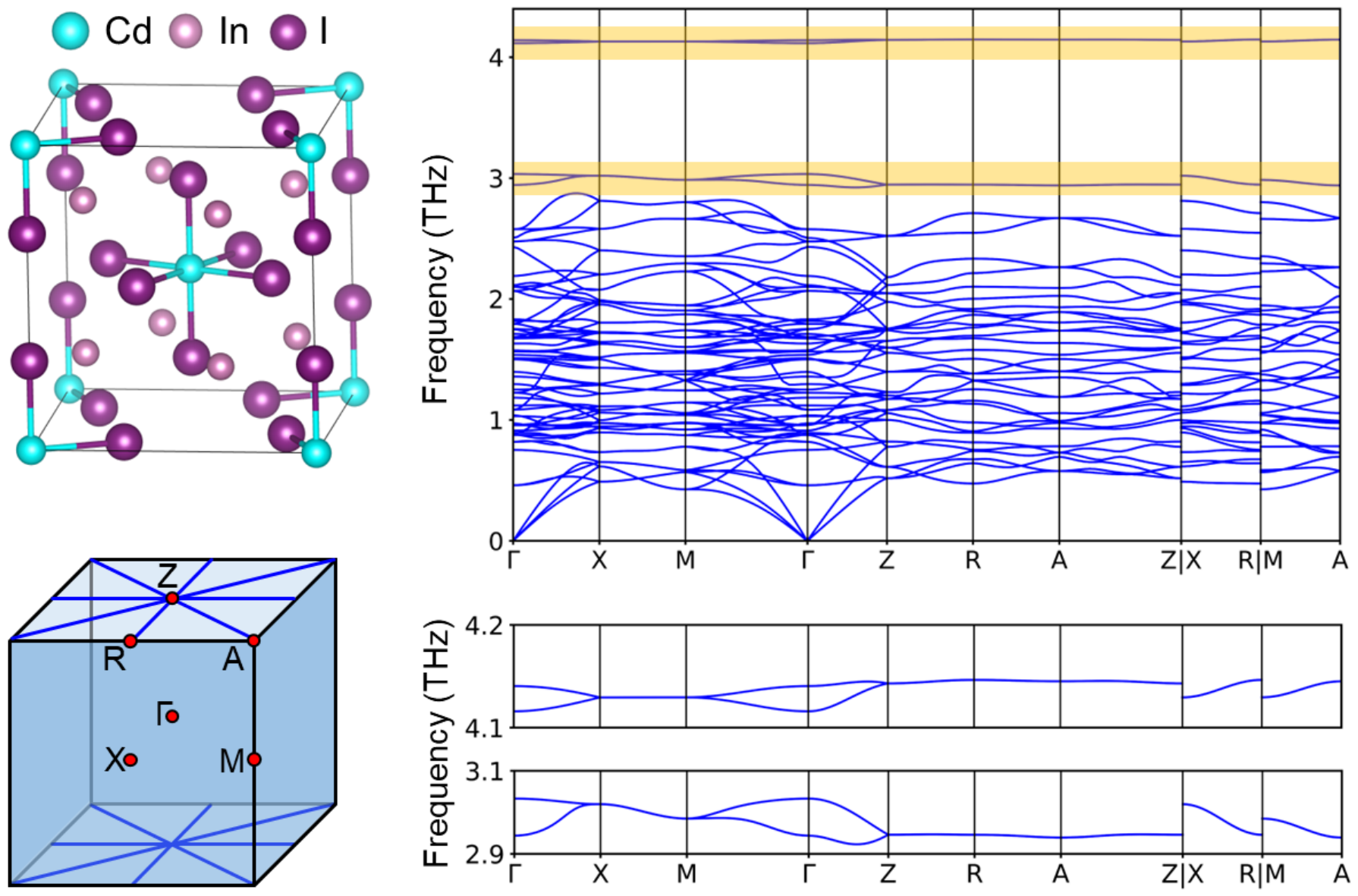

**Figure S18.** The 2NS+WNN phonons of mp-616218 $CdIn_4I_6$ (SG 128), without LOTO.

Figure S19 shows the 2NS+HNN phonon in SG 136 $ZnSb_2O_6$. Here, HNN means the nodal net contains both QNL and WNL. The orbital basis should be in-plane $p_{x,y}$ orbitals locating at 2a or 2b Wyckoff positions. In Figure S20(b-d), the drumhead SSs from WNLs and torus SSs from QNL are found in $ZnSb_2O_6$.

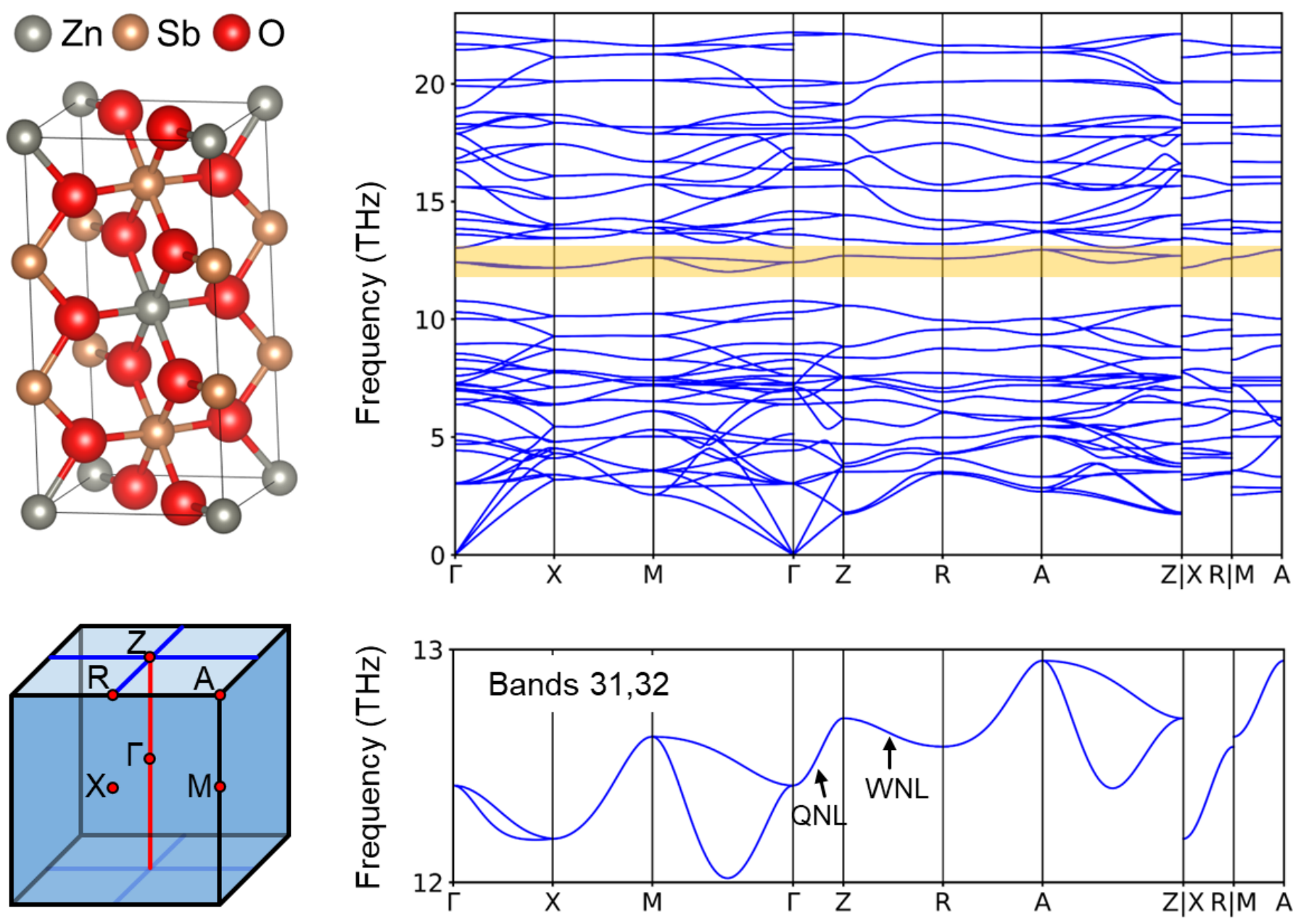

**Figure S19.** The 2NS+HNN phonon of mp-3188 $ZnSb_2O_6$ (SG 136), with LOTO.

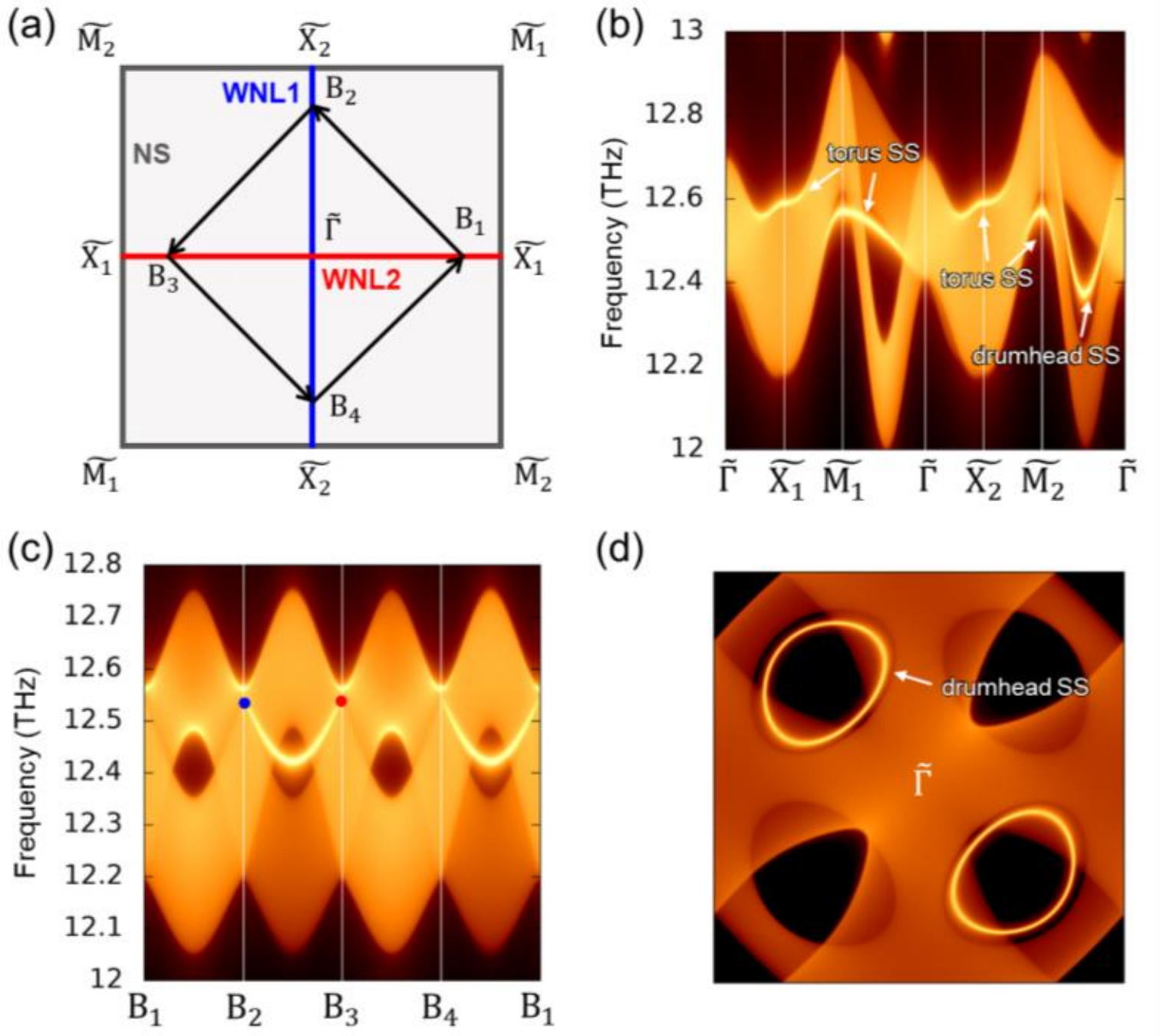

**Figure S20.** (a) Schematic of the (001) surface Brillouin zone of $ZnSb_2O_6$. (b) Surface spectrum of $ZnSb_2O_6$ along high-symmetry lines. The torus SSs from QNL and topologically nontrivial drumhead SSs from WNLs are marked. (c) Surface spectrum of $ZnSb_2O_6$ along selected paths as shown in (a). (d) The iso-frequency contour of (001) surface at 12.46 THz.

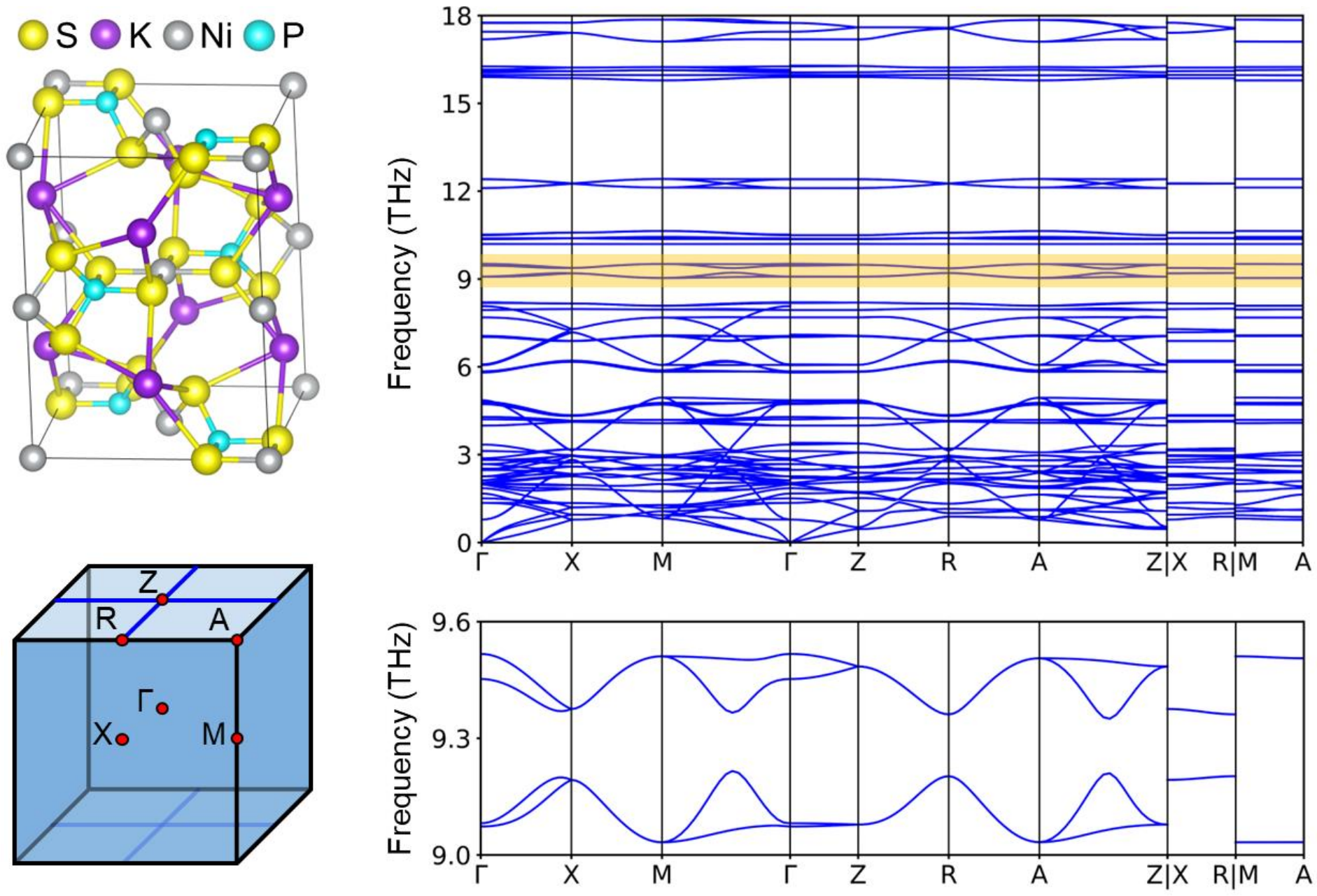

**Figure S21.** The 2NS+WNN phonons of mp-662530 $KNiPS_4$ (SG 136), with LOTO.

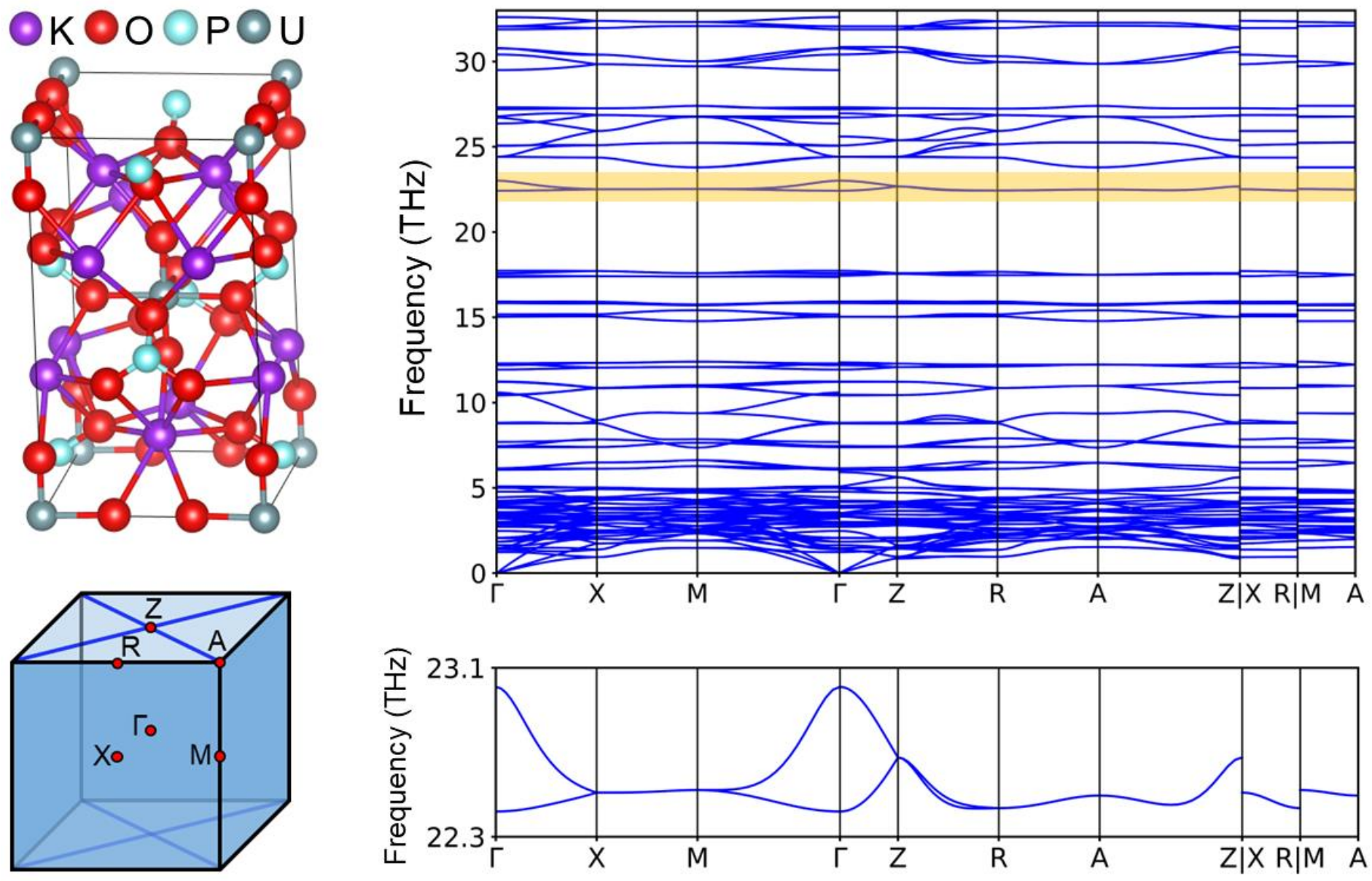

**Figure S22.** The 2NS+WNN phonon of mp-559639 $K_4UP_2O_{10}$ (SG 137), with LOTO.

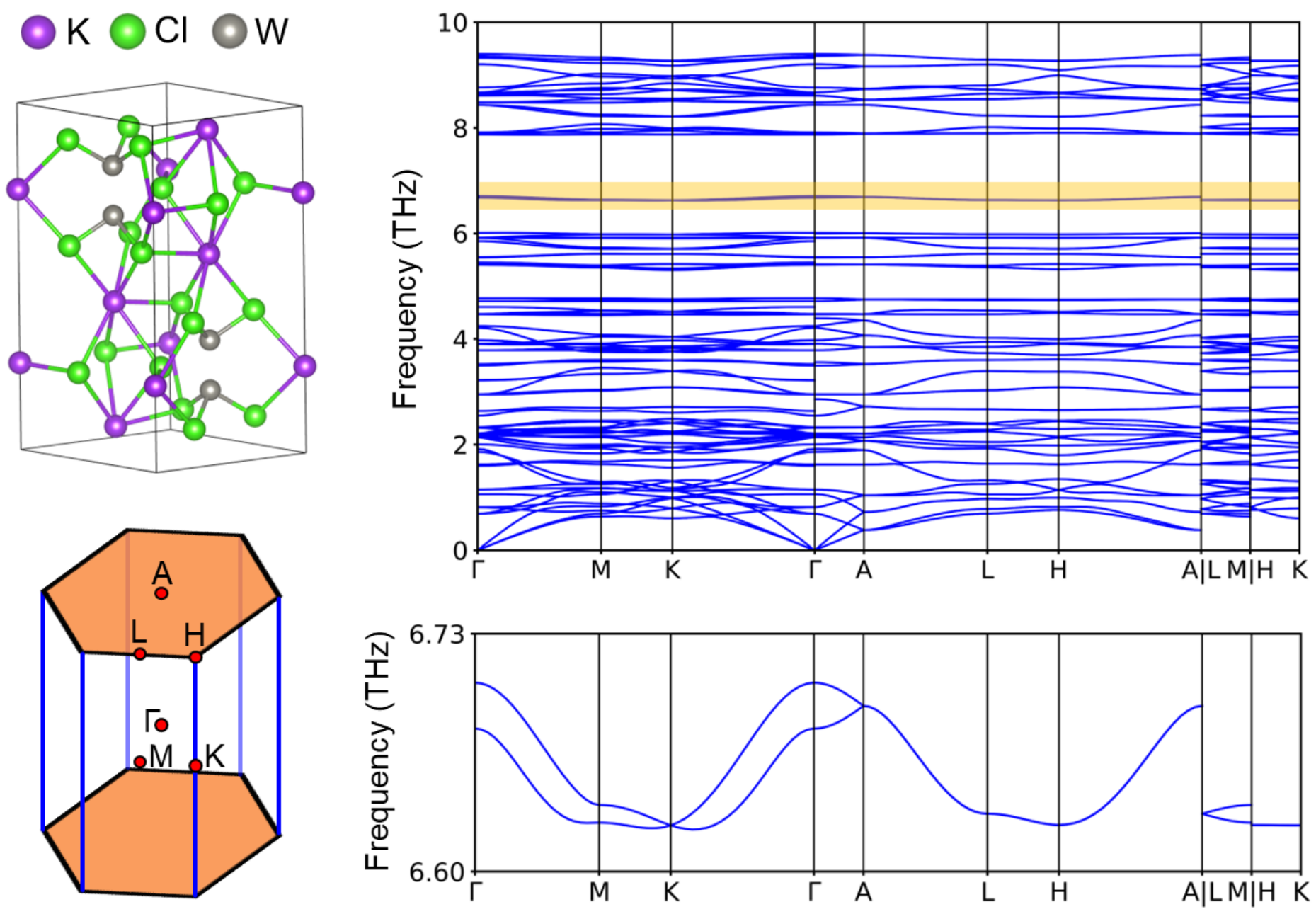

**Figure S23.** The 1NS+WNL phonon of mp-27506 $K_3W_2Cl_9$ (SG 176), with LOTO.

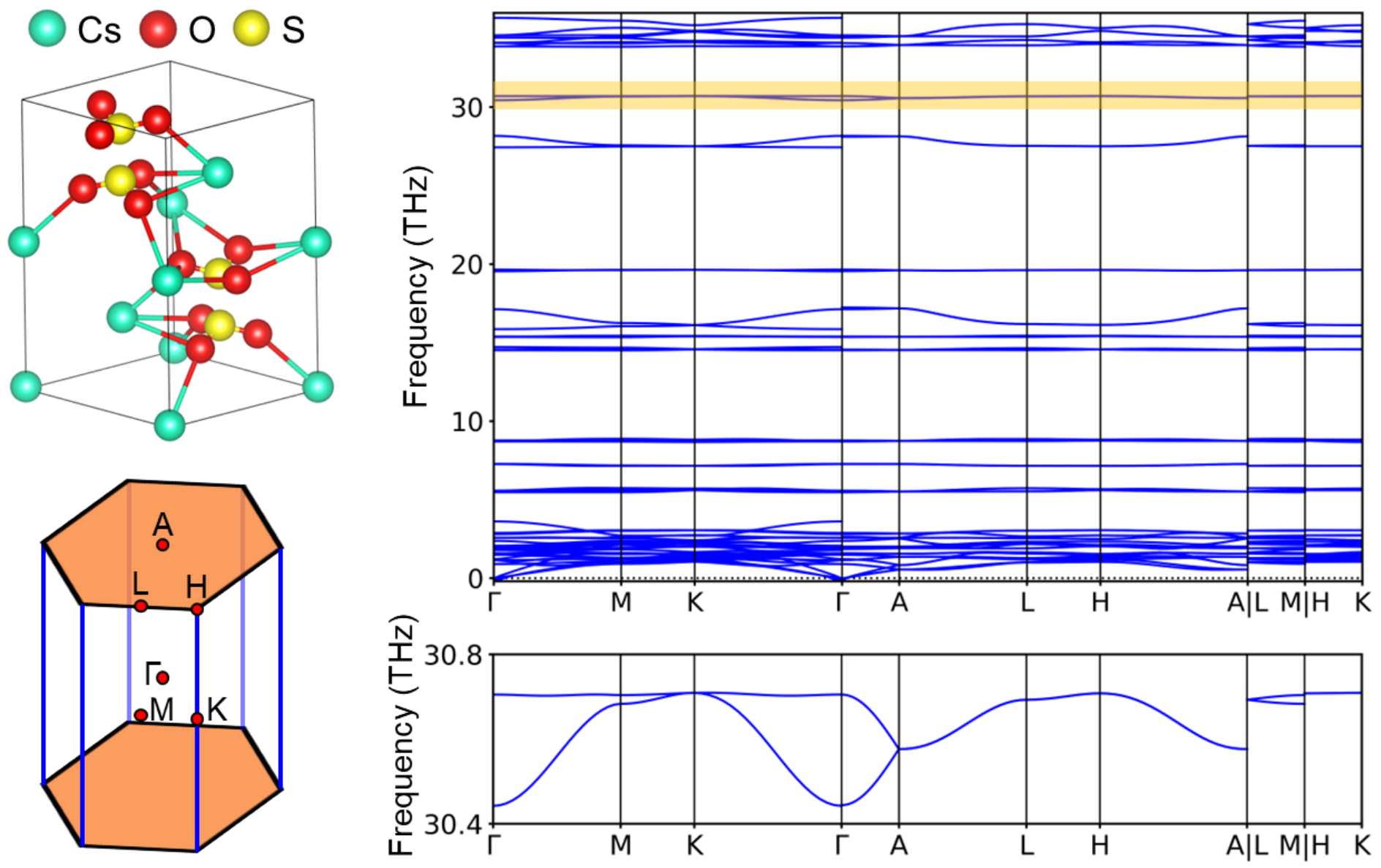

**Figure S24.** The 1NS+WNL phonon of mp-561681 $CsSO_3$ (SG 186), with LOTO.

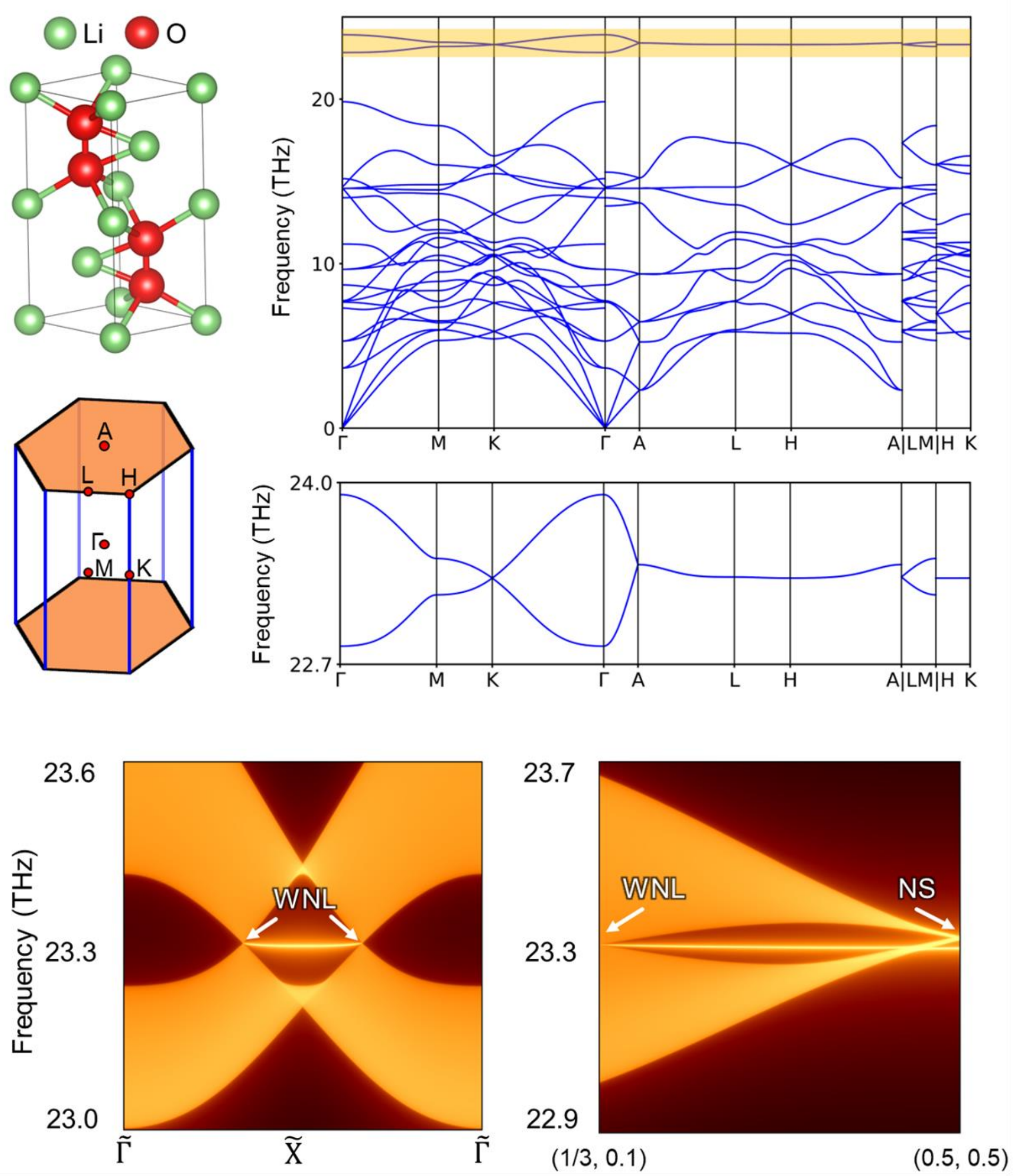

**Figure S25.** The 1NS+WNL phonon of mp-841 $Li_2O_2$ (SG 194), with LOTO. The bottom two panels are (100) surface spectra of $Li_2O_2$. One should notice there only exist WNL-WNL drumhead SSs. The NS cannot serve as termination of SS.

# IV. Four-Band Situations of NS+NL Phonons

This section discusses about NS+NL phonons embedded in four bands. According to the causes of four-band, this section is divided into the following three parts.

## 1. Symmetry-enforced DPs or DNLs

| SG | Symbol | NS | DP or DNL | WNL | QNL | Candidate |
|---|---|---|---|---|---|---|
| 29* | $Pca2_1$ | $NS_{ZURT}$ | U $\{U_1^2 \oplus U_1^2\}$, R $\{R_1^2 \oplus R_1^2\}$ | XS $\{D_1^1 \oplus D_2^1\}$, XU $\{G_1^2\}$, SR $\{Q_1^2\}$ | | mp-721312 $Na_2H_2CO_4$ |
| 33* | $Pna2_1$ | $NS_{ZURT}$ | U $\{U_1^2 \oplus U_1^2\}$ | XS $\{D_1^1 \oplus D_2^1\}$, YS $\{C_1^1 \oplus C_2^1\}$, XU $\{G_1^2\}$, YT $\{H_1^2\}$ | | mp-16996 $Li_3GaSiO_5$ |
| 52 | Pnna | $NS_{ZURT}$ | S $\{S_1^2 \oplus S_2^2\}$ | XS $\{D_1^2\}$, YS $\{C_1^1 \oplus C_4^1$ or $C_2^1 \oplus C_3^1\}$, XU $\{G_1^2\}$ YT $\{H_1^2\}$, SR $\{Q_1^1 \oplus Q_3^1$ or $Q_2^1 \oplus Q_4^1\}$ | | mp-27398 $TlBr_2$ |
| 54 | Pcca | $NS_{ZURT}$ | U $\{U_1^2 \oplus U_2^2\}$, R $\{R_1^2 \oplus R_2^2\}$ | XS $\{D_1^2\}$, SR $\{Q_1^1 \oplus Q_4^1$ or $Q_2^1 \oplus Q_3^1\}$, XU $\{G_1^1 \oplus G_4^1$ or $G_2^1 \oplus G_3^1\}$ | | mp-556666 $CaBiCO_4F$ |
| 56 | Pccn | $NS_{UXS}$, $NS_{TYS}$ | U $\{U_1^2 \oplus U_2^2\}$, T $\{T_1^2 \oplus T_2^2\}$ | ZU $\{A_1^2\}$, ZT $\{B_1^2\}$ | | mp-558350 $Zn_2TeBr_2O_3$ |
| 60 | Pbcn | $NS_{UXS}$, $NS_{TYS}$ | T $\{T_1^2 \oplus T_2^2\}$, UR $\{P_1^2 \oplus P_1^2\}$ | ZU $\{A_1^2\}$, ZT $\{B_1^1 \oplus B_3^1$ or $B_2^1 \oplus B_4^1\}$ | | mp-1666 $Th_2S_5$ |
| 130 | P4/ncc | $2\times NS_{XRM}$ | A $\{A_1^2 \oplus A_2^2$ or $A_3^2 \oplus A_4^2\}$, R $\{R_1^2 \oplus R_2^2\}$ | ZR $\{U_1^2\}$, ZA $\{S_1^2\}$ | **ΓZ $\{LD_5^2\}$** | mp-23947 $Cs_3MgH_5$ |
| 135 | $P4_2/mbc$ | $2\times NS_{XRM}$ | A $\{A_1^2 \oplus A_2^2$ or $A_3^2 \oplus A_4^2\}$ | ZA $\{S_1^2\}$ | **ΓZ $\{LD_5^2\}$** | mp-9880 $YB_2C$ |
| 138 | $P4_2/ncm$ | $2\times NS_{XRM}$ | R $\{R_1^2 \oplus R_2^2\}$ | ZR $\{U_1^2\}$ | **ΓZ $\{LD_5^2\}$** | mp-754761 $Sr_2HfO_4$ |
| 205 | Pa-3 | $3\times NS_{XRM}$ | MR $\{T_1^2 \oplus T_1^2\}$ | ΓR $\{LD_2^1 \oplus LD_3^1\}$ | | mp-730 $PtP_2$ |

**Table S1. Four-band situation-I: symmetry-enforced DPs or DNLs.** The asterisks label non-centrosymmetric SGs. The QNLs in bold lines are not symmetry-enforced, their emergence depends on particular basis functions.

Table S1 shows the first situation that there exist symmetry-enforced DPs or DNLs, which means the elementary band representations for listed SGs are at least four-band.

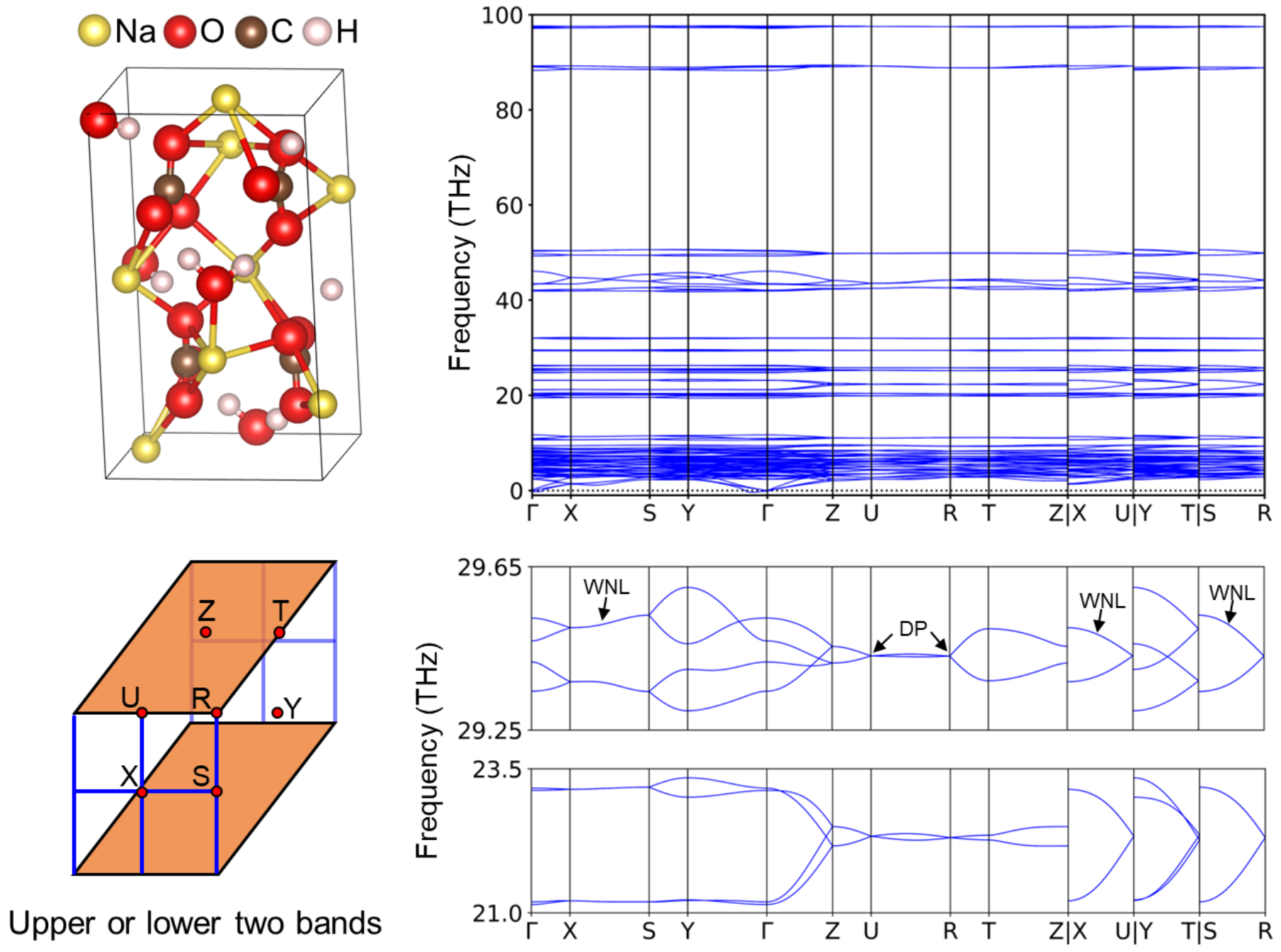

**Figure S26.** The 1NS+WNN phonons in mp-721312 $Na_2H_2CO_4$ (SG 29), without LOTO.

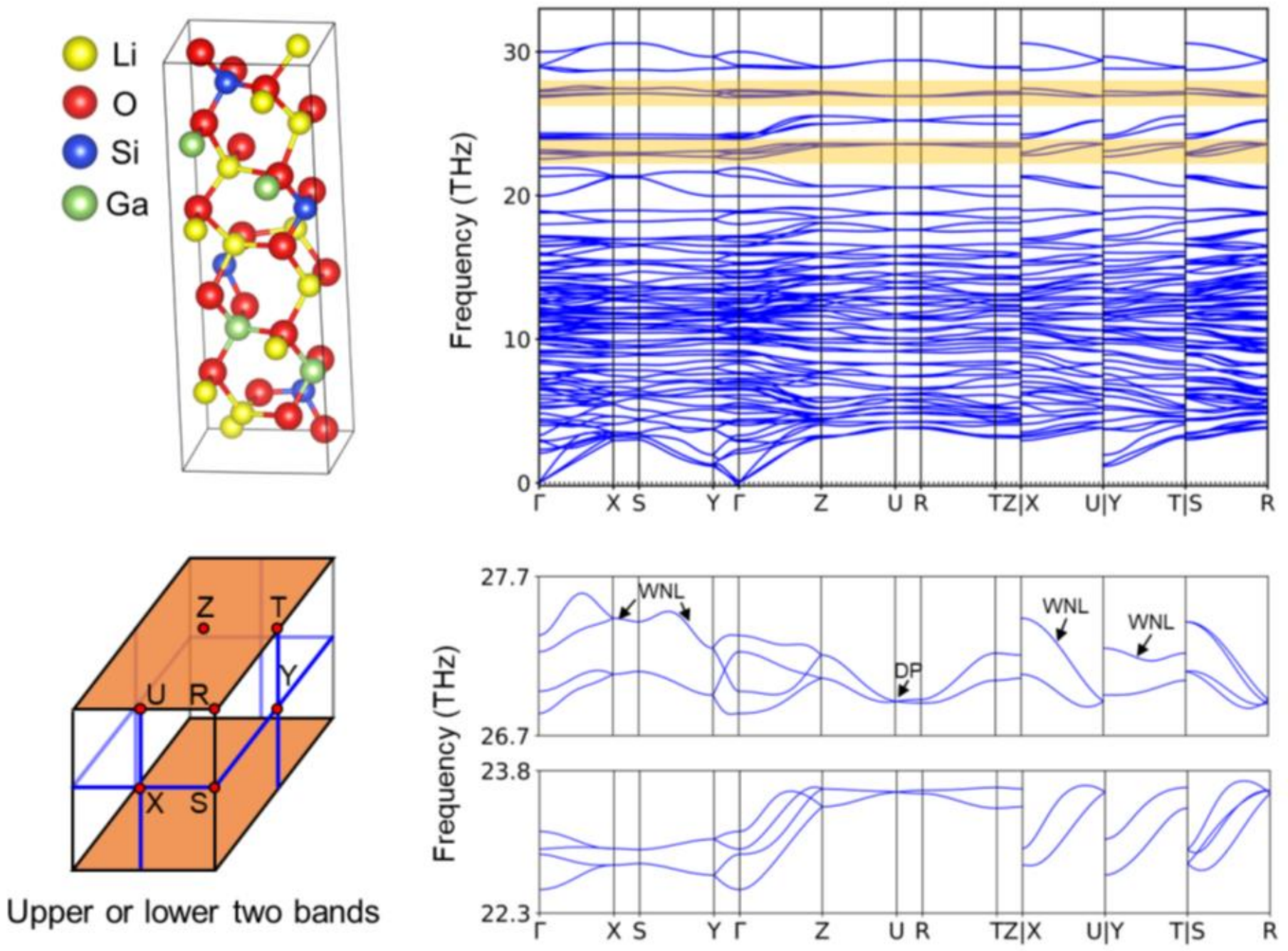

**Figure S27.** The 1NS+WNN phonons in mp-16996 $Li_3GaSiO_5$ (SG 33), without LOTO.

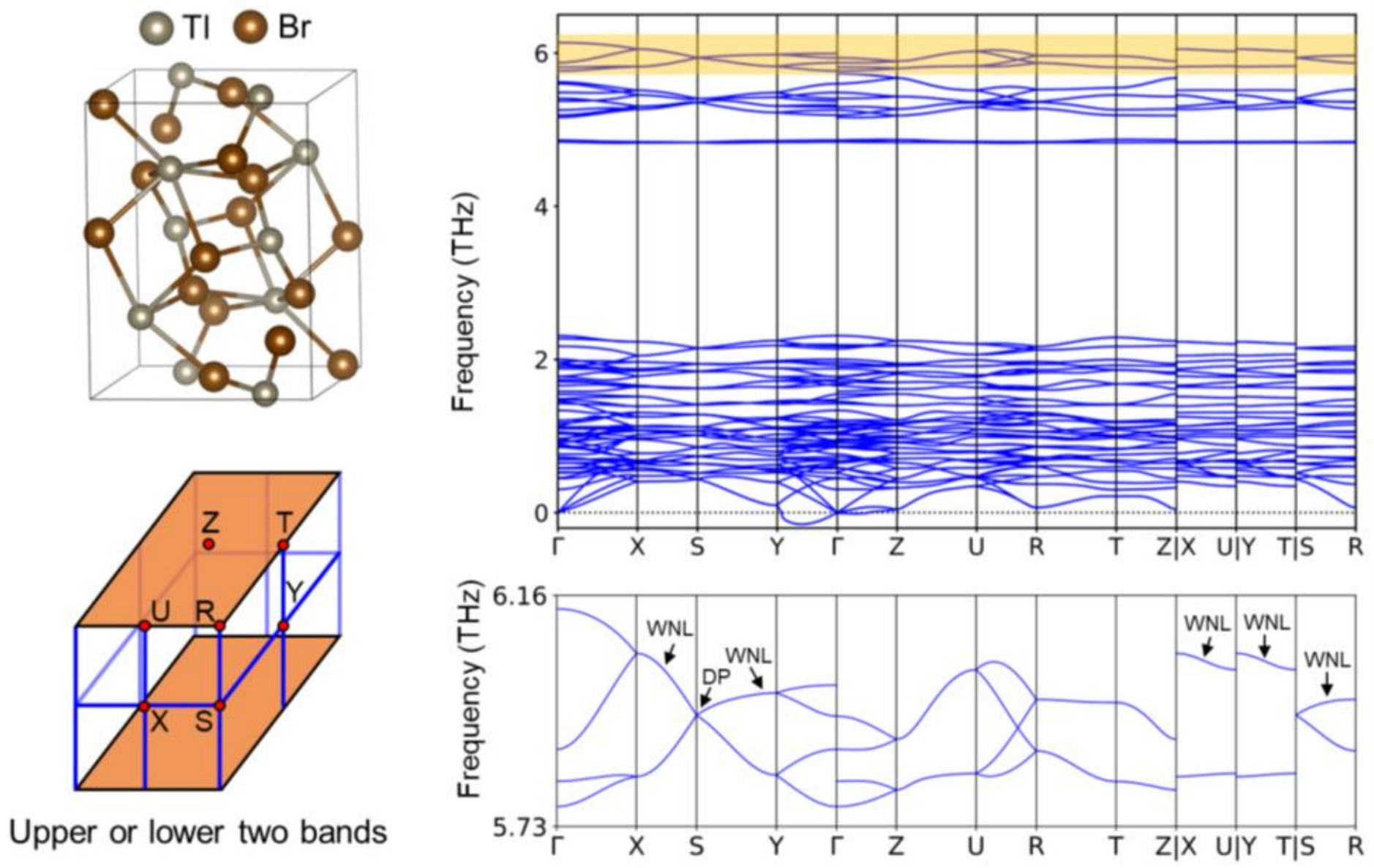

**Figure S28.** The 1NS+WNN phonons in mp-27398 $TlBr_2$ (SG 52), with LOTO.

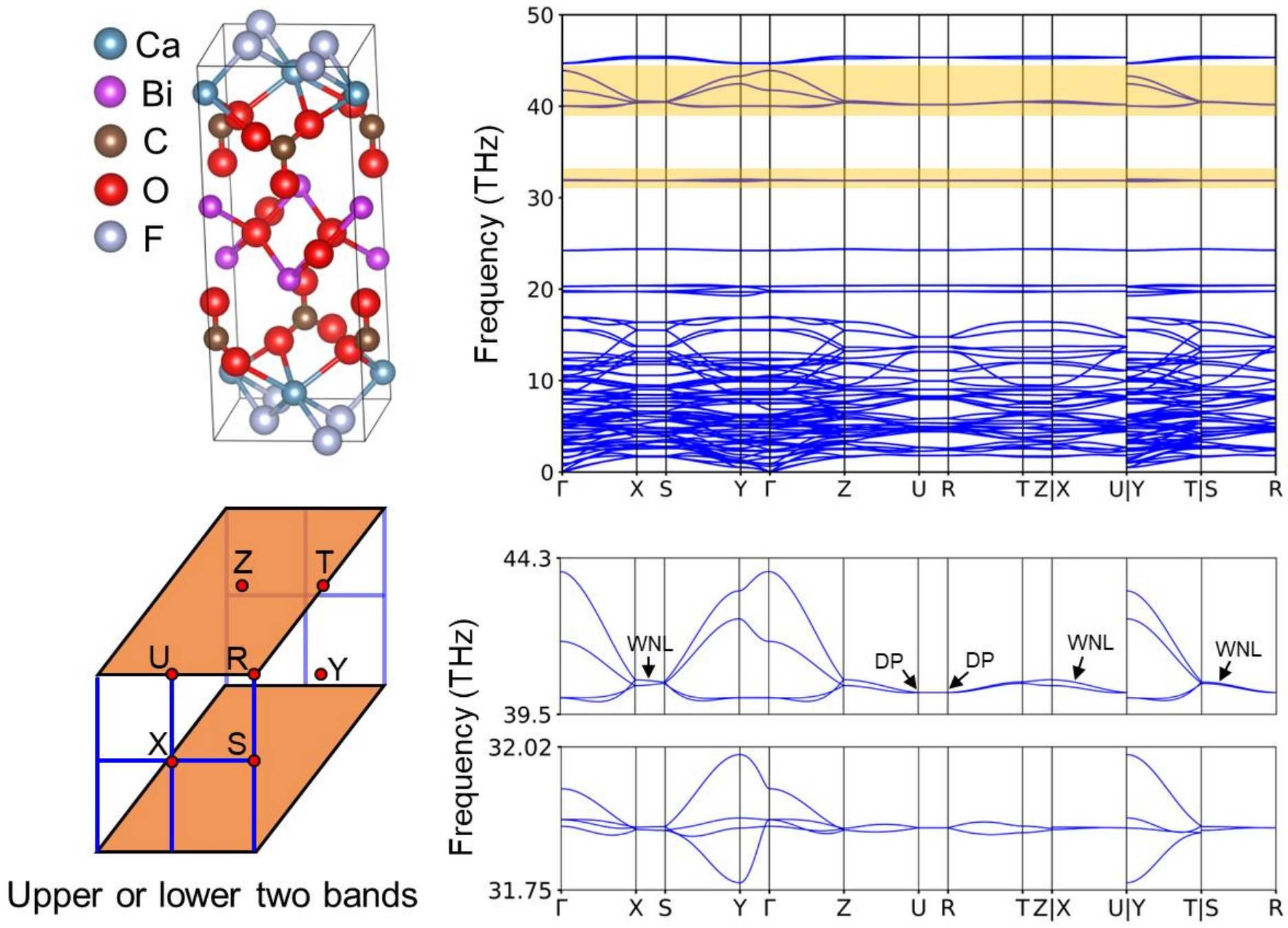

**Figure S29.** The 1NS+WNN phonons in mp-556666 $CaBiCO_4F$ (SG 54), without LOTO.

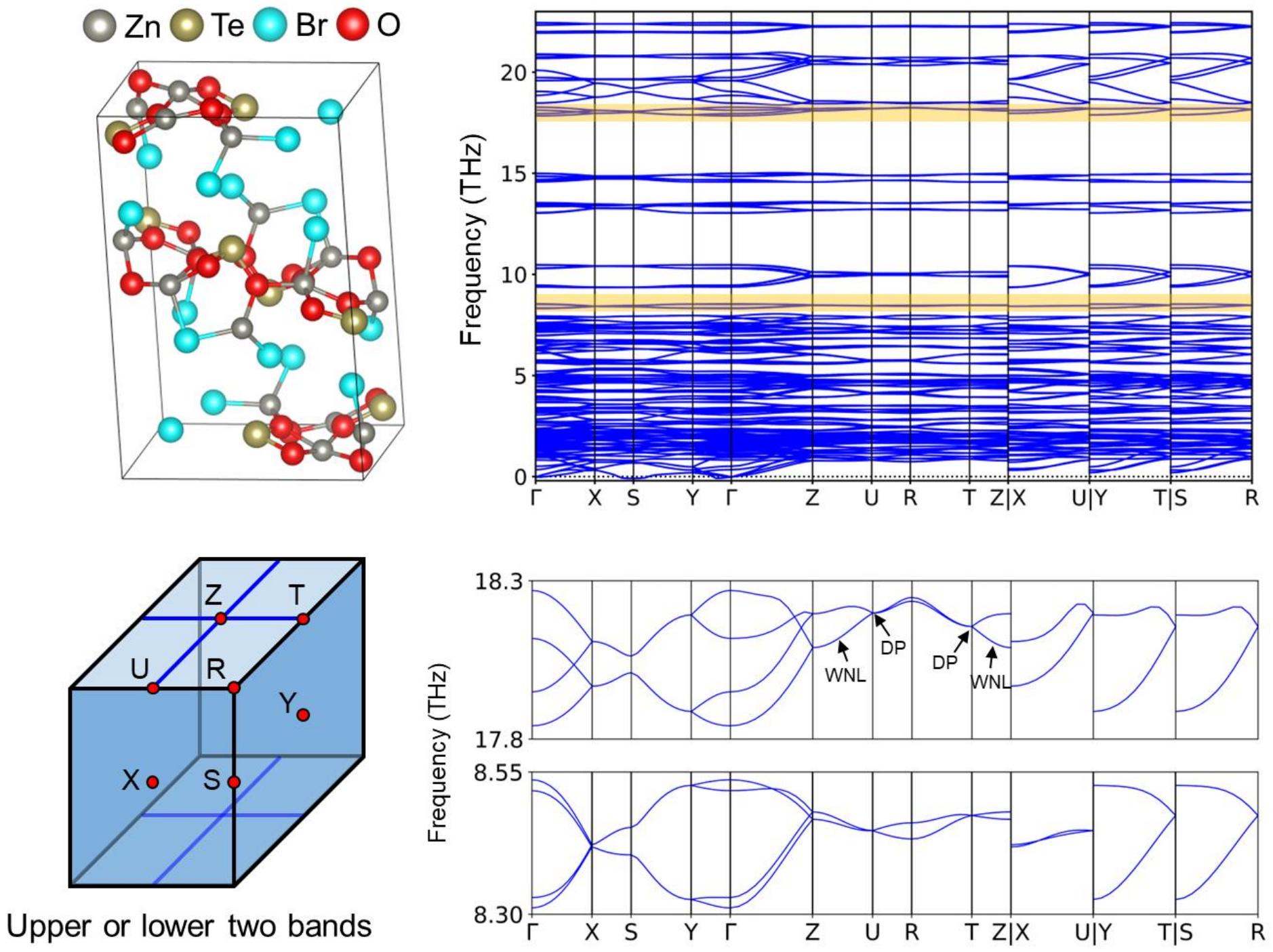

**Figure S30.** The 2NS+WNN phonons in mp-558350 $Zn_2TeBr_2O_3$ (SG 56), without LOTO.

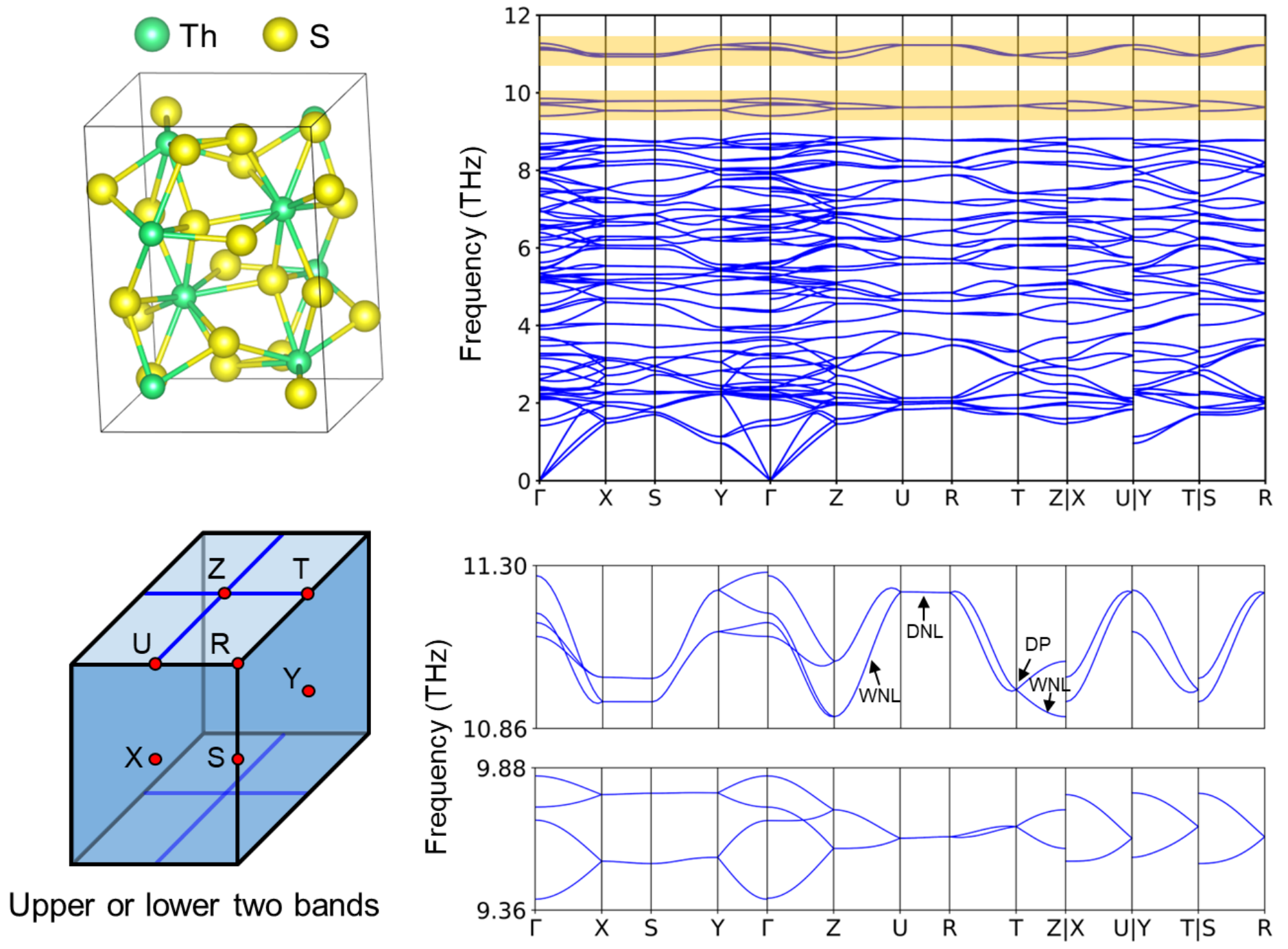

**Figure S31.** The 2NS+WNN phonons in mp-1666 $Th_2S_5$ (SG 60), with LOTO.

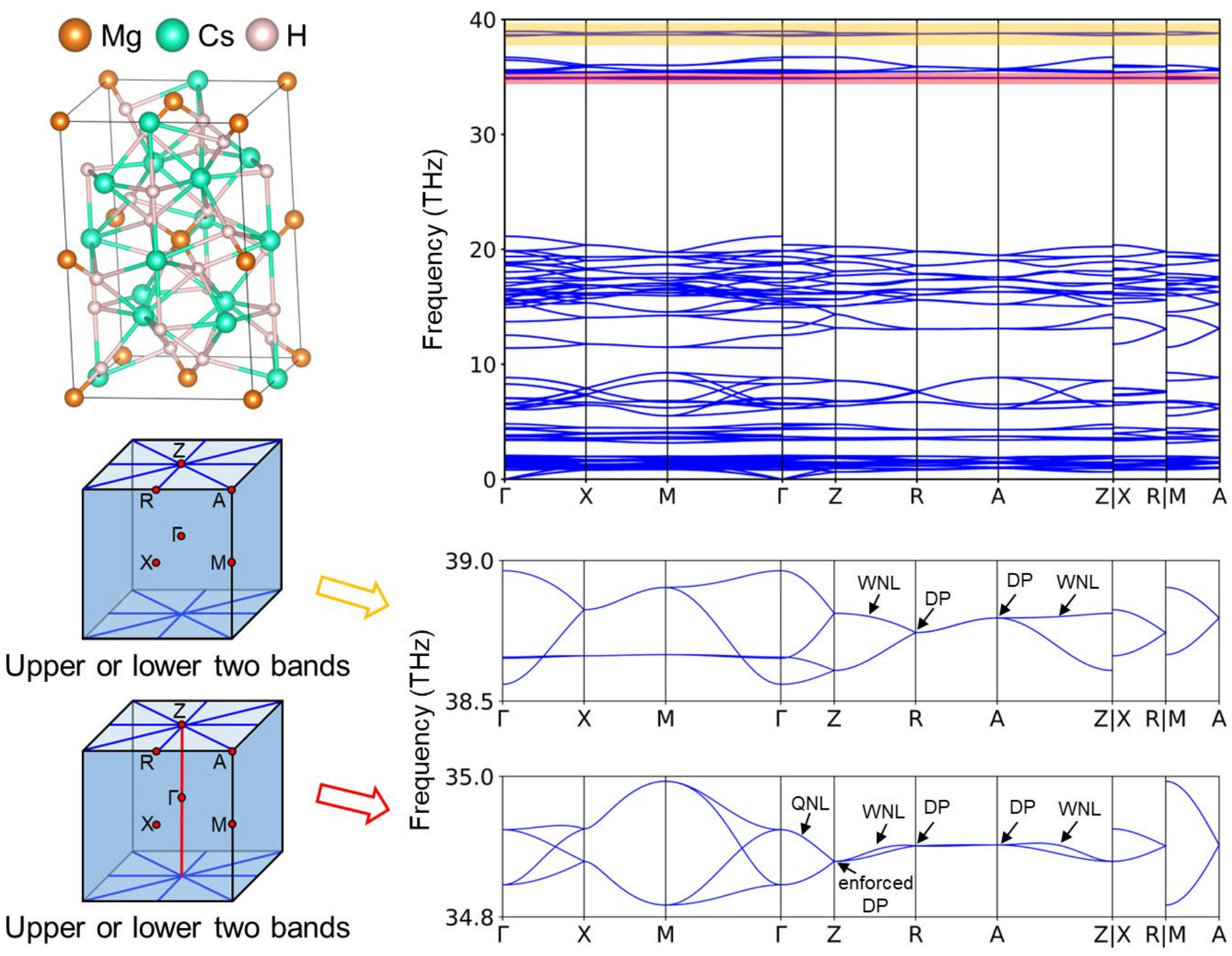

**Figure S32.** The 2NS+WNN and 2NS+HNN phonons in mp-23947 $Cs_3MgH_5$ (SG 130), with LOTO.

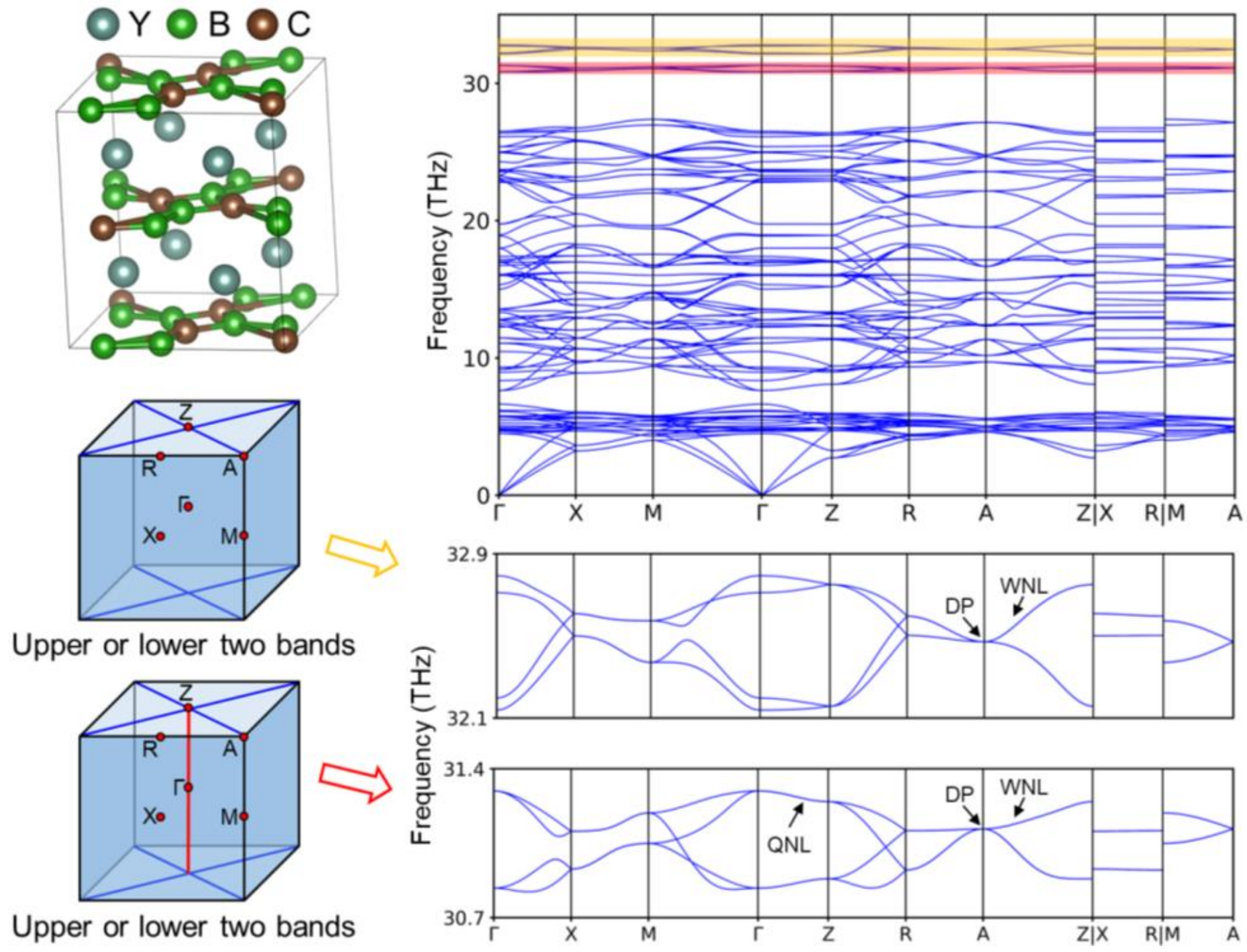

**Figure S33.** The 2NS+WNN and 2NS+HNN phonons in mp-9880 $YB_2C$ (SG 135), without LOTO.

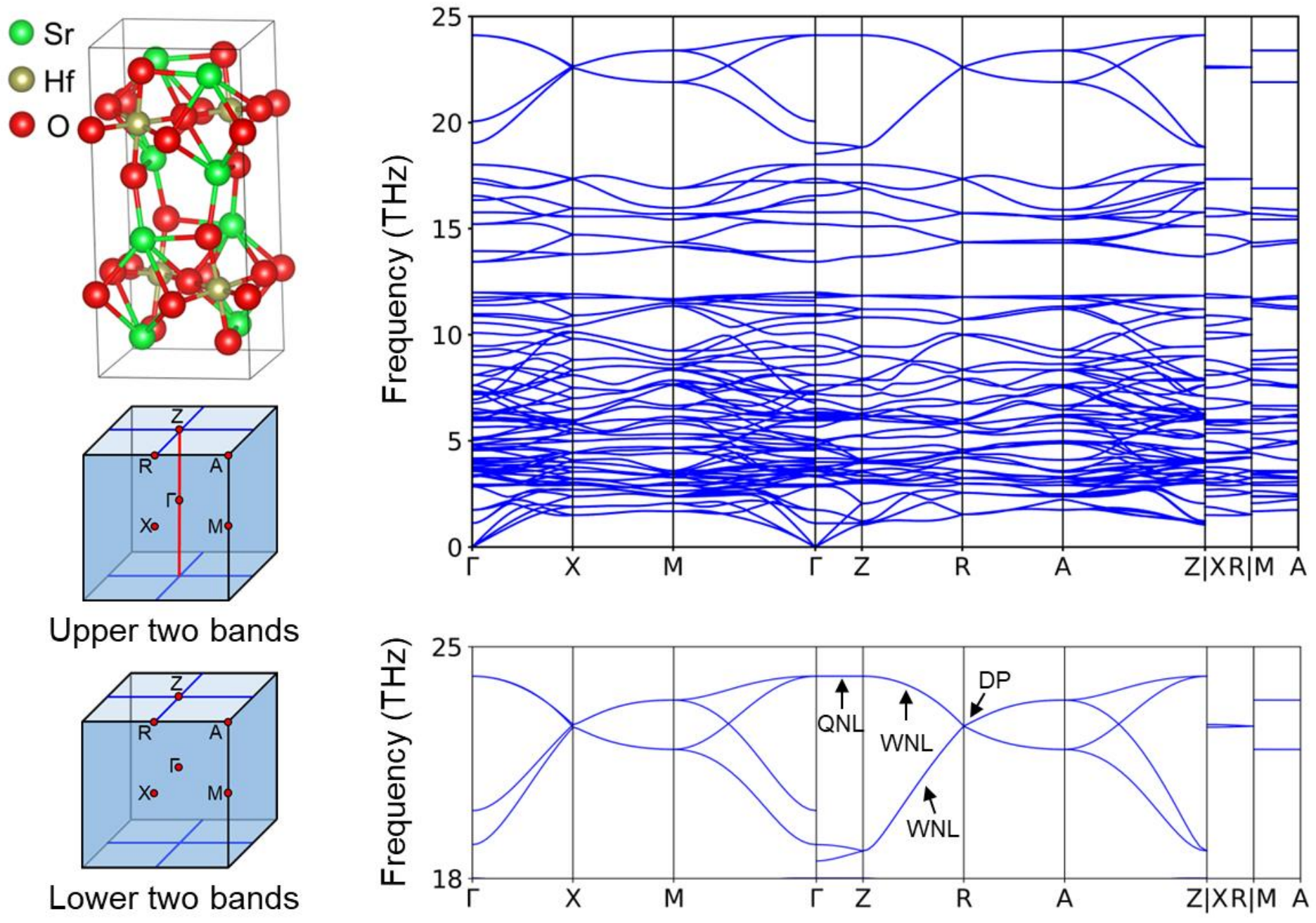

**Figure S34.** The 2NS+WNN and 2NS+HNN phonons in mp-754761 $Sr_2HfO_4$ (SG 138), with LOTO.

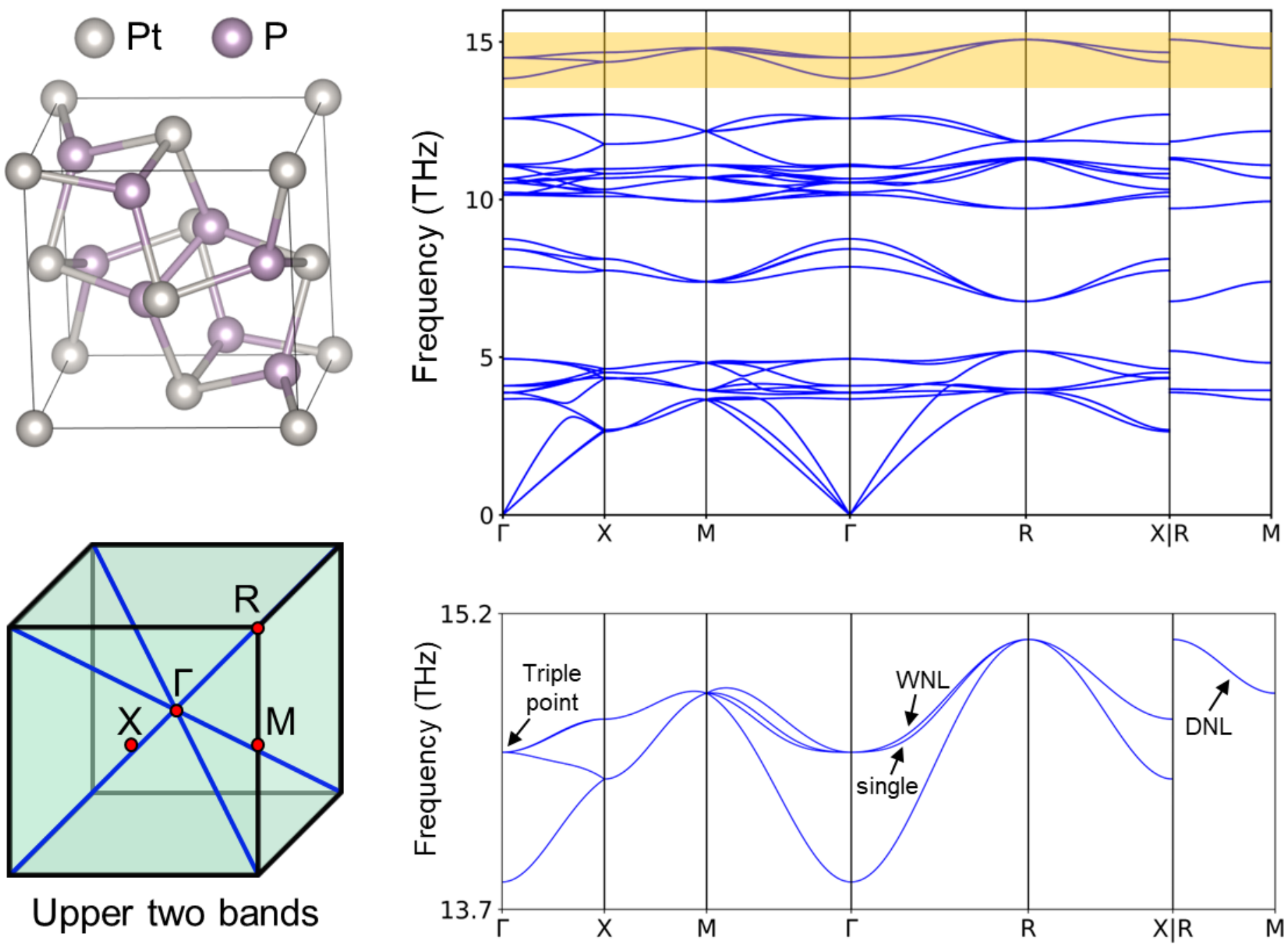

**Figure S35.** The 3NS+WNN phonons in mp-730 $PtP_2$ (SG 205), with LOTO.

## 2. DPs induced by particular NLs

| SG | Symbol | NS | WNL with enforced DP | QNL with enforced DP | Candidate |
|---|---|---|---|---|---|
| 128 | P4/mnc | 2×$NS_{XRM}$ | | ΓZ {$LD_5^2$ and $LD_5^2$} with Z {$Z_3^2 \oplus Z_4^2$} | mp-24822 $Na_5Al_3H_{14}$ |
| 176 | $P6_3/m$ | $NS_{ALH}$ | | ΓA {$DT_3^1 \oplus DT_5^1$ and $DT_2^1 \oplus DT_4^1$} with A {$A_2^2 \oplus A_3^2$} | mp-27506 $K_3W_2Cl_9$ |
| 185* | $P6_3cm$ | $NS_{ALH}$ | KH {$P_3^2$ and $P_3^2$} with H {$H_3^2 \oplus H_3^2$} | ΓA {$DT_5^2$ and $DT_6^2$} with A {$A_5^2 \oplus A_6^2$} | mp-14385 $Ba_3TmB_3O_9$ |
| 186* | $P6_3mc$ | $NS_{ALH}$ | | ΓA {$DT_5^2$ and $DT_6^2$} with A {$A_5^2 \oplus A_6^2$} | mp-29116 $Ta_3SeI_7$ |
| 193 | $P6_3/mcm$ | $NS_{ALH}$ | KH {$P_3^2$ and $P_3^2$} with H {$H_5^2 \oplus H_6^2$} | ΓA {$DT_5^2$ and $DT_6^2$} with A {$A_3^4$} | mp-16595 $K_3DySi_2O_7$ |
| 194 | $P6_3/mmc$ | $NS_{ALH}$ | | ΓA {$DT_5^2$ and $DT_6^2$} with A {$A_3^4$} | mp-862693 $BaNaH_3Pd$ |

**Table S2. Four-band situation-II: NL-induced DPs.** The asterisks label non-centrosymmetric SGs. Specific WNLs or QNLs should emerge in pairs (linked by "and" in the brackets), so as to generate additional four-fold DPs.

For SGs listed in Table S2, when particular NL appears, this NL must come in pairs with another conjugated NL, and produce quadruple degenerate DP at one endpoint of the NL. Thus, this second four-band situation is called as NL-induced DPs.

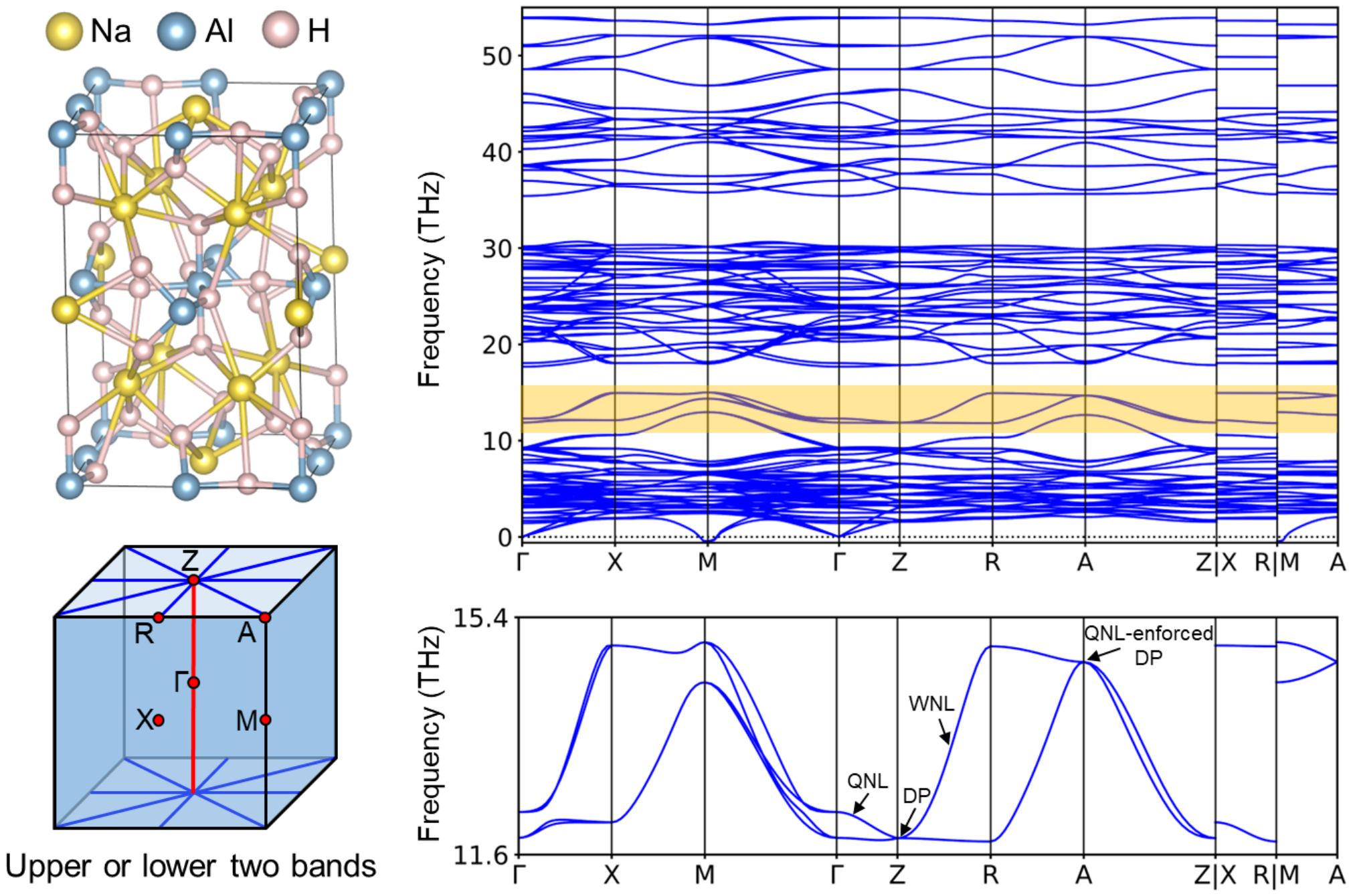

**Figure S36.** The 2NS+HNN phonons in mp-24822 $Na_5Al_3H_{14}$ (SG 128), without LOTO.

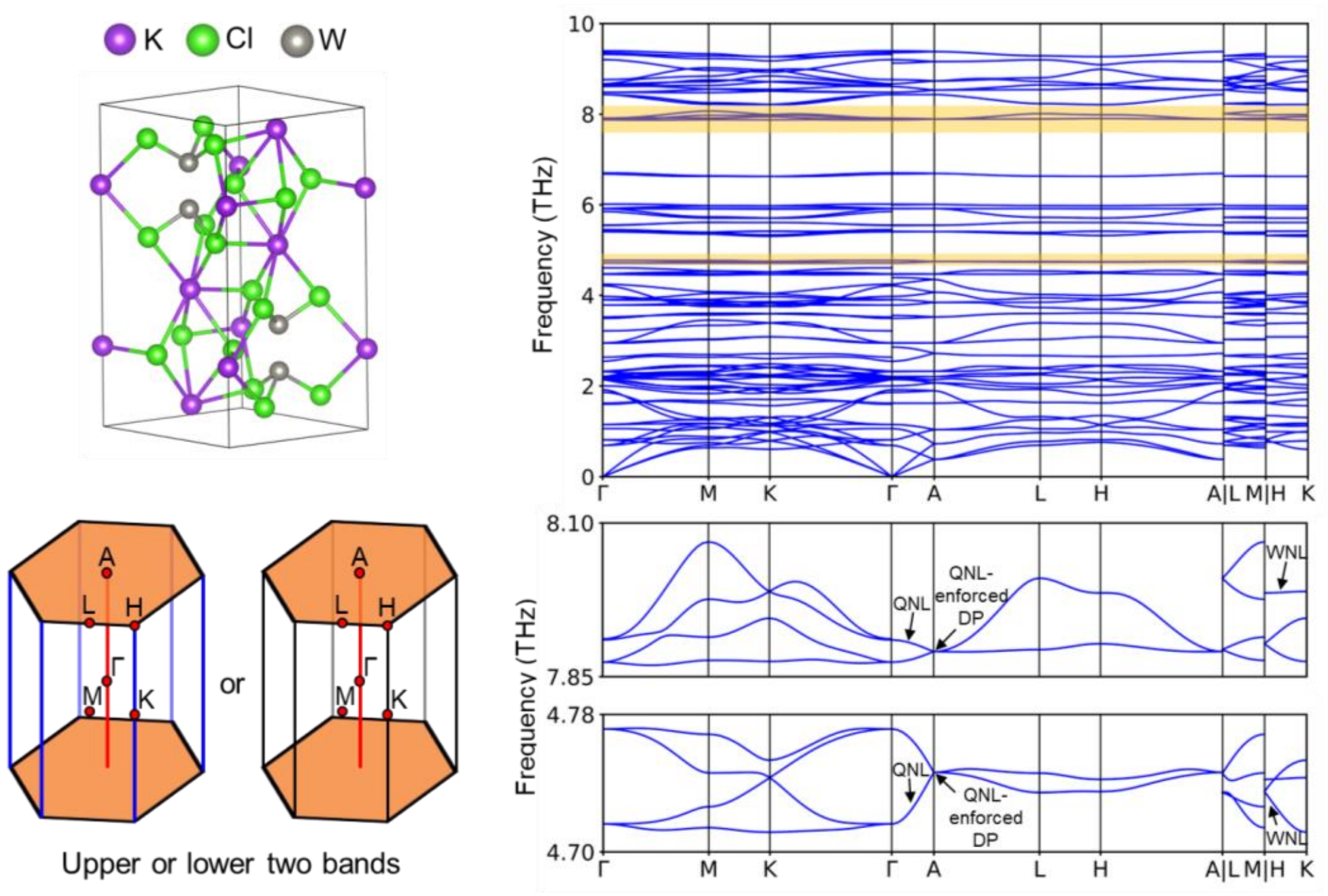

**Figure S37.** The 1NS+QNL+WNL and 1NS+QNL phonons in mp-27506 $K_3W_2Cl_9$ (SG 176), with LOTO.

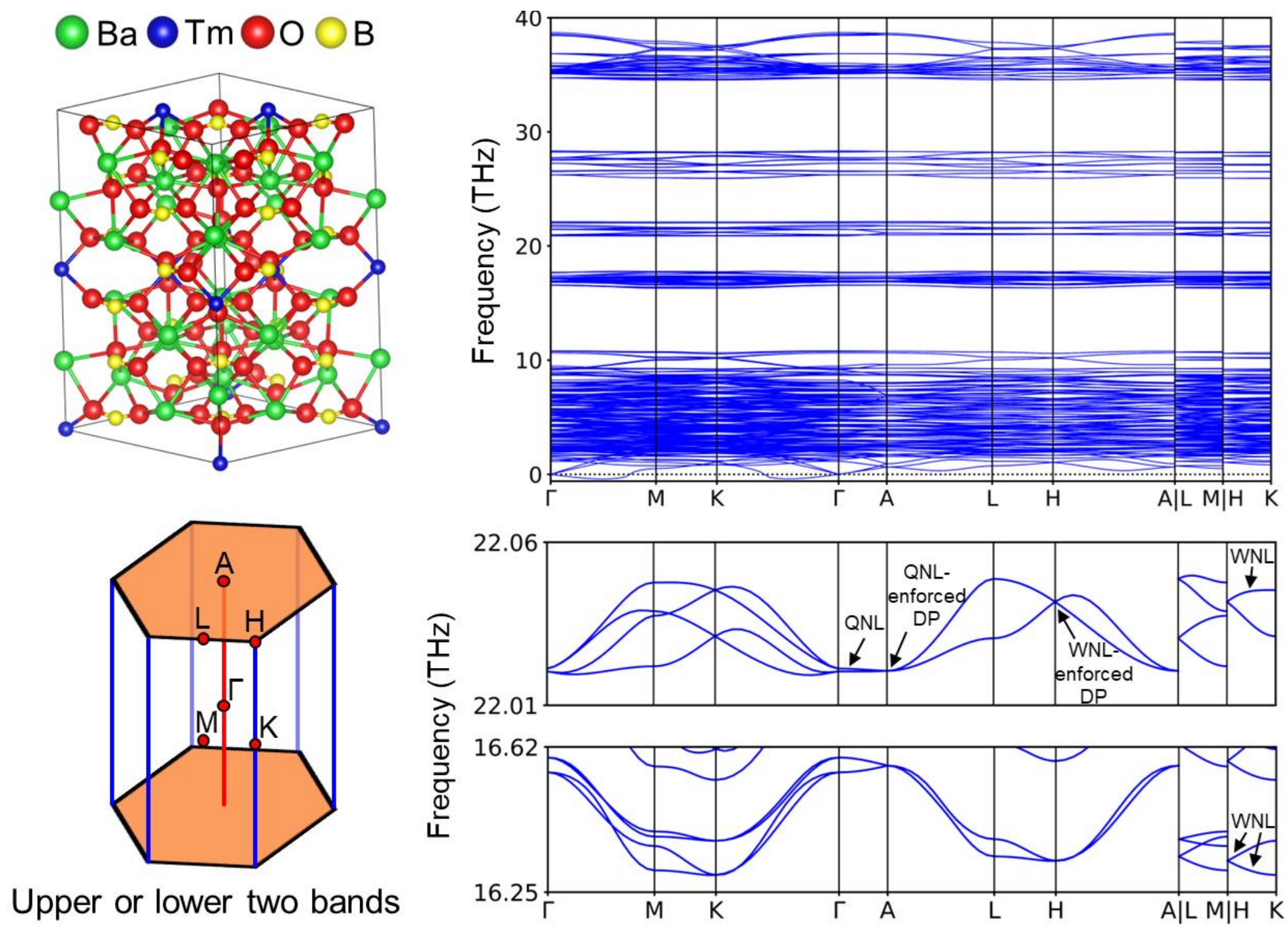

**Figure S38.** The 1NS+QNL+WNL phonons in mp-14385 $Ba_3TmB_3O_9$ (SG 185), without LOTO.

Figure S39 shows the NL-induced DP of SG186 $Ta_3SeI_7$. For phonon bands 61-64 of $Ta_3SeI_7$, a pair of QNLs, i.e., QNL1 and QNL2 along Γ-A path, induce the DP at A point. The irreducible representation of A point is of $A_5^2 \oplus A_6^2$. In (100) surface, the DP is projected to $\widetilde{A}$ point, and QNL1/QNL2 are projected to $\widetilde{\Gamma}$-$\widetilde{A}$ path. Figure S39(e) shows the double surface arcs of the DP. Figure S39(f) and Figure S39(g) shows the drumhead SSs of QNL1 and QNL2, respectively.

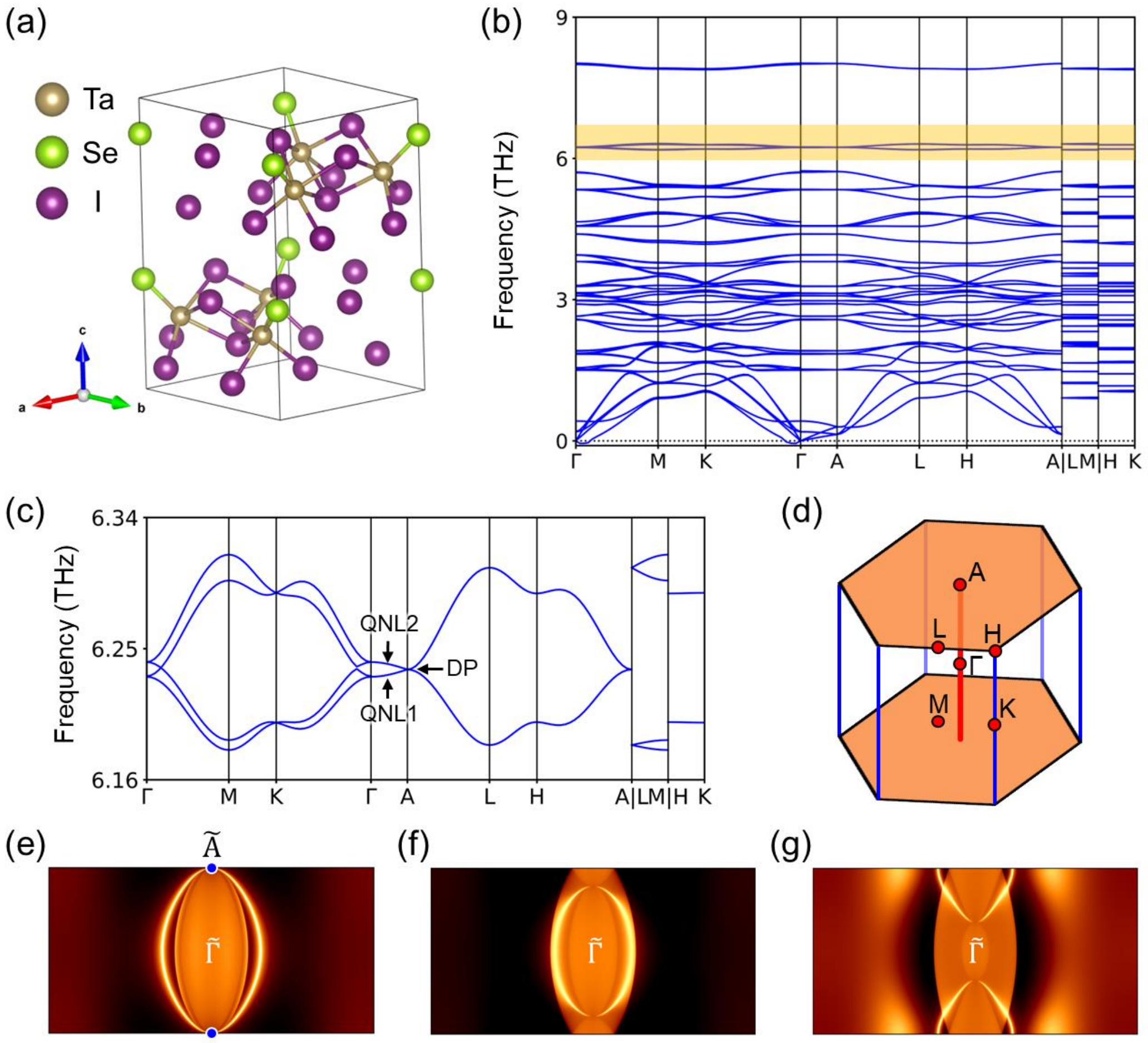

**Figure S39.** The 1NS+QNL+WNL phonons in mp-29116 $Ta_3SeI_7$ (SG 186). (a) Crystal structure of $Ta_3SeI_7$. (b) Phonon dispersions of $Ta_3SeI_7$ with LOTO correction. (c) Bands 61-64, as marked by yellow in (b). Here, QNL1/QNL2 stand for the lower/upper QNL. (d) Schematic of the Brillouin zone and the 1NS+WNL+QNL phonon of bands 61,62 (or bands 63,64). The iso-frequency contours of (100) surface at (e) 6.23564 THz; (f) 6.234 THz and (g) 6.24 THz.

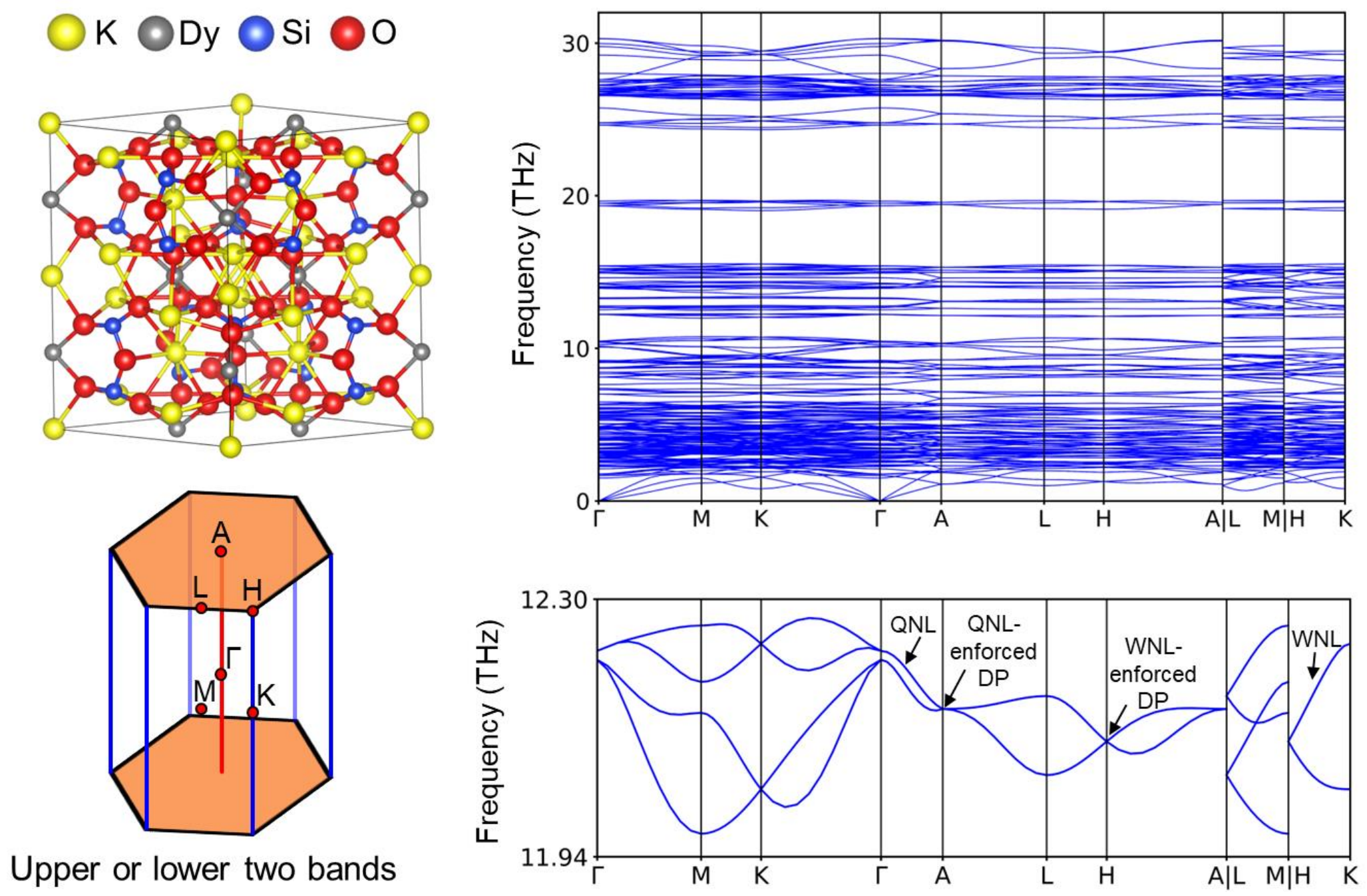

**Figure S40.** The 1NS+QNL+WNL phonons in mp-16595 $K_3DySi_2O_7$ (SG 193), without LOTO.

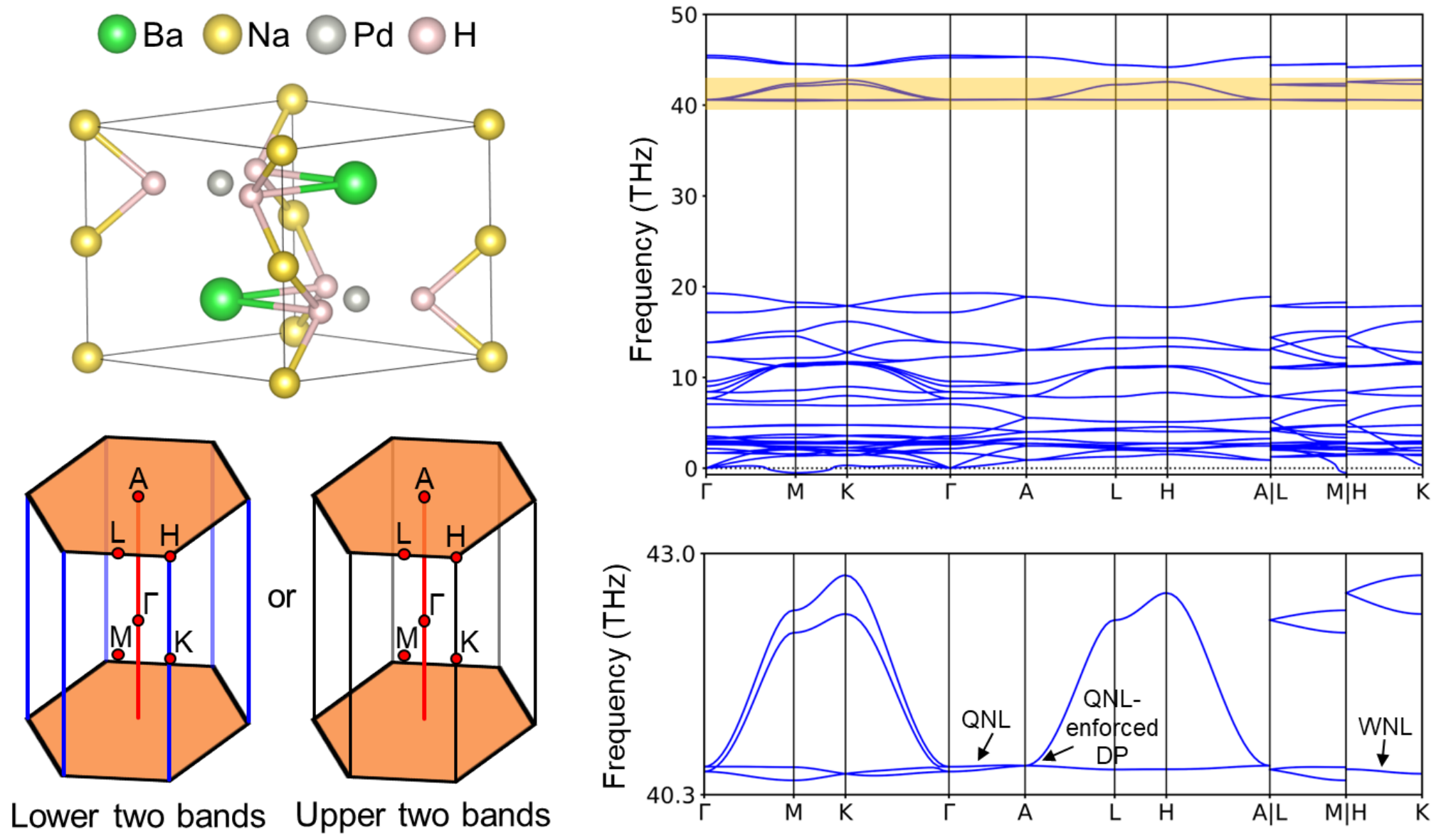

**Figure S41.** The 1NS+QNL+WNL and 1NS+QNL phonons in mp-862693 $BaNaH_3Pd$ (SG 194), without LOTO.

## 3. Hourglass phonons caused by CR violations

| SG | Symbol | NS | QNL | Violated CRs | Candidate |
|---|---|---|---|---|---|
| 114* | $P\text{-}42_1c$ | 2×$NS_{XRM}$ | ΓZ $\{LD_3^2 \oplus LD_4^2$ and $LD_3^2 \oplus LD_4^2\}$ with Z $\{Z_1^1 \oplus Z_2^1$ and $Z_3^1 \oplus Z_4^1\}$ | **ZR**: $R_1^2 \Rightarrow U_1^1 \oplus U_2^1$; $Z_1^1 \oplus Z_2^1 \Rightarrow 2U_2^1$ or $Z_3^1 \oplus Z_4^1 \Rightarrow 2U_1^1$ | mp-27144 $TlPO_3$ |
| 129 | P4/nmm | 2×$NS_{XRM}$ | ΓZ $\{LD_5^2$ and $LD_5^2\}$ with Γ $\{GM_{5+}^2$ and $GM_{5-}^2\}$ and Z $\{Z_{5+}^2$ and $Z_{5-}^2\}$ | **ΓX**: $X_1^2 \Rightarrow DT_1^1 \oplus DT_3^1$ or $X_2^2 \Rightarrow DT_2^1 \oplus DT_4^1$; $GM_{5+}^2 \Rightarrow DT_2^1 \oplus DT_3^1$ or $GM_{5-}^2 \Rightarrow DT_1^1 \oplus DT_4^1$ **ΓM**, **ZR** and **ZA** as well | mp-15844 $TaPO_5$ |
| 137 | $P4_2/nmc$ | 2×$NS_{XRM}$ | ΓZ $\{LD_5^2$ and $LD_5^2\}$ with Γ $\{GM_{5+}^2$ and $GM_{5-}^2\}$ and Z $\{Z_3^2$ and $Z_4^2\}$ | **ΓX**: $X_1^2 \Rightarrow DT_1^1 \oplus DT_3^1$ or $X_2^2 \Rightarrow DT_2^1 \oplus DT_4^1$; $GM_{5+}^2 \Rightarrow DT_2^1 \oplus DT_3^1$ or $GM_{5-}^2 \Rightarrow DT_1^1 \oplus DT_4^1$ **ΓM**, **ZR** as well | mp-559639 $K4UP_2O_{10}$ |

**Table S3. Four-band situation-III: hourglass phonons due to the violations of two-band CRs.** The asterisks label non-centrosymmetric SGs. The fourth column shows the QNLs and particular irreducible representations of Γ point or Z point which induce the violation of CRs. The fifth column gives the high-symmetry lines (marked in bold) where hourglass phonons emerge. Here, the CRs (⇒) of both two endpoints are listed and "or" links all possible cases.

Table S3 shows the last four-band situation, i.e., hourglass phonons caused by CR violations. For SG 114, SG 129 and SG 137, when QNL appears, the two-band CR should be violated along some high symmetry paths, resulting in the hourglass phonon. Taking the Γ-X path of SG 129 as an example, if the Γ-Z path is a QNL, then CR allowed decompositions are: $GM_{5+}^2 \Rightarrow DT_2^1 \oplus DT_3^1$ or $GM_{5-}^2 \Rightarrow DT_1^1 \oplus DT_4^1$ for Γ point, and $X_1^2 \Rightarrow DT_1^1 \oplus DT_3^1$ or $X_2^2 \Rightarrow DT_2^1 \oplus DT_4^1$ for X point. Since the CRs for Γ point and X point are inconsistent, a higher four-band representation is needed. Besides, we notice a recent work has reported the hourglass phonons in SG 129 [9].

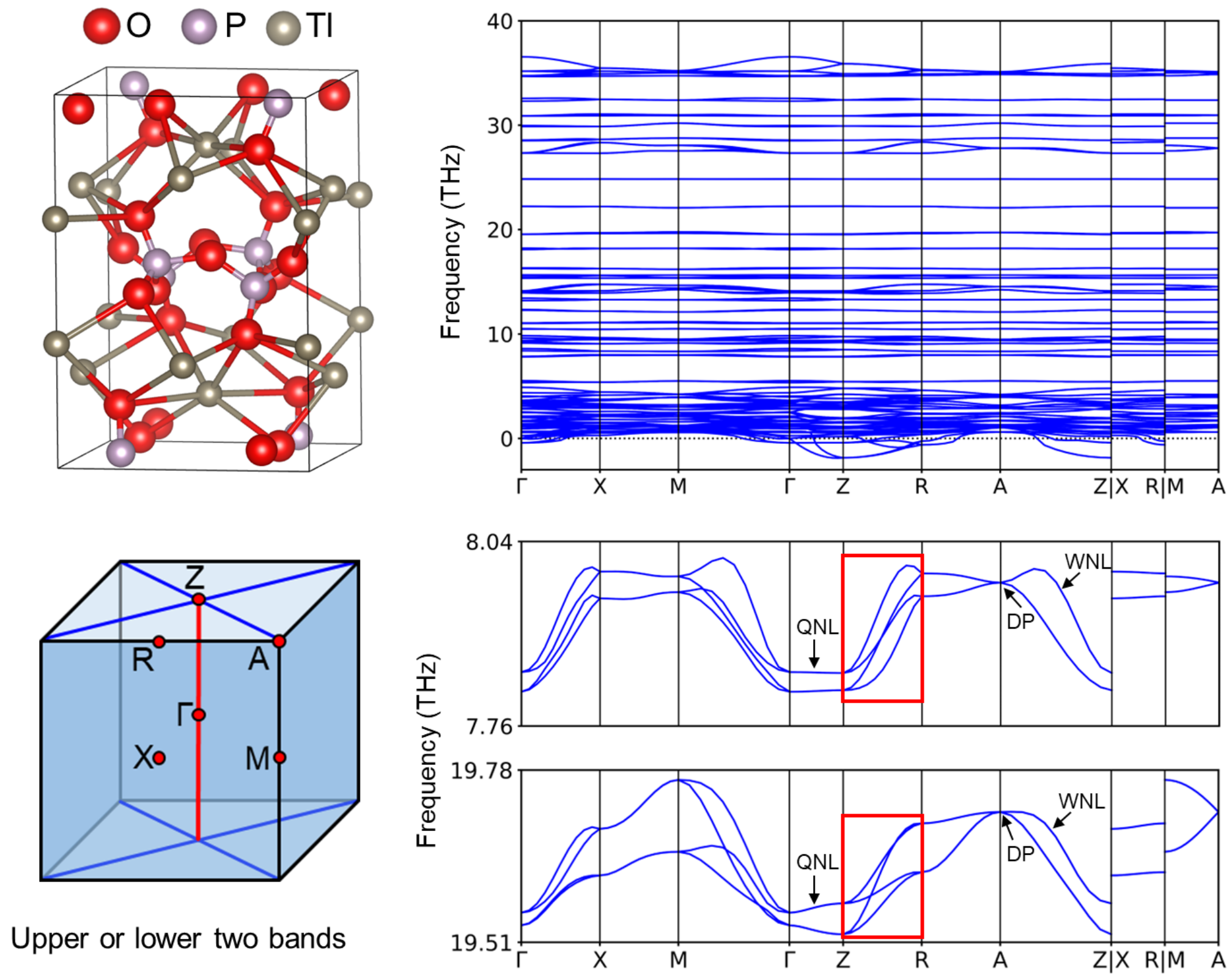

**Figure S42.** The 2NS+HNN phonons in mp-27144 $TlPO_3$ (SG 114), without LOTO. The red boxes mark the hourglass phonons. The DPs are caused by an irreducible representation of $A_5^2 \oplus A_5^2$.

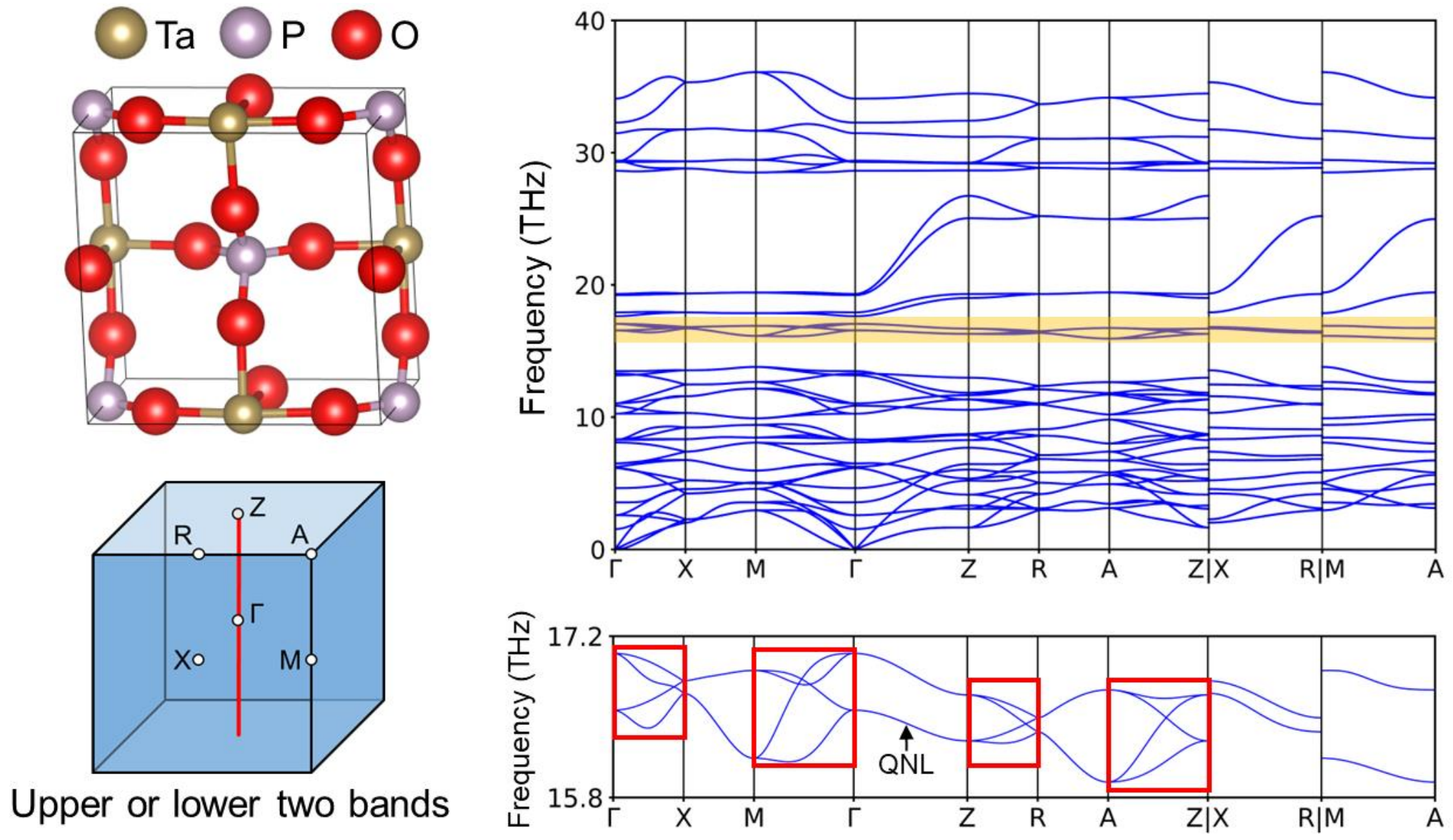

**Figure S43.** The 2NS+QNL phonons in mp-15844 $TaPO_5$ (SG 129), without LOTO. The red boxes mark the hourglass phonons.

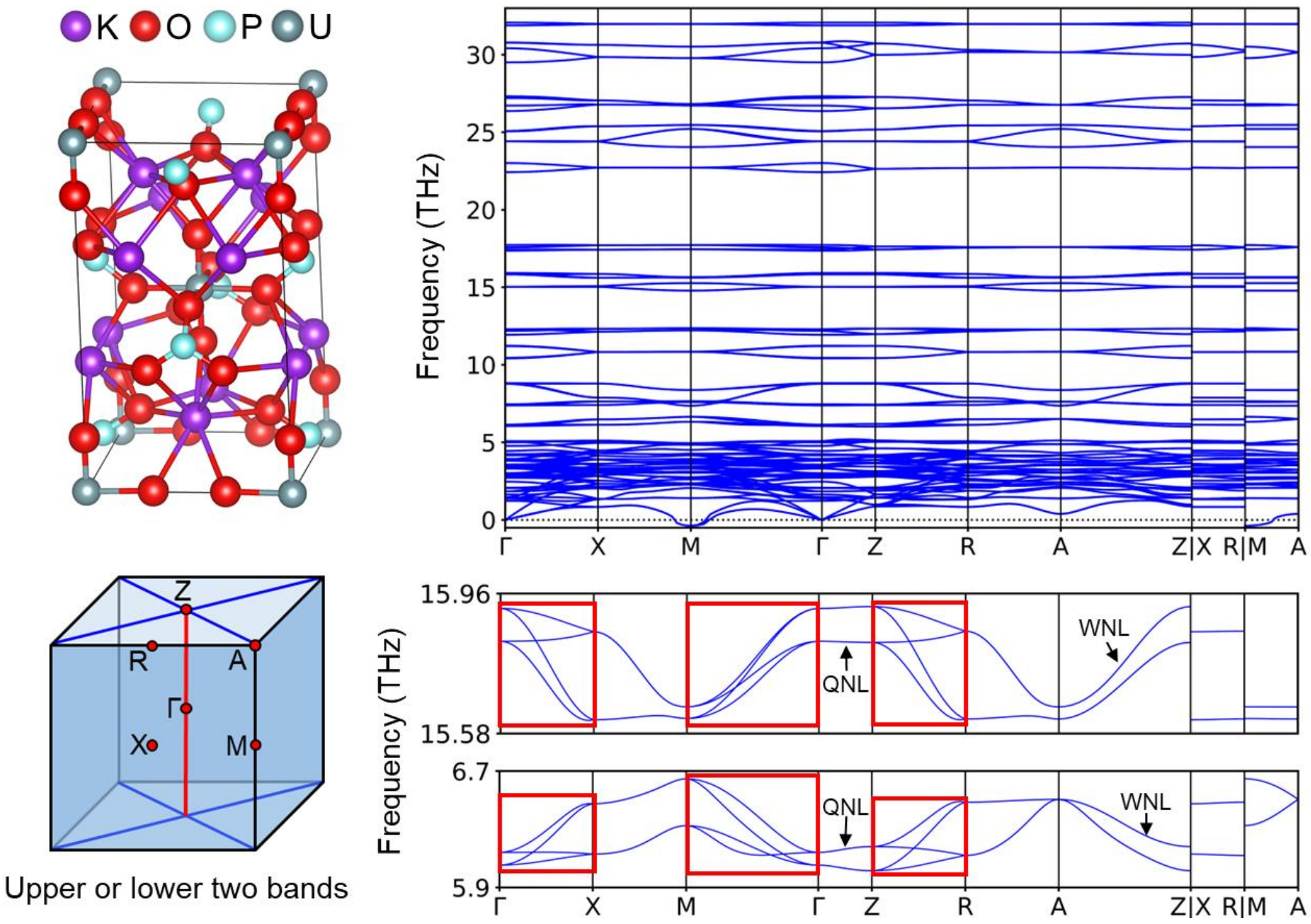

**Figure S44.** The 2NS+HNN phonons in mp-559639 $K_4UP_2O_{10}$ (SG 137), without LOTO. The red boxes mark the hourglass phonons.